\documentclass{aa}
\usepackage[varg]{txfonts}
\usepackage{natbib}
\def\gta{\;\lower 0.5ex\hbox{$\buildrel > \over \sim$}}
\def\lta{\;\lower 0.5ex\hbox{$\buildrel < \over \sim$}}
\def\ebf{}
\usepackage[section]{placeins} 
\usepackage{hyperref}
\usepackage{subfi gure}
\defcitealias{Nichols2014}{NRJ14}
 
\titlerunning{Post-infall evolution of satellites}
\authorrunning{Nichols~et~al.}

\begin{document}

\title{The post-infall evolution of a satellite galaxy}
\author{Matthew~Nichols \inst{\ref{inst1}\thanks{\email{matthew.nichols@epfl.ch}}}
\and
Yves~Revaz \inst{\ref{inst1}}
\and
Pascale~Jablonka \inst{\ref{inst1},\ref{inst2}}}

\institute{Laboratoire d'Astrophysique, \'Ecole Polytechnique F\'ed\'erale de Lausanne (EPFL), 1290, Sauverny, Switzerland \label{inst1}
    \and
    GEPI, Observatoire de Paris, CNRS UMR 8111, Universit\'e Paris Diderot, 92125, Meudon Cedex, France \label{inst2}}
\date{}
\abstract{
We describe a new method that only focuses  on the local region surrounding an infalling dwarf in an effort to understand how the hot baryonic halo  alters the chemodynamical evolution of dwarf galaxies.
Using this method, we examine how a dwarf, similar to Sextans dwarf spheroidal, evolves in the corona of a Milky Way-like galaxy.
We find that even at high perigalacticons the synergistic interaction between ram pressure and tidal forces transform a dwarf into a stream, suggesting that Sextans was much more massive in the past to have survived its perigalacticon passage.
In addition, the large confining pressure of the hot corona allows gas that was originally at the outskirts to begin forming stars, initially forming stars of low metallicity compared to the dwarf evolved in isolation.
This increase in star formation eventually allows a dwarf galaxy to form more metal rich stars compared to a dwarf in isolation, but only if the dwarf retains gas for a sufficiently long period of time.
In addition, dwarfs that formed substantial numbers of stars post-infall have a slightly elevated $[$Mg$/$Fe$]$ at high metallicity ($[$Fe$/$H$]$$\sim$$-1.5$).
}

\maketitle
\section{Introduction} 
The concordant cosmological model of dark energy and cold dark matter, $\Lambda$CDM, predicts that galactic halos contain numerous subhalos that have accreted over cosmic time {\ebf \citep[e.g.][]{Klypin1999,Moore1999}}.
A portion of these accreted subhalos are thought to host the dwarf galaxies that are visible today as satellite galaxies, although exactly which subhalos would host galaxies is an open question\citep{Boylan-Kolchin2011a}.

Regardless of the details of early galaxy formation, only the most massive satellite dwarf galaxies orbiting close in to the host galaxy are able to retain gas  to the present day \citep{Grcevich2009,Chiboucas2013}, with the other dwarfs being gas-poor dwarf spheroidals (dSphs).
Further out, however, dwarf galaxies of lower mass are able to retain a detectable amount of gas, which is able to fuel ongoing star formation, suggesting that environmental effects are the cause of this gas-poor gas-rich dichotomy \citep{Grcevich2009}.

{\ebf As galaxy simulations increase in resolution more attention is being paid towards the evolution of dwarf galaxies and how  simulations compare to observations.}
Previously we have examined the chemodynamical evolution of dSphs as they interact tidally with a larger galaxy \citep[][hereafter NRJ14]{Nichols2014}.
Considering only tidal effects, dwarfs lose substantial quantities of gas because of  a synergistic effect of supernova induced expulsion and tidal forces.
However, all dwarfs were able to retain quantities of gas that would be readily detectable today unless they were completely destroyed by tides.
A potential solution to the removal of gas, without tidally destroying the dwarfs, lies in ram pressure stripping.

Previous work has shown that the interaction of the hot corona and satellite galaxies is likely to remove all gas from them \citep{Mayer2006,Nichols2011,Zolotov2012,Nichols2013,Gatto2013}, however, problems with Kelvin-Helmholtz instabilities  \citep{Mayer2006,Zolotov2012}, or the treatment of star formation \citep{Nichols2011,Nichols2013,Gatto2013} justify new simulations.

The stripping of galaxies close in $(<50$~kpc$)$ is likely to occur independent of any internal process within the galaxy \citep{Mayer2006}, but further out, internal feedback resulting from star formation is needed to remove the gas \citep{Gatto2013}.
In previous work on the topic, the star formation has only been treated  with simplistic models for analytic ease \citep{Nichols2011,Nichols2013} or assumed based on the observationally derived star formation history \citep{Gatto2013}.
These approaches gloss over the wide range of observational properties that exist within the dSphs and what that might say about their evolution post-infall.

{\ebf \citet{Sawala2012} and \citet{Zolotov2012} simulated dwarf galaxies within a cosmological framework, running simulations of Milky Way sized galaxies at high resolution.
Here, using traditional smoothed particle hydrodynamic (SPH) simulations with self-consistent star formation and metallicity injection, \citeauthor{Zolotov2012} found that baryons can assist in the tidal stripping of dwarf galaxies and can dramatically change the inner region of a dwarf through supernova feedback \citep[see also][]{Governato2010,Pontzen2012}.
Known problems with the traditional SPH models \citep{Agertz2007} however, allowed many dwarf spheroidal analogues to retain gas to the present day and potentially underestimated the impact of ram pressure stripping.

\citeauthor{Sawala2012} found that many dwarfs are likely to enter the halo already gas-free, and tidal stripping combined with UV heating and supernova feedback can remove gas from many of the remaining dwarfs.
 With the  use of traditional SPH models and at low resolution, however,  ram pressure stripping may have been overestimated for some dwarfs, which may accelerate the stripping and cause some dwarfs to be tidally destroyed through this baryon-tide synergy.

Ideally large cosmological simulations, such as found in \citet{Sawala2012,Zolotov2012}, could be run for any change in the initial conditions or numerical methods (e.g. meshless or updated SPH methods).
Because of the large computational cost, however, more efficient methods must be considered for the bulk of simulations.}

{\ebf The need for star formation, metallicity evolution, and tidal forces to be handled simultaneously can be understood by considering the wide range of properties impacted by internal star formation and tidal or ram-pressure stripping.
For example, present day gas-rich dwarfs show constant density cores \citep{Oh2015}, a process thought to be caused by supernova feedback \citep{Governato2010,Pontzen2012}.
On the scales of the dwarf spheroidals the existence of cores is slightly less clear \citep[see e.g.][]{deBlok2008,Walker2011,Breddels2013} and, as the inner profile can alter the tidal evolution \citep{Kazantzidis2013}, simulating the internal star formation is vitally important to understanding the evolution of a dwarf.
In addition to the internal star formation changing  its profile, the removal of gas also impacts the inner density \citep{Arraki2014} and the inclusion of this baryon loss may lead to a lower fraction of surviving dwarfs than expected from simulations that retain gas.
In addition the metallicity evolution is intricately linked with the star formation history (and contributes towards its regulation through cooling), and with high-resolution studies of dwarf spheroidals now available \citep[e.g.][just for Sextans dwarf spheroidal]{Shetrone2001,Aoki2009,Tafelmeyer2010,Kirby2011,Theler2015} the metallicity evolution can provide a sanity test for any model aiming to reproduce small satellite galaxies.}

In this paper, we describe modifications to the chemodynamical tree-SPH code \texttt{GEAR} \citep{Revaz2012}, which allows us to {\ebf efficiently} simulate the post-accretion history of a galaxy and discuss how a dwarf galaxy, which has properties similar to an observed dwarf (excepting of course the excess in gas) changes as it evolves  into a host galaxy after infall.
We also pay particular attention to the chemical evolution of the dwarf galaxies and how this evolution is influenced by ram pressure stripping.
We describe the method in \S\ref{sec:mods}, the initial conditions and set up of our simulations in \S\ref{sec:sims}, the post-infall evolution by comparison with an isolated evolution in \S\ref{sec:results}, and conclude in \S\ref{sec:conclusion}.

\section{Simulating ram pressure stripping} \label{sec:mods}

We have previously used the chemodynamical code \texttt{GEAR} \citep{Revaz2012}, which is an extension of the public \texttt{Gadget-2} code \citep{Springel2005a}, to examine dwarfs in isolation \citep{Revaz2012} and those that are tidally interacting \citepalias{Nichols2014}.
Along with the latest additions to the code \citep{Revaz2015}, including the pressure-entropy SPH formulation of \citet{Hopkins2013}, we extend this code  to simulate the ram pressure and tidal forces simultaneously upon an infalling dwarf.

To avoid the computationally expensive simulation of an entire hot halo, we performed wind-tunnel simulations, where only the region of the halo that interacts with the dwarf is considered.
The dwarf is placed inside a gas-filled periodic box, where particles are injected from one side and deleted when crossing the opposite side.
The use of the pressure-entropy SPH formulation allows us to account for Kelvin-Helmholtz instabilities \citep{Hopkins2013} a known problem with the traditional SPH models \citep{Agertz2007}.
To properly describe the tidal forces acting upon the dwarf, we rotate this periodic box along the orbit of the dwarf. This kind of approach allows us to capture any synergy between the tidal and ram pressure forces upon the dwarf.
The moving of the periodic box is conceptually similar to the method used by \citet{Ploeckinger2014} to study the evolution of tidal dwarf galaxies, but with an evolving halo density.
We describe this rotation scheme in \S\ref{subsec:rot}.

\begin{figure*}
        \centering
        \includegraphics[width=\textwidth]{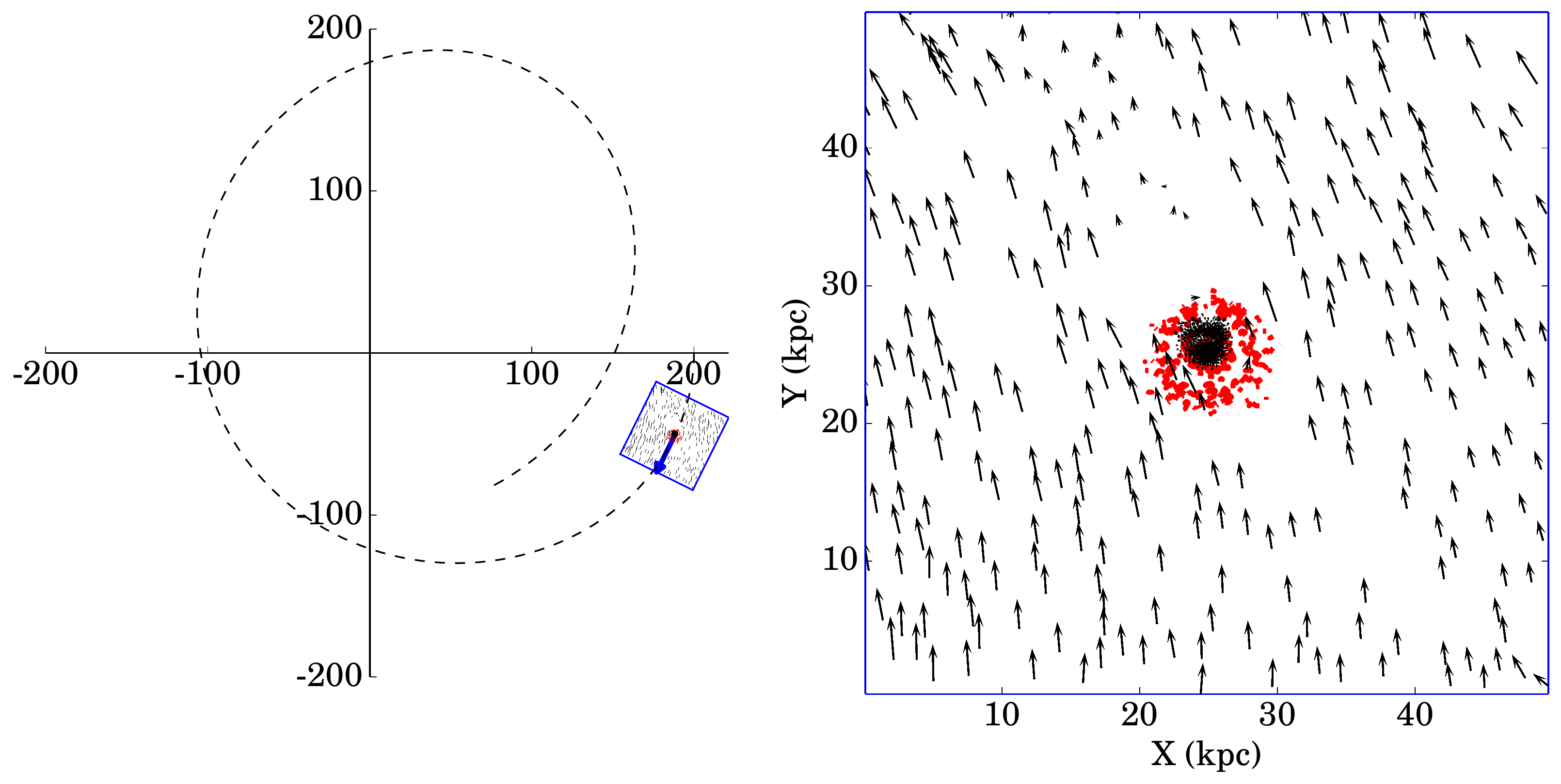}
        \caption{A schematic illustration of the method. This is a representation of the simulation RPS-100 described below $\sim$$400$~Myr after infall.
        The left-hand panel shows the predicted orbit for $7$~Gyr and the position of the rotating frame at the snapshot.
        The arrow shows the direction of the dwarf's orbit.
        The right-hand panel is a quiver plot showing the position and velocity of a $0.5$~kpc central slice of gas, with contours of the projected dark matter density  shown in red.
        For conceptual ease, the velocity in an inertial frame is shown, which is related by a transformation to the rotating frame described in Eq. \ref{eq:v}.}
        \label{fig:schematic}
\end{figure*}

In addition to the tidal forces changing over the orbit of the dwarf, the changing density of the galactic corona should be reflected in the density of gas within the periodic box.
In addition, the wake behind a dwarf should not be able to cross the periodic boundary to prevent it from interacting with the dwarf.
These issues can be addressed through the deletion of particles at the boundary behind the dwarf, and an insertion of new gas in front of the dwarf.
This kind of  approach has already been used successfully in grid codes to study the hot halo density \citep{Gatto2013} and the wake produced by galaxies infalling into a cluster \citep{Roediger2015}.
We describe a similar adaption for the SPH code \texttt{GEAR} here, described in \S\ref{subsec:PC} and \S\ref{subsec:PD}, for the creation and deletion, respectively.
A schematic illustration of both processes is shown in Fig. \ref{fig:schematic} in which we show how the periodic box moves along the orbit and the effect of the rotation scheme within the periodic box.

\subsection{Rotation scheme and the reference frame} \label{subsec:rot}

To simulate the tidal forces upon the dwarf galaxy, we apply the standard fictitious forces throughout the periodic box to simulate a solid rotation of the box  \citep[a non-solid rotation is also possible, although this introduces
further issues, e.g.][]{Huber2001}. 

To apply the fictitious forces, we apply the standard equations of motion in a rotating reference frame, describing the rotation through the use of quaternions\footnote{We write quaternions in the vector form $\vec{q}=(\vec{x},w)=(x_1,x_2,x_3,w)=x_1i+x_2j+x_3k+w$, where $i^2=j^2=k^2=ijk=-1$, $w$ is the real part of the quaternion $\vec{q}$ and $x_1,x_2,x_3$ are real numbers corresponding to the magnitude of the imaginary components. The addition of two quaternions $\vec{q}_1=(\vec{x}_1,w_1)$ and $\vec{q}_2=(\vec{x}_2,w_2)$ is given by $\vec{q}_1+\vec{q}_2 = (\vec{x}_1+\vec{x}_2,w_1+w_2)$. The (non-commutative)multiplication of $\vec{q}_1$ and $\vec{q}_2$ is given by $\vec{q}_1\vec{q}_2 = (w_1\vec{x}_2+w_2\vec{x}_1 + \vec{x}_1\times\vec{x}_2,w_1w_2-\vec{x}_1\cdot\vec{x}_2)$. The inverse of a quaternion $\vec{q}=(\vec{x},w)$ is given by $\vec{q}^{-1}=(-\vec{x},w)/||\vec{q}||^2$, where $||\vec{q}|| = \sqrt{x_1^2+x_2^2+x_3^2+w^2}$ is the norm of $\vec{q}$. A rotation of a vector $\vec{x}$ by an angle $\theta$ about an axis $\vec{y}$ (resulting in vector $\vec{z}$), can be represented by $(\vec{z},0) = \vec{q}(\vec{x},0)\vec{q}^{-1}$, where $\vec{q} = (\vec{y}\sin[\theta/2],\cos[\theta/2])$.} around a pivot point inside the box.
Although not commonly seen in astronomy, quaternions are extensively used to describe rotations in programming.
Quaternions, which are an extension of  complex numbers,  are able to describe $3$D rotations with the useful properties that they can be easily updated with time and are reasonably forgiving of the gradual accumulation of errors. As the rotation involves renormalisation the errors only accumulate in the rotation direction/angle  and do not change the magnitude of any vector.
Throughout we write equations that involve the rotation of a vector by a quaternion. Here when written in the same equation as a quaternion, a vector $\vec{x}$ should be understood to be the quaternion $(\vec{x},0)$.

\subsubsection{Reference frames}

We locate the fixed pivot point within the periodic box at $\vec{x}_p=(x_p,y_p,z_p)$, which represents the (moving) point $\vec{X}_p$ relative to the host galaxy.
For clarity, we use upper case variables when discussing things in the reference frame of the host galaxy and lower case within the periodic box.
At the start of the simulation, this pivot point is set to be the position of the particle with the minimum potential inside the periodic box.
As many equations arising from the rotation of the box depend upon the relative distance to the pivot, for clarity we use an asterisk to denote the difference between a vector and the pivot, e.g. $\vec{X}^* = \vec{X} - \vec{X}_p$ where necessary throughout the paper.
The position of any other point, $\vec{x}$, within the periodic box corresponds to the point $\vec{X}$ in the reference frame of the host galaxy by
\begin{equation}
  \vec{X}^* = \vec{q}^{-1}\vec{x}^*\vec{q},
\end{equation}
where $\vec{q}$ is the relevant quaternion rotating a vector from the host galaxy frame to the periodic box.
We choose $\vec{q}$ to be the unit quaternion that maps the unit velocity vector of the pivot $\hat{{\vec{V}}}_p$ to one of the axis in the periodic box, which we arbitrarily choose to be $-\hat{\vec{y}}$.

\subsubsection{Equations of motion}
Within the periodic box the velocity is calculated in the usual way according to the standard equations of motion for the rotating reference frame
\begin{equation}
        \vec{v} = \vec{v}^*= \vec{q}\vec{V}^*\vec{q}^{-1} - \vec{\omega}\times\vec{x}^*,\label{eq:v}
\end{equation}
where $\vec{\omega}$ is the angular velocity within the rotating reference frame.
Because of the dependence on the position, the velocity of particles that cross any boundary need to be recalculated. 
This recalculation is not hugely important and, in  practice, the gas pressure has a tendency to correct the particle velocity; gas that crosses a boundary is unlikely to interact with the dwarf, but correcting the gas in this way reduces the error associated with the periodic box.

Each particle within the box experiences an acceleration force $\vec{a}$, dependent upon the potential the potential of the host galaxy, $\Phi$; the angular velocity; the acceleration, present on a particle inside the periodic box at a position $\vec{x}$ and velocity $\vec{v}$, is given by
\begin{equation}
  \vec{a} = \vec{a}^* = \vec{q}\left[-\nabla\Phi\left(\vec{X}\right)-\vec{A}_{p} \right]\vec{q}^{-1} - 2\vec{\omega}\times\vec{v} - \vec{\omega}\times\left(\vec{\omega}\times\vec{x}^*\right) - \dot{\vec{\omega}}\times\vec{x}^*,\label{eq:acc}
\end{equation}  
where $\Phi\left(\vec{X}\right)$ is the potential at the point $\vec{X}$, $\vec{A}_p$ the acceleration of the pivot and the standard Coriolis, centrifugal, and Euler forces.

The acceleration of the pivot, $\vec{A}_{p} = -\nabla\Phi(\vec{X}_{p}) + \vec{A}_{p,\rm{C}}$, comprises the gravitational acceleration due to the potential of the host galaxy plus an added correction term, $\vec{A}_{p,\rm{C}}$, to account for the drag the dwarf experiences through the orbit. This drag and the effect on the dwarf's orbit is discussed further in \S\ref{subsec:orbchange}.

The correction term is ad hoc, intended to disappear when not needed and be sufficiently small when needed to minimise numerical errors that occur when $\vec{A}_{p, {\rm C}}$ is large (most of these errors  manifest as issues with particle creation/deletion, discussed below).
The correction manifests as a moving of the box off its original orbit following the new trajectory of the dwarf, modified by ram pressure and tidal stripping.
The correction term is calculated by finding the average position $\vec{X}_{64}$ of the $64$ stellar or dark matter particles that had the lowest potential at infall.
The choice of $64$ particles balances the risks between just taking the average of all particles, which results in a poor approximation when the galaxy becomes tidally disrupted, and choosing only a handful, which can focus on tidal tails if any of the particles are ejected from the core.
In the simulations discussed here we use
\begin{equation}
  \vec{A}_{p,{\rm C}} = 0.54~{\rm Gyr}^{-1}\vec{V}^*_{64} +0.022~{\rm Gyr}^{-2}f\left(\vec{X}^*_{64} ,\vec{V}^*_{64}\right)\vec{X}^*_{64},
\end{equation}
where $f(x,v)$ is a function evaluated in each Cartesian dimension that returns $1$ if, $xv>0$ or $|v| < 2$~km~s$^{-1}$, and otherwise $0$ .

\subsubsection{Computing the rotation of the reference frame}
As $\vec{q}$ is not static, it must also be updated every time step.
To do so, at every time step we first calculate the the angular velocity, $\vec{\Omega}$ in the host galaxy frame, and $\vec{\omega}= \vec{q}^{-1}\vec{\Omega}\vec{q}$ in the periodic box.
The parameter~$\vec{\Omega}$ is calculated whenever the velocity is updated and is given by the standard relation
\begin{equation}
        \vec{\Omega} = \frac{\vec{V}_p\times\vec{A}_p}{||\vec{V}_p||^2}.
\end{equation}

Once the angular velocity is known, $\dot{\vec{q}}$ is given by the useful relation
\begin{equation}
        \dot{\vec{q}} = 0.5\vec{\Omega}\vec{q},
\end{equation} 
which allows the rotation to be updated each time step in time with the velocity.
After each update in $\vec{q}$ it is renormalised, ensuring $||\vec{q}|| = 1$, and ensuring that errors only accumulate in the angle and do not artificially increase the magnitude of the vectors.

That the rotation quaternion changes every time step and that the velocity of a particle appears in the calculation of the acceleration is problematic.
\texttt{Gadget-2} and by extension \texttt{GEAR} relies on a standard Kick-Drift-Kick integrator, which relies on the Hamiltonian of the system being separable into independent position and momentum coordinates.
Unfortunately,  the standard Hamiltonian for this system breaks this condition for rotating reference frames and, as such, we lose the symplectic nature of the leapfrog integrator.
In an effort to minimise the error resulting from this, we make a slight alteration to the calculation of the acceleration when implemented in the code.

Normally, the velocity is updated by an acceleration term, which only depends upon the position
\begin{equation}
\vec{v}_{t+1/2} = \vec{v}_{t-1/2} + \vec{a}(\vec{x}_{t})\Delta{}t.\end{equation}
However, in the rotating frame it is clear from eqn \ref{eq:acc} that the acceleration term also includes a velocity dependence and terms dependent upon the changing rotation quaternion and angular velocity.
We instead update the velocity through an iterated approximation of the velocity at the time step, which performs reasonably well in simulations of the tidal force (tests of this method and others are provided in appendix \ref{app:tide-test}).
Care should be taken to note that this scheme is not the best performer at large step sizes, such as in dark matter only simulations.
First, we assume that the change in rotation quaternion is negligible across any time step, secondly we assume that the acceleration in $\vec{\omega}$ is slow compared to the time step size.
These two assumptions hold well at time steps determined by the Courant condition inside the simulation.
The first assumption leads to $\vec{q}_{t,{\rm pred}} = \vec{q}_{t-1/2}$ and the latter assumption means that we can approximate $\vec{\omega}$ at a time $t$ by
\begin{equation}
\vec{\omega}_{t,{\rm pred}} = \vec{\omega}_{t-1/2} +  \dot{\vec{\omega}}_{\rm old}\Delta{}t,
\end{equation}
where $\dot{\vec{\omega}}_{\rm old}$ is the change in $\vec{\omega}$ at the previous time step.
The velocity is then updated from a time $t-1/2\Delta{}t$ to a time $t+1/2\Delta{}t$, according to
\begin{eqnarray}
        \vec{v}_{t,{\rm pred}} &=&  \vec{v}_{t-1/2} + \vec{a}(\vec{q}_{t-1/2},\vec{x}_{t},\vec{v}_{t-1/2},\vec{\omega}_{t-1/2})\Delta{}t/2,\\
        \vec{v}_{t,{\rm pred}} &=& \vec{v}_{t-1/2} + \vec{a}(\vec{q}_{t-1/2},\vec{x}_{t},\vec{v}_{t,{\rm pred}},\vec{\omega}_{t,{\rm pred}})\Delta{}t/2,\\
        \vec{v}_{t+1/2} &=& \vec{v}_{t-1/2} + \vec{a}(\vec{q}_{t-1/2},\vec{x}_{t},\vec{v}_{t,{\rm pred}},\vec{\omega}_{t,{\rm pred}})\Delta{}t.
\end{eqnarray}

\subsection{Creation}\label{subsec:PC}

To simulate the increase in density over the orbit of the dwarf, particles are created and inserted into the box through the bottom side, we describe the exact mechanism of creation in appendix \ref{app:partcre}.
The velocity of the particle is then determined by the velocity of the pivot point and the correction to the rotating frame of reference,
\begin{equation}
  \vec{v} = (0,||\vec{V}_p||,0) -\vec{\omega}\times\left[\vec{x'}-(0,y_p,0)\right]. \label{eq:partvel}
\end{equation}
As the orbit can change over the scale of the box this injection is not perfect though and the gas hits the dwarf at a slight angle.
 
The remaining properties, temperature, metallicity, pressure, entropy, etc., are determined according to the properties of the chosen halo profile.
Our choice of profile is a hot gaseous halo in hydrostatic equilibrium described in \S\ref{subsec:host}, but in practice can be any function, with the caveat that a profile that rapid changes in pressure results in pressure waves.
These errors and simple tests of the changing density profile are discussed further in Appendix \ref{app:halo}.
Notably, these errors depend on the sign of $\dot{R}_p$ and therefore partially cancel out, the results therefore may have an error in timing, but qualitatively are reduced over a full orbit because of this sign dependence.

\subsection{Deletion}\label{subsec:PD}

To prevent particles from reappearing in the periodic box and interacting with the dwarf, they are deleted when leaving the rear of the box, or in the case of  dark matter and stellar particles, from the sides or front of the box.
As gaseous particles possess a pressure, we consider them separately from dark matter and stellar particles when considering which particles to delete.
Any gas particles (halo or dwarf), which upon moving drift within $L/100$ of the rear box edge on that time step, are deleted.
The choice of deleting them slightly before the edge of the box serves two purposes, one is to minimise the amount of neighbour particles occurring across the rear boundary, and secondly, to provide a small boost to particles about to leave the box.
This slight vacuum does produce an increase in density waves of particles entering the box since they see the absence of particles on the other side, but these oscillations are greatly damped by the time the particles reach the dwarf.
The lack of neighbours across the edges minimises errors that can occur near perigalacticon when the dwarf's orbit changes rapidly (and so the particles may have minimal velocity in the $y$ direction).

Any dark matter or stellar particles are deleted if they drift within $L/500$ of any box edge or $L/100$ of the rear box edge. 
This extra deletion prevents a spurious build-up of particles at the edges of the box, which can occur when tidal tails pass the boundaries of the box.

\begin{figure}
\centering
\includegraphics[width=0.5\textwidth]{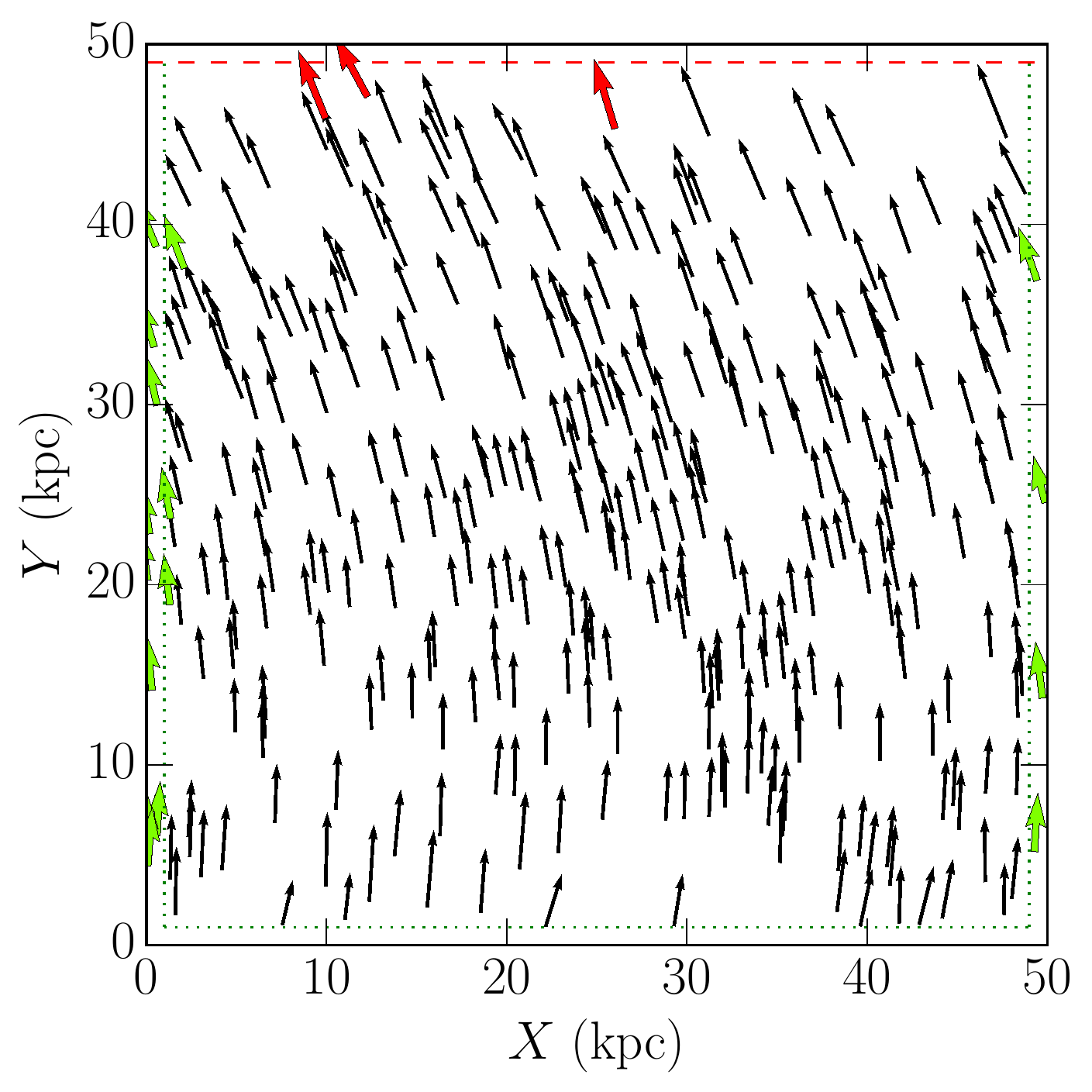}
\caption{Schematic of particle deletion. Shown in a $50\times50$~kpc box. The deletion regions (exaggerated by a factor of $10$ from that used) are shown with a red dashed line (for deletion of all particles) and green dotted lines (deletion of stellar/dark matter particles). Particle positions and velocities are shown with arrows.
Particles that would be deleted on the next time step if they were stellar or dark matter particles are shown in green.
Particles that would be deleted on the next time step regardless of their type (gas, stellar, or dark matter) are shown in red.
The remaining particles, which are not deleted regardless of type, are shown in black.
}\label{fig:partdel}
\end{figure}

To assist in understanding, we illustrate this mechanism in Fig. \ref{fig:partdel} in which we show which particles would be deleted in future time steps if they were gas, dark matter, or stellar particles.

\section{Simulations}\label{sec:sims}
To investigate the impact of ram pressure and tidal forces upon a dwarf galaxy, we investigate two Sextans-like dwarf galaxies, one with a pseudo-isothermal dark matter profile reminiscent of today's dwarf \citep{Battaglia2011} and one with a denser NFW profile.
In both cases, we use the same pseudo-isothermal gas profile.
We investigate three scenarios, the dwarf in isolation and as it undergoes ram pressure throughout two different orbits, both of which have an apogalacticon of $200$~kpc and differing perigalacticons of $50$~kpc and $100$~kpc.

\begin{figure*}
\centering
\includegraphics[width=\textwidth]{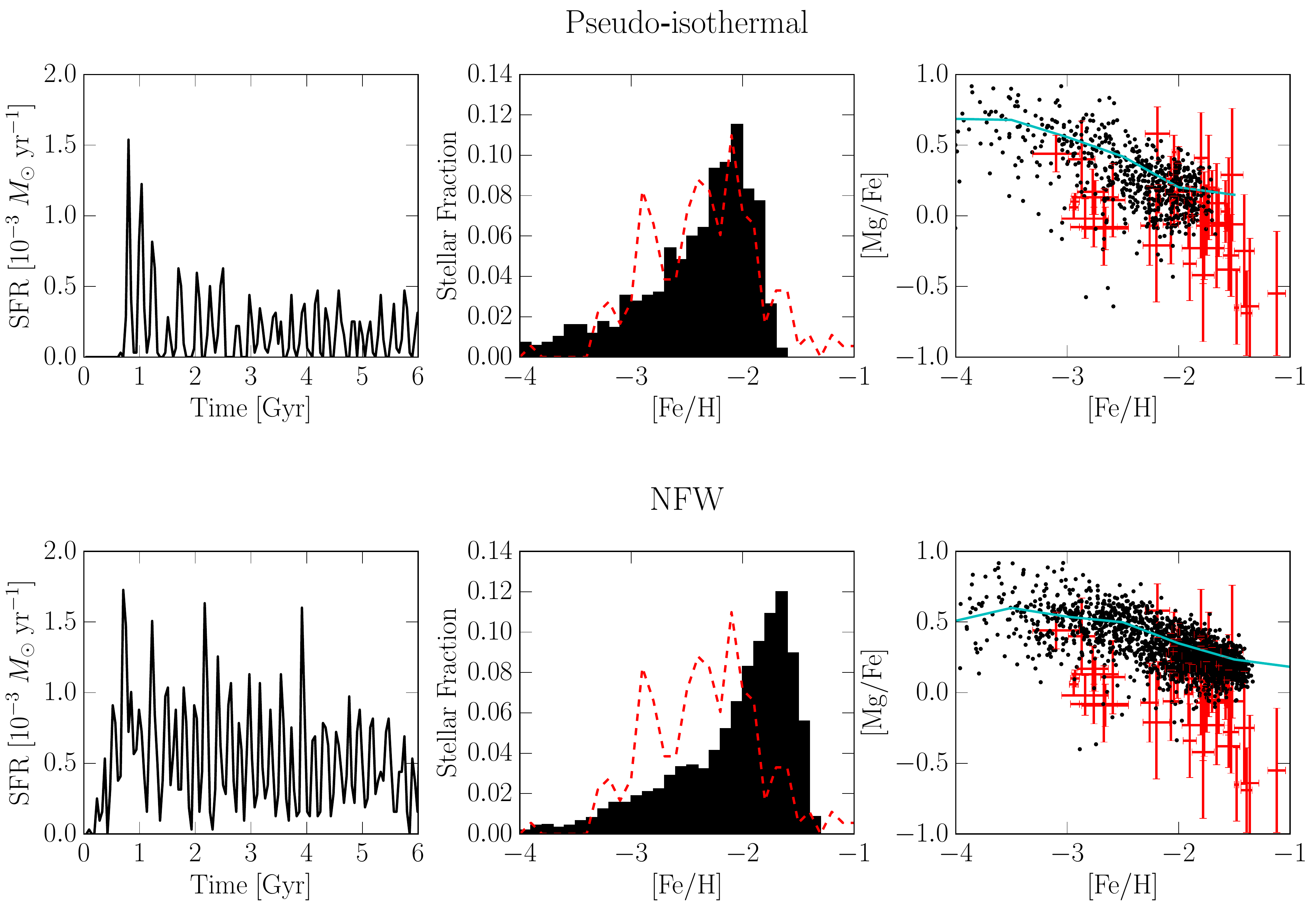}
\caption{Properties of the pseudo-isothermal dwarf (top) and NFW dwarf (bottom) in isolation after $6$~Gyr of evolution.
The properties of the Sextans dSph are shown with a red dashed line in the $[$Fe$/$H$]$ histogram (middle panel) \citep{Battaglia2011}.
The $[$Mg$/$Fe$]$ v $[$Fe$/$H$]$ of available RGB stars \citep{Shetrone2001,Aoki2009,Tafelmeyer2010,Kirby2011,Theler2015}, selected as per \citet{Theler2015}, are shown with points in the third panel. The Mg values are derived from one line and show much larger scatter at low $\alpha$ that is not observed \citep[see][]{Theler2015}. The median $[$Mg$/$Fe$]$ in each $[$Fe$/$H$]$ bin of the stellar particles is shown with a cyan line in the same panel.
This choice of parameters achieves the approximate metallicity distribution (although missing the most metal rich stars) and the turnover in [$\alpha$/Fe] of Sextans dSph.}\label{fig:params}
\end{figure*}

The dark matter and gas profile parameters (see Table \ref{table:params}) and parameters for star formation and feedback are calibrated such that this isolation model produces a dwarf with chemodynamical properties similar to Sextans dSph after 6 Gyr total.
Based on this calibration throughout all models, we use a star formation efficiency of $c_\star=0.01$, a supernova energy of $\epsilon_{\rm SN}=0.5\times10^{51}$~erg,  a density threshold for star formation of $\rho_{\rm th}=0.01$~cm$^{-3}$ {\ebf, and temperature threshold of $T_{\rm th}=3\times10^4$~K. Changing the density or temperature thresholds has minimal impact on the final results at the resolutions considered here \citep[see e.g.][]{Revaz2012}}.
{\ebf As per \citet{Revaz2012} and \citetalias{Nichols2014} below $T=10^4$~K, we use the metallicity-dependent cooling according to \citet{Sutherland1993}. Below this, we also account for H$_2$ and HD cooling and cooling due to fine structure line transitions of oxygen, carbon, silicon, and iron, according to \citet{Maio2007}.

The initial mass function (IMF) is chosen to match that of \citet{Kroupa2001}, and we carefully track the number of Type Ia and Type II supernovae that  explode at each time step calculating the rates according to the random scheme of \citet{Revaz2015}.
This treatment has shown good energy conservation and works well over a large variety of masses reproducing accurate stellar populations when compared to observations \citep{Revaz2012,Revaz2015}.
Metals are diffused throughout the nearby gas with a smooth metallicity scheme \citep{Wiersma2009} whereby metal contents are calculated via the SPH scheme at any individual time step, which reduces the dispersion in $[\alpha/$Fe$]$ at low metallicity.}
We show the resulting star formation history and metallicity profiles in Fig. \ref{fig:params}.

\subsection{Initial density profiles}

In half of our models, the dwarf's dark matter profile is initially described by a pseudo-isothermal profile,\begin{equation}
        \rho(r)=\frac{\rho_c}{1+r^2/r_c^2}\,,\label{eq:pseudo}
\end{equation}
where $\rho_c$ is the core density and $r_c$ is the core radius.
This choice of profile has much observational support \citep[e.g.][]{deBlok2008,Walker2011} although if the true profile of dwarf spheroidals is closer to NFW \citep[which can not be ruled out;][]{Breddels2013}, it is likely to overestimate the effects of tidal interactions \citep{Kazantzidis2013}.
To examine the effects of the profile choice, we also consider an NFW profile  \citep{Navarro1997}  for the dwarf's dark matter,
\begin{equation}
        \rho(r) = \rho_{s}\frac{r/r_s}{(1+r/r_s)^2r/r_s}\,,\label{eq:NFW}
\end{equation}
where $\rho_{s}$ is a scale density and $r_s$ a scale radius.
In both cases, the velocity of the dark matter is determined from the Jeans equations.

The gas associated with the dwarf is initially set at the same position as the dark matter and slightly rotated to avoid overlapping particles, with a temperature determined by converting all the kinetic energy. The kinetic energy is determined by the Jeans equations into thermal energy, thereby giving the gas zero velocity initially.
We show the simulation parameters in Table \ref{table:params}  and the model differences in Table \ref{table:models}.

\begin{table}
 \caption{Initial density parameters}
 \centering
 \begin{tabular}{lc}
 \hline\hline
 {Parameter} & {Value}\\
 \hline
 $M_{\rm tot}$ & $3.5\times10^8$~M$_\odot$\\
 $M_{\rm DM,tot}$ & $2.98\times10^8$~M$_\odot$\\
 $M_{\rm gas,tot}$ & $0.53\times10^8$~M$_\odot$\\
 $M_{\rm DM,part}$ & $39.5\times10^3$~M$_\odot$\\
 $M_{\rm gas,part}$ &$6.0\times10^3$~M$_\odot$\\
 $r_{c,{\rm gas}}$ & $2$~kpc\\
 $\rho_{c,{\rm gas}}$ & $5.4\times10^5$~M$_\odot$~kpc$^{-3}$\\ 
 \hline
 Pseudo-isothermal models & \\
 \hline
 $r_{c,{\rm DM}}$ & $2$~kpc\\
 $\rho_{c,{\rm DM}}$ &  $3.0\times10^6$~M$_\odot$~kpc$^{-3}$\\
 \hline
 NFW models &\\
 \hline
 $r_{s,{\rm DM}}$ & $1$~kpc\\
 $\rho_{s,{\rm DM}}$ & $2.3\times10^7$~M$_\odot$~kpc$^{-3}$\\
 \hline
 \end{tabular}
 \tablefoot{Simulation parameters for the dwarf galaxy. The parameter $M_{\rm tot}$ is the total mass of the system (excluding the low-density medium), $M_{\rm RM,tot}$ the total dark matter mass, $M_{\rm gas,tot}$ is the total gas mass (excluding the low-density medium), $N_{\rm part}$ is the total particle number (excluding the low-density medium), $M_{\rm DM,part}$ is the dark matter particle mass, $M_{\rm gas,part}$ is the gas particle mass (of all gaseous particles), $c_\star$ is the star formation parameter, $\epsilon_{\rm SN}$ is the energy released per supernova, $r_{c,{\rm gas}}$ is the core radius of the gas in the dwarf, $\rho_{c,{\rm gas}}$ is the core gas density of the dwarf.}\label{table:params}
 \end{table}

\begin{table}
 \caption{Simulation parameters}
 \centering
 \begin{tabular}{lccccc}
 \hline\hline
 {Model} &Profile type& {Perigalacticon} & {Apogalacticon} \\
&&{$R_{\rm peri}$} & {$R_{\rm apo}$}\\
&& $[$kpc$]$ & $[$kpc$]$\\ 
 \hline
 ISO & Pseudo  & - & - \\
 RPS-50 & Pseudo & $200$ & $50$\\
 RPS-100 & Pseudo &$200$ & $100$\\
 NISO & NFW & - & - \\
 NRPS-50 & NFW &  $200$ & $50$\\
 NRPS-100 & NFW & $200$ & $100$\\
 \hline
 \end{tabular}
 \tablefoot{Model peri- and apogalacticons and whether they have a pseudo-isothermal or NFW dark matter profile.}\label{table:models}
 \end{table}

To test convergence, we  also ran all simulations at higher resolutions with particle masses a factor of four lower.
 We find very little difference dynamically, with gas being removed at the same time and the final state of the dwarfs being dynamically similar.
However, as increased resolution can cause a spurious increase in the scatter of $[\alpha/$Fe$]$ \citep{Revaz2015} because of scatter in the IMF, we use the larger particle gas mass.

\subsection{Host galaxy}\label{subsec:host}

The host galaxy is described by a multi-component static potential within which a dwarf experiences tidal effects and a hot gas halo, which provides ram pressure stripping.
The potential of the host galaxy consists of a bulge and disk modelled as Plummer profiles with parameters chosen to match the Milky Way rotation curve of \citet{Xue2008} and an NFW dark matter halo with a virial mass of $8\times10^{11}$~M$_\odot$  and concentration $c=21$ as per \citet{Kafle2014}.
We  model the disk as a spherical potential, because  large distances are involved (at least $50$~kpc) and this approximation makes little difference to the final form.
{\ebf The potentials of the bulge and disk have minimal effect for most of the orbit and could be left off, however, the inclusion of their mass does result in a slight increase in velocity at perigalacticon and because of the strong dependence of ram pressure on the velocity, we include them for the sake of completeness.}
The forms and parameters of which are given in Table \ref{table:MW}.

\begin{table*}[hbt]
 \caption{Host halo parameters}
 \centering
 \begin{tabular}{lccc}
 \hline\hline
 {} & {$\phi(R)$ Form} & {Parameters} & {Reference}\\
 \hline
 Bulge & $-GM/\sqrt{R^2+a^2}$ & $M=1.3\times10^{10}~\mathrm{M}_\odot$ & [1]\\
 &&$a=0.5$~kpc&\\
 Disk & $-GM/\sqrt{R^2+a^2}$ & $M=5.8\times10^{10}~\mathrm{M}_\odot$ & [1]\\
 &&$a=5$~kpc&\\
 &&$M_{\rm vir} = 8\times10^{11}~\mathrm{M}_\odot$&\\Halo & $-GM_{\rm vir}\ln(1+Rc/R_{\rm vir})/[\ln(1+c) - c/(1+c)]$&$c=21$&[2]\\&&$R_{\rm vir} = 240$~kpc&\\
 \hline
 \end{tabular}
 \tablefoot{\footnotesize{{\bf References:} [1] \citet{Xue2008} [2] \citet{Kafle2014}}}\label{table:MW}
 \end{table*}

\begin{figure}
\centering
\includegraphics[width=0.5\textwidth]{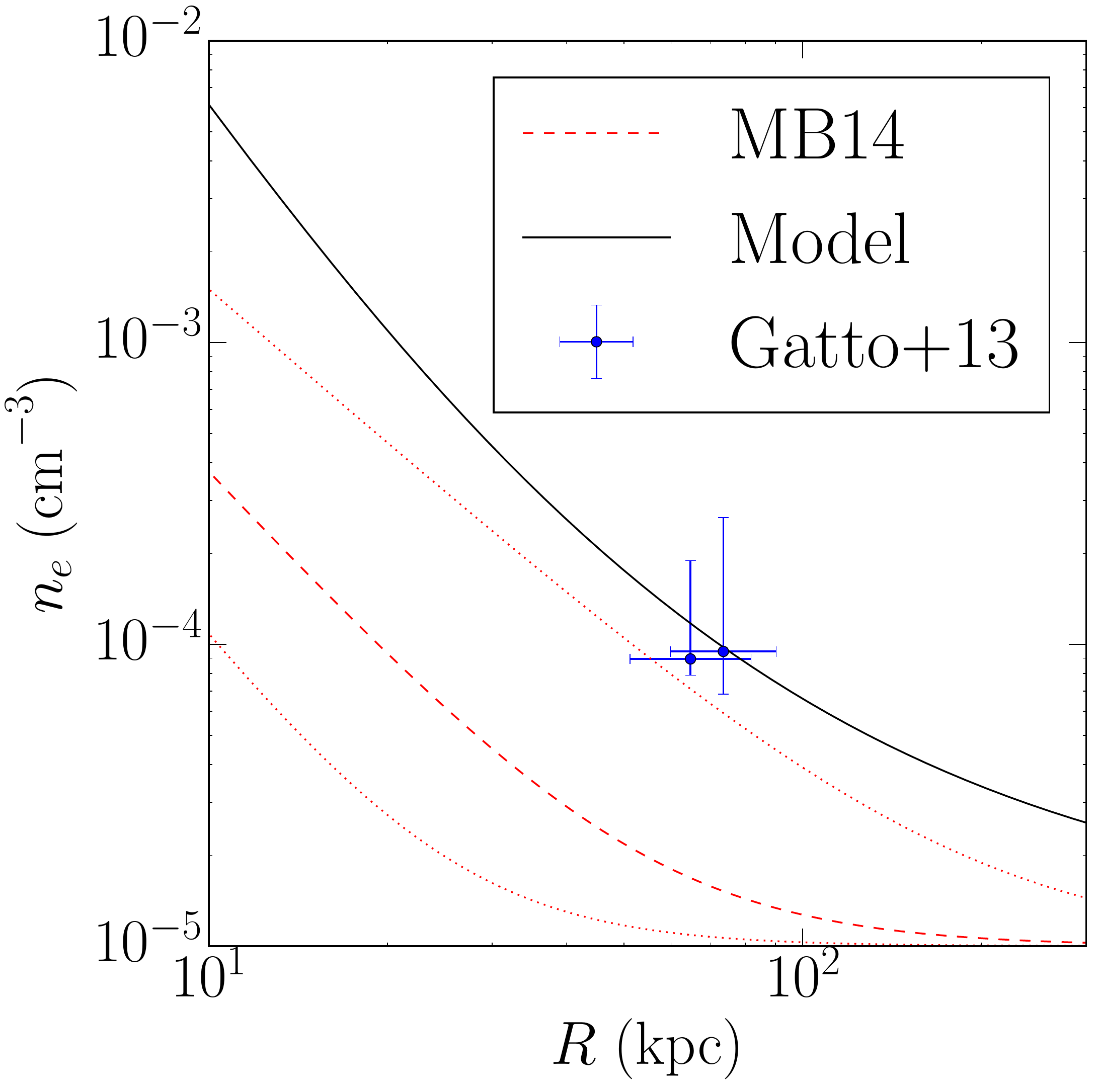}
\caption{Halo density profile for our chosen model (black line).
The best-fit $\beta$ profile of \citet{Miller2013} is shown with a red dashed line, with dotted lines indicating the error in $\beta$.
The density of the halo determined by the simulations of \citet{Gatto2013} for Carina and Sextans to be stripped in one passage are shown with blue points.}\label{fig:halo}
\end{figure}

The hot halo of the galaxy is considered to be an isothermal halo of $2\times10^6$~K in hydrostatic equilibrium with the potential.
This choice only requires  a scale density, which we set by determining that the electron number density at $50$~kpc is given by $n_e = 2\times10^{-4}$~cm$^{-3}$.
The profile is shown in Fig. \ref{fig:halo} and compared with the profile of \citet[red dashed line, with $\beta$ errors in red dotted line]{Miller2013} and \citet{Gatto2013}.
This profile gives a high density compared to \citet{Miller2013} profile, particularly at high galactocentric radius ($R$$\sim$$100$~kpc).
{\ebf This profile is approximately the same as that of \citet[in which the blue points]{Gatto2013}} are necessary to strip the Sextans and Carina in one perigalactic passage, and is slightly below that used by \citet{JBH2007} to reproduce the H$\alpha$ emission along the Magellanic Stream.

\subsection{Model Initialisation}

For the first two gigayears, all simulations are identical and evolved without any external potential but with a surrounding warm medium.
The dwarf is evolved in the centre of a periodic box of side length $L=50$~kpc.
Outside the dwarf a warm ($10^4$~K) low-density ($3.7\times10^{-5}$~cm$^{-3}$) medium fills the box, this density is equal to the density of the hot halo at $200$~kpc but of a much cooler temperature and therefore lower pressure.
This kind of medium is used to simulate the effect of the intergalactic medium.
The isolation model is then evolved for a further $12$~Gyr in this state.
\subsection{Insertion into the hot halo}
For the RPS models, after $2$~Gyr the dwarfs are extracted from this periodic box by selecting only particles within a radius of $10$~kpc from the centre. Only a handful of particles originally associated with the dwarf are beyond this radius, and the exact choice does not impact the results appreciably.
This removal of particles outside this radius minimises the slight over-pressurisation of the hot halo that would result if the particles are kept.
The dwarf is then reinserted in the centre of a new periodic box of the same size $L=50$~kpc filled with a hot gaseous medium with a relative velocity according to the halo density profile at $R=200$~kpc ($n_e=3.7\times10^{-4}$~cm$^{-3}$) and corresponding orbital velocity.
We do not remove any of this medium when inserting the dwarf.

Not removing this medium produces an overlap between the hot medium and the dwarf, however, at the slow (but non-negligible) velocity the dwarf has at apogalacticon, and because of the hot medium having extremely inefficient cooling, the dwarf quickly reaches pressure equilibrium with the external medium without losing gas or accreting the hot medium.

\section{Results}\label{sec:results}

As the dwarf undergoes extended evolution inside a hot halo, its chemodynamical properties post-infall are dramatically altered.
The gaseous profile is affected almost immediately, the dark matter profile is altered as gas is evacuated, and tides take place.
These profile changes also alter the star formation and chemical evolution of the dwarf, altering the properties that would be observed today.
Drag as a result of the ram pressure also changes the orbit of the result, reducing the perigalacticon and apogalacticon of the dwarfs, suggesting that dwarfs may arrive slightly later than pure kinematic studies may suggest \citep[e.g.][]{Rocha2012}.

\subsection{Centre calculation}
For every property except the global star formation (shown in \S\ref{subsec:SF}), we calculate the centre of the dwarf by applying a Gaussian smoothing function to the three-dimensional density histogram of 64 pixels (originally over the box) with a standard deviation of the Gaussian of 6 pixels.
We then refine around the densest point, halving the size of each dimension considered in each iteration until a region of $(6$~kpc$)^3$ is considered, and then perform the same operation again.
From this densest point we do a simple mass weighted average position over all particles within $3$~kpc of this point.
This procedure has been found to be relatively stable around disrupting dwarfs in \citetalias{Nichols2014}, and successfully avoids dense points in tidal tails in favour of the centre of the dwarf.

\subsection{Star formation}\label{subsec:SF}

\begin{figure*}[t]
        \subfigure[Pseudo-isothermal]{\includegraphics[width=0.495\textwidth]{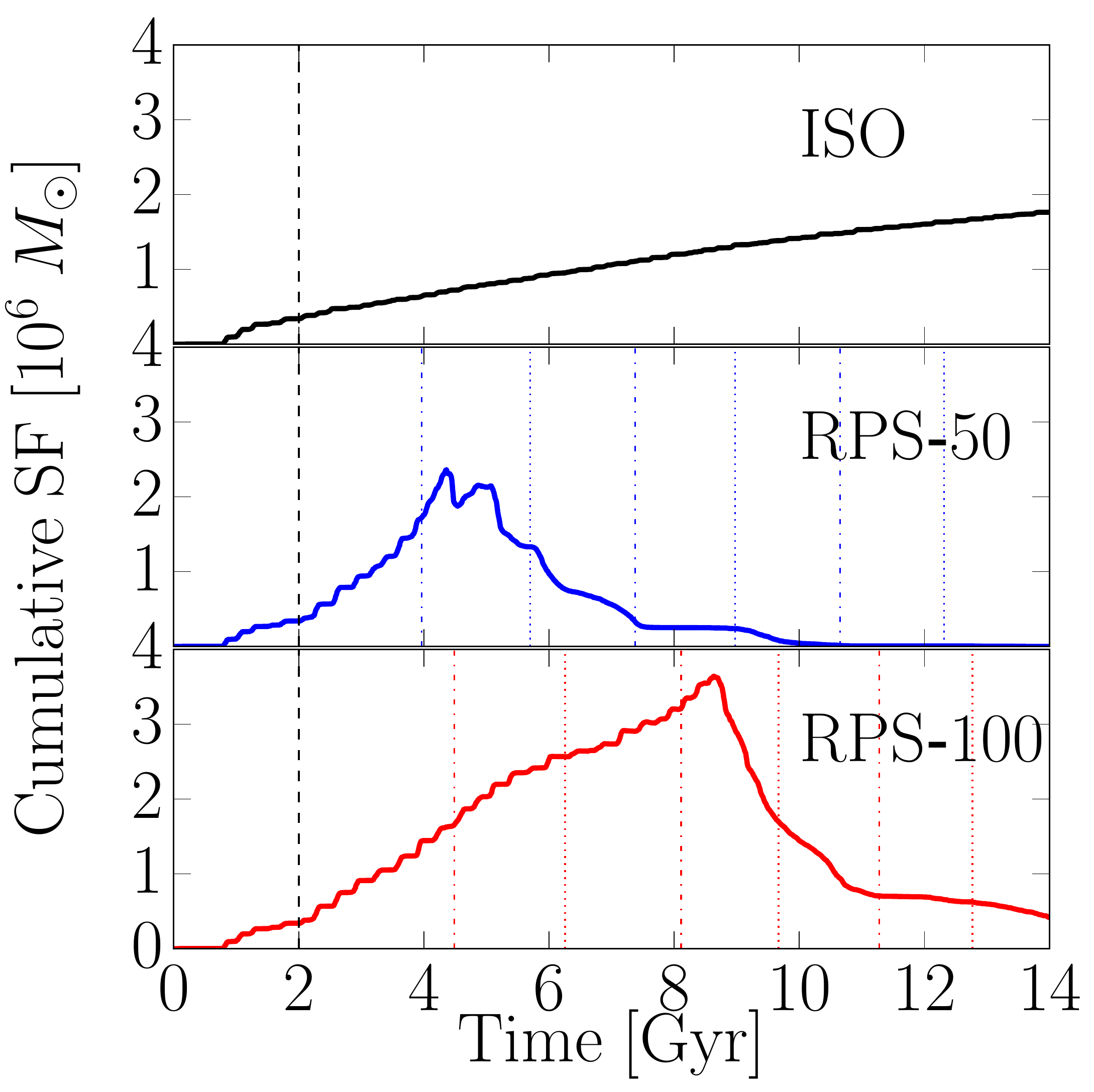}}
        \subfigure[NFW]{\includegraphics[width=0.495\textwidth]{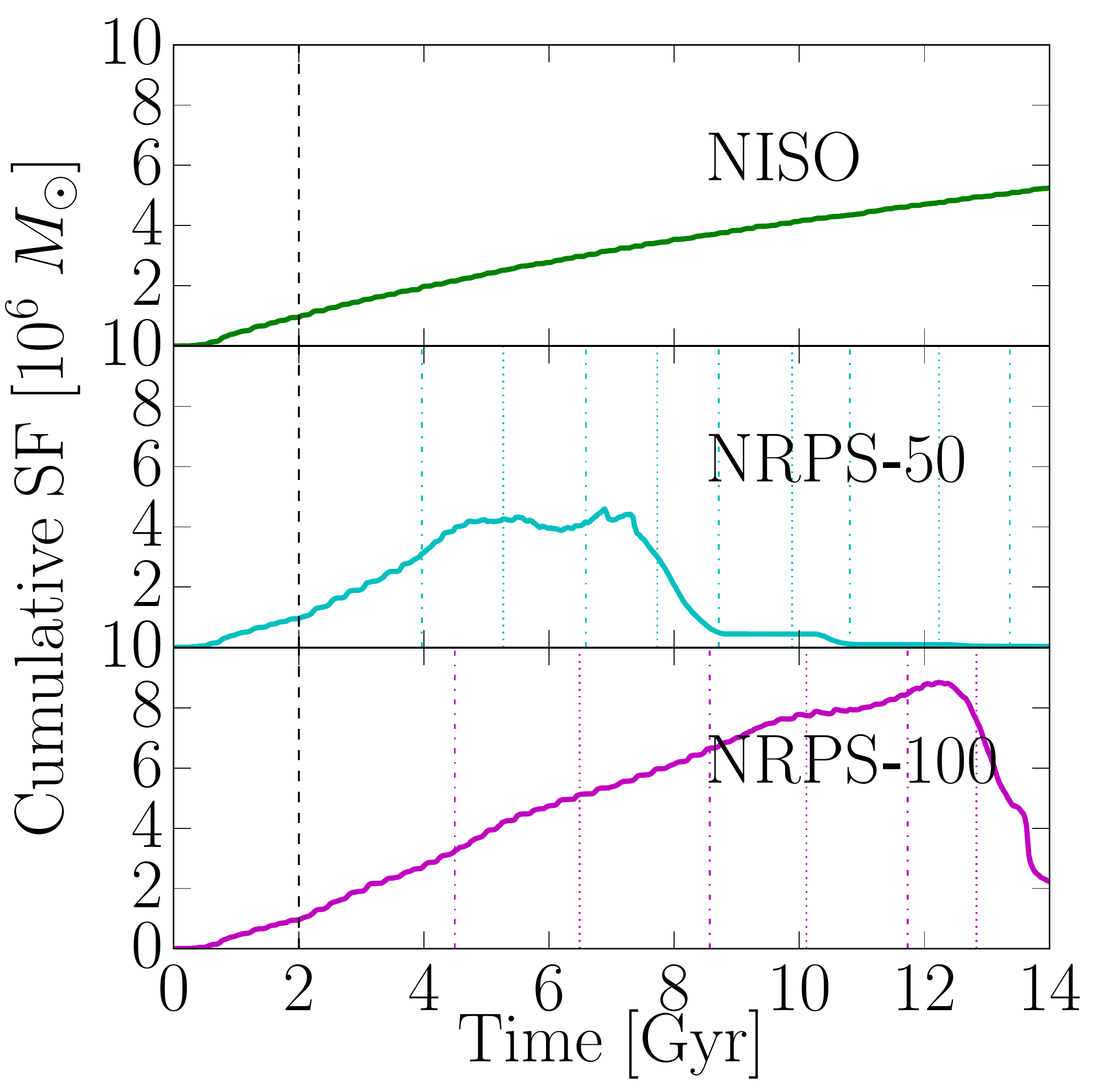}}

        \caption{Mass of stellar particles inside the simulations at any time (cumulative star formation minus stellar mass lost to tides and supernova explosions).  ISO is the pseudo-isothermal model in isolation, RPS-50 with a perigalacticon of $50$~kpc, and RPS-100 with a perigalacticon of $100$~kpc. NISO is the NFW model in isolation, NRPS-50 with a perigalacticon of $50$~kpc, and NRPS-100 with a perigalacticon of $100$~kpc. The time of infall (before which all simulations are identical) is shown with a black dashed line, approximate perigalacticons are shown with a dot-dash line in the relevant colour and apogalacticons in a dotted line.
        In all cases, the peri-/apogalacticon shown is of the pivot and the actual peri-/apogalacticons of the dwarf may differ slightly in time.}\label{fig:SFH}
\end{figure*}

Upon entering the confining medium of the halo where temperature greatly exceeds that used to simulate the intergalactic medium, star formation increases dramatically as a consequence of the increase in pressure. It has been suggested that this increase in star formation happens in the Milky Way satellites, based on comparisons between kinematic properties and star formation histories \citep{Rocha2012}.
The total stellar mass is shown in Fig. \ref{fig:SFH}, which includes all stellar particles within the box at any time, including any that  are formed in gas clouds as they are stripped from the dwarf.
Here any drop in the stellar mass represents the tidal stripping of stars from the dwarf, which can take place even while star formation continues within the box.

\begin{figure*}
        \subfigure[Pseudo-isothermal]{\includegraphics[width=0.5\textwidth]{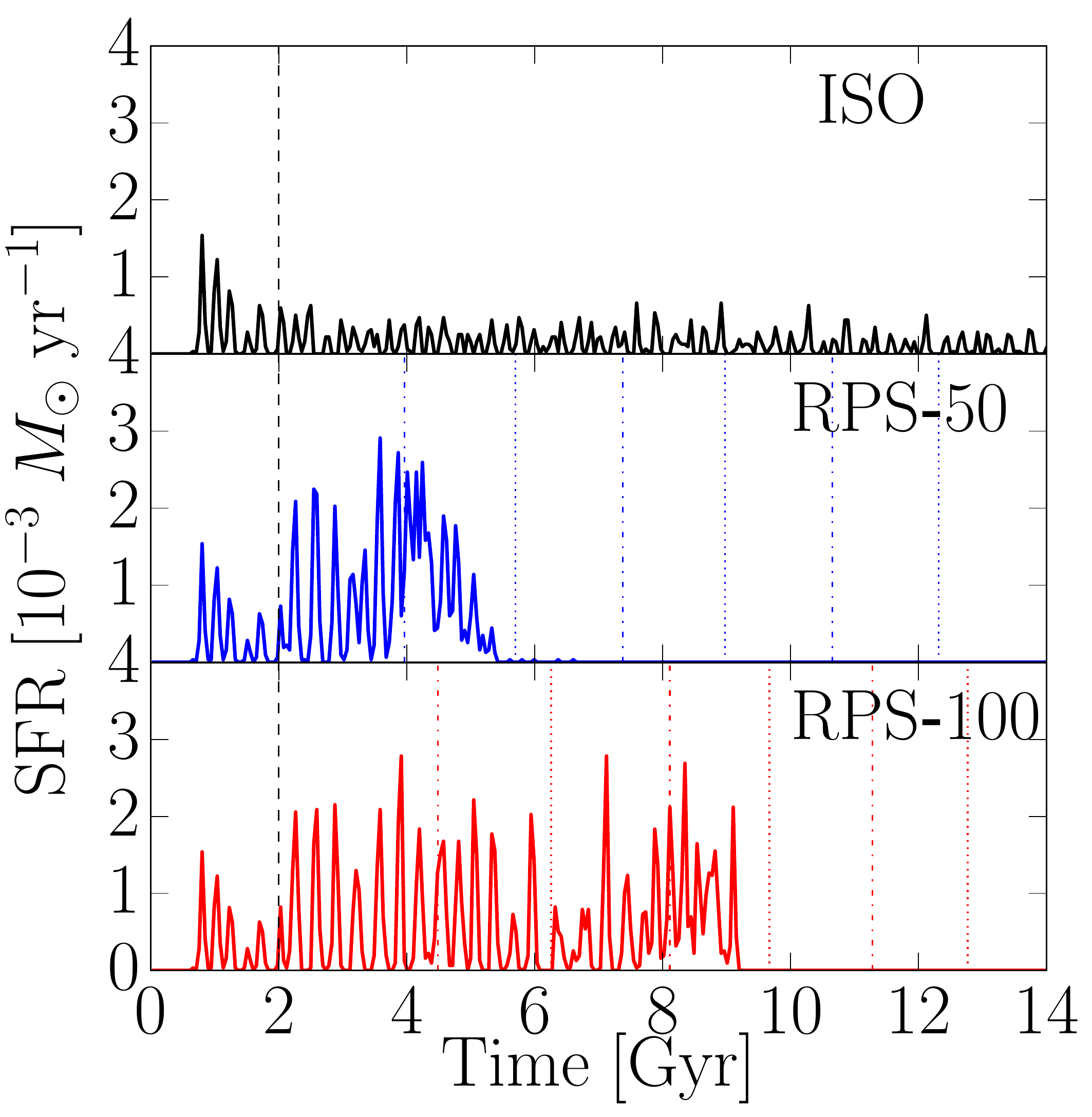}}
        \subfigure[NFW]{\includegraphics[width=0.5\textwidth]{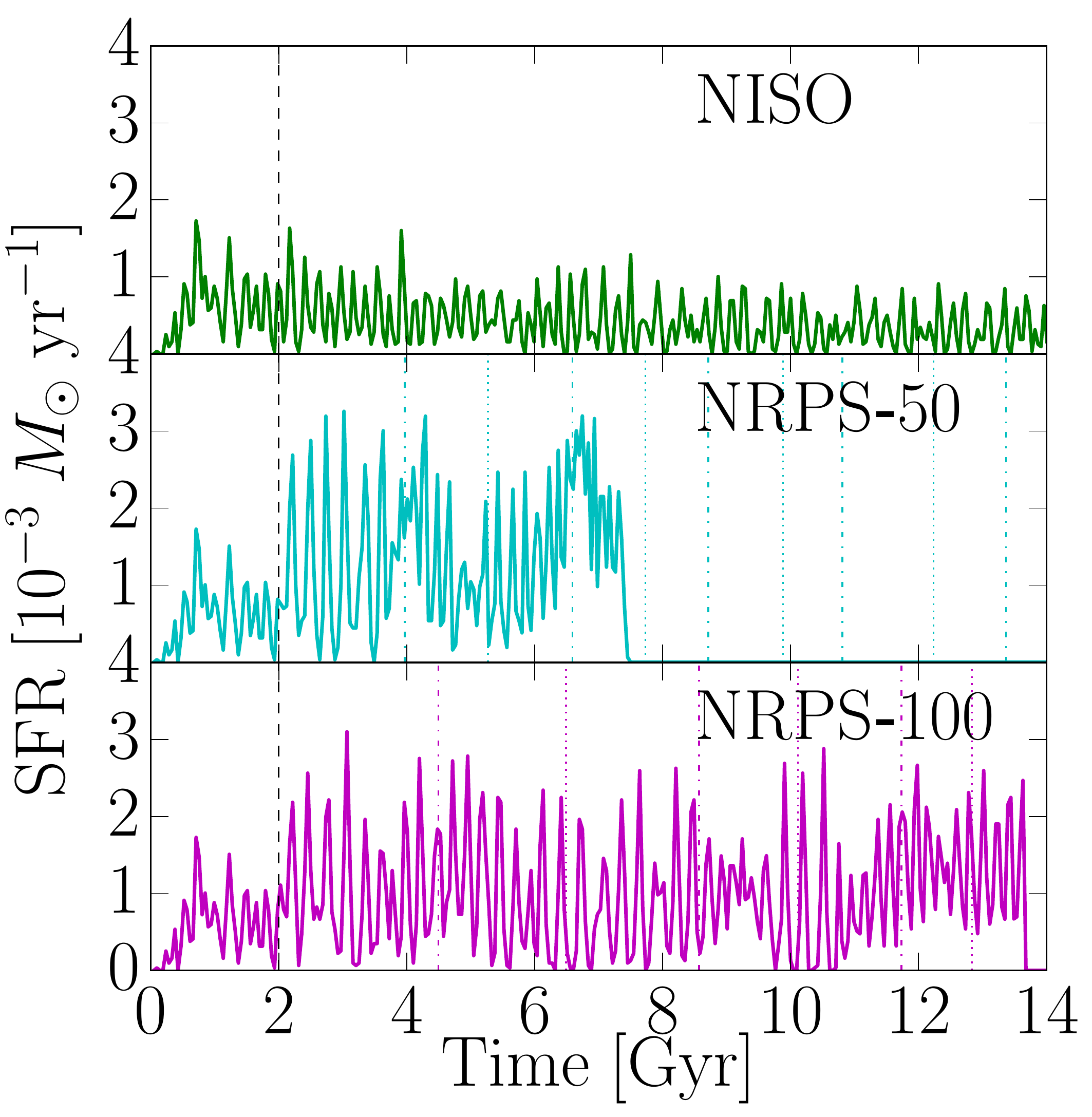}}
        \caption{The star formation rate inside the simulations. Lines and colours are the same as in Fig. \ref{fig:SFH}.}\label{fig:sfr}
\end{figure*}

For the pseudo-isothermal models, the isolation case, ISO, continues a steady approximate continuous increase in stellar mass.
Post-infall, the RPS-50 model begins to experience a large jump in star formation mass, before rapidly dropping just after the first perigalacticon passage (a sign of extensive tidal stripping as stellar particles are removed from the simulation).
This bursty behaviour is reflected in the star formation rate shown in Fig. \ref{fig:sfr}.

The RPS-100 model continues star formation for a much longer period, but undergoes two separate phases.
Firstly, the star formation is enhanced similar to that of RPS-50, but after the first perigalacticon passage, a slight decrease is seen corresponding to the change in gradient in Fig. \ref{fig:SFH}, before the dwarf is tidally disrupted after the second perigalacticon passage.

For the NFW models, because of the denser dark matter profile, the stellar mass tends to be higher than in the pseudo-isothermal models.
Again, the isolation case, NISO, continues a steady approximately continuous increase in stellar mass.
For those post-infall, however, the star formation is much more continuous than in the pseudo-isothermal models.
Still enhanced, the models continue star formation until in the case of NRPS-50 it is suddenly halted, and the rate of star formation in NRPS-100 drops after each perigalacticon passage.

Here, the star formation is enhanced in NRPS-50 before being shut down around $6$~Gyr as gas is finally removed from the dwarf towards near the second perigalacticon (but see also \S\ref{subsubsec:gasmass}).
In NRPS-100, the enhancement is again seen, but after perigalacticon the star formation rate decreases slightly, corresponding to the change in gradient in Fig. \ref{fig:SFH}.
Before finally being stopped only a few hundred $Myr$ before the end of the simulation.
That a rapid change in star formation at perigalacticon may allow the determination of the time of perigalacticon passage, particularly for recently accreted dwarfs where the time resolution of the star formation history is higher.

\subsection{Mass evolution over time}\label{subsec:Mass}

\begin{figure*}
        \centering
        \leavevmode
        \subfigure[Pseudo-isothermal DM mass]{\includegraphics[width=0.495\textwidth]{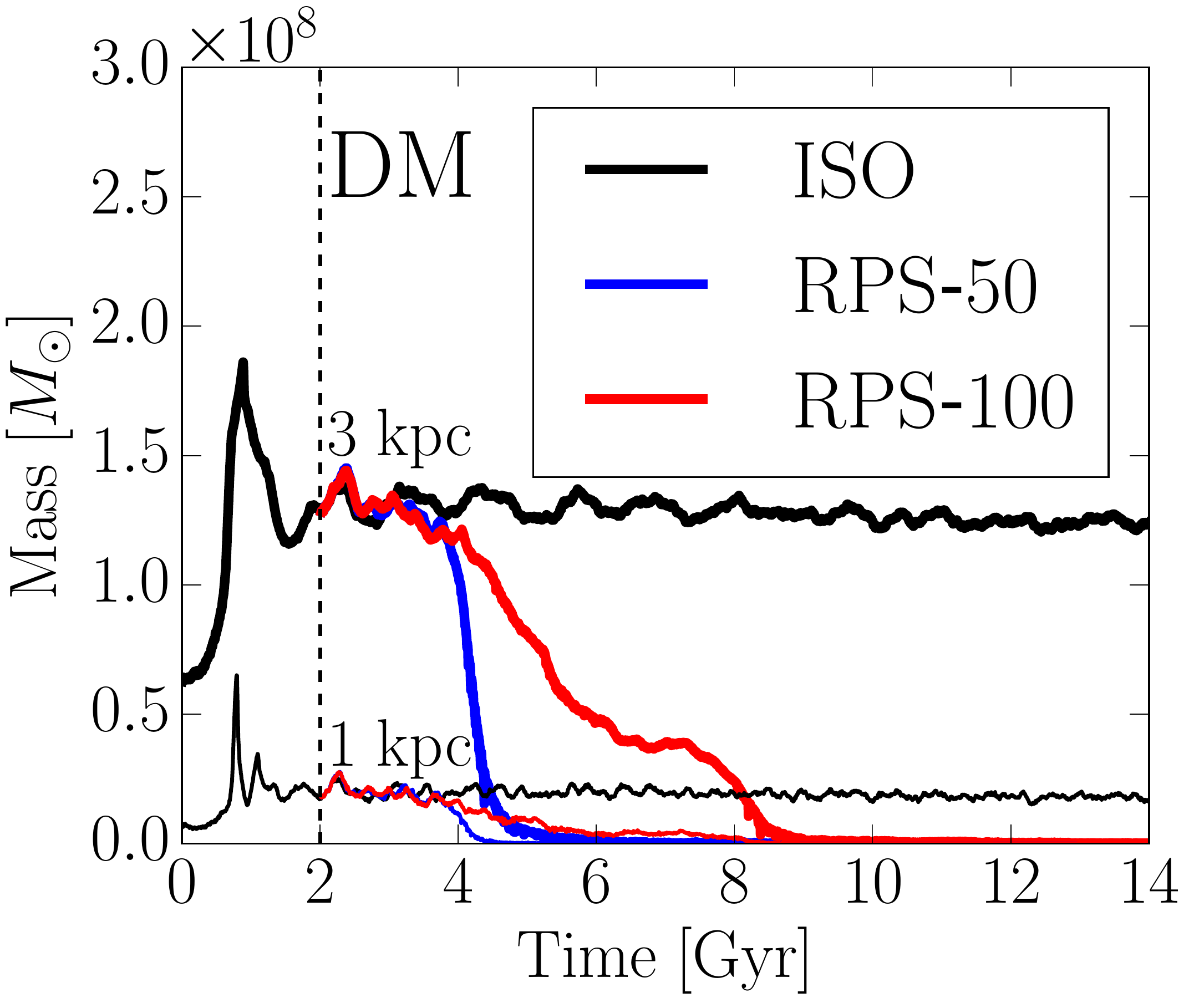}}
        \subfigure[NFW DM mass]{\includegraphics[width=0.495\textwidth]{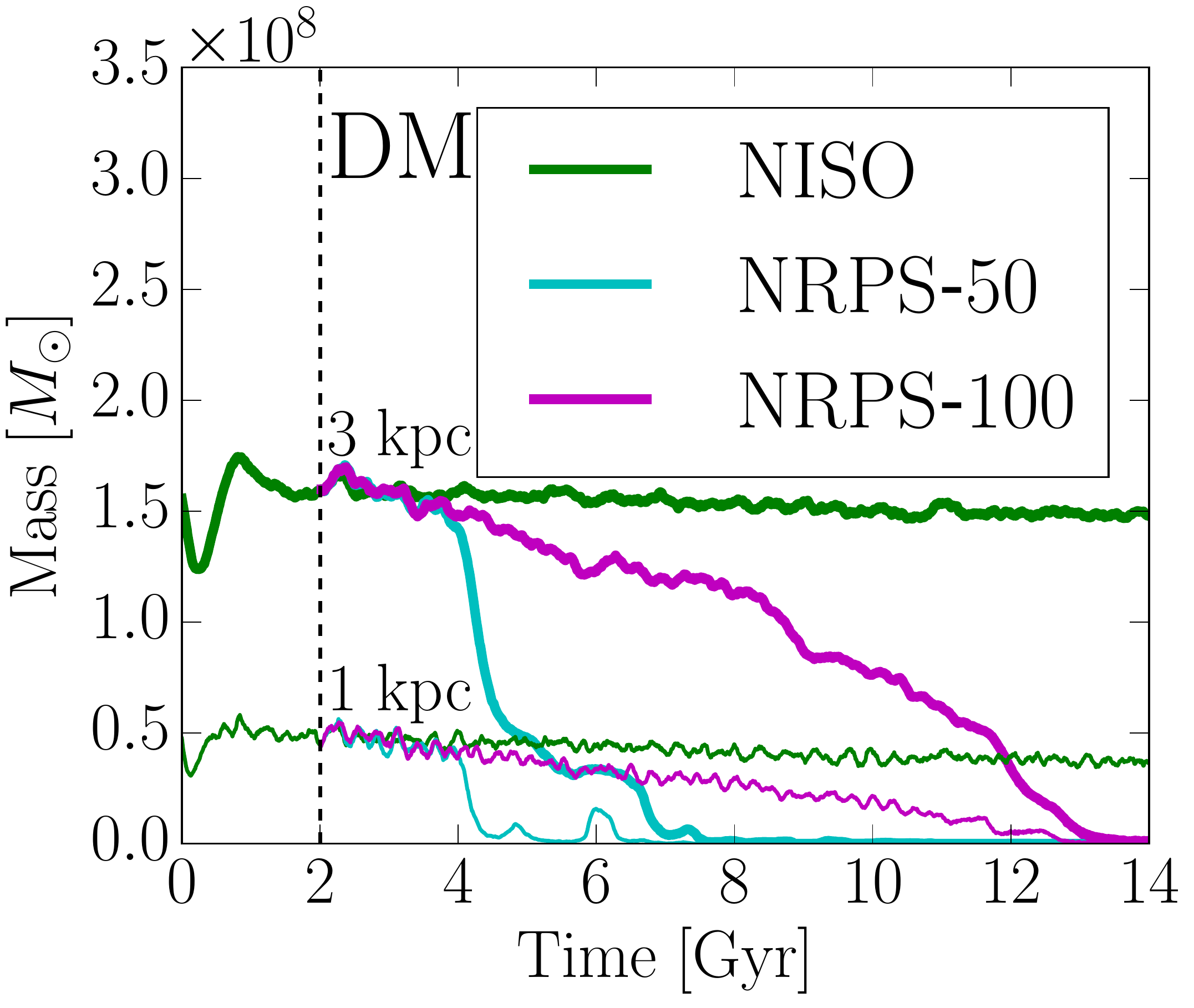}}
        \subfigure[Pseudo-isothermal gas mass]{\includegraphics[width=0.495\textwidth]{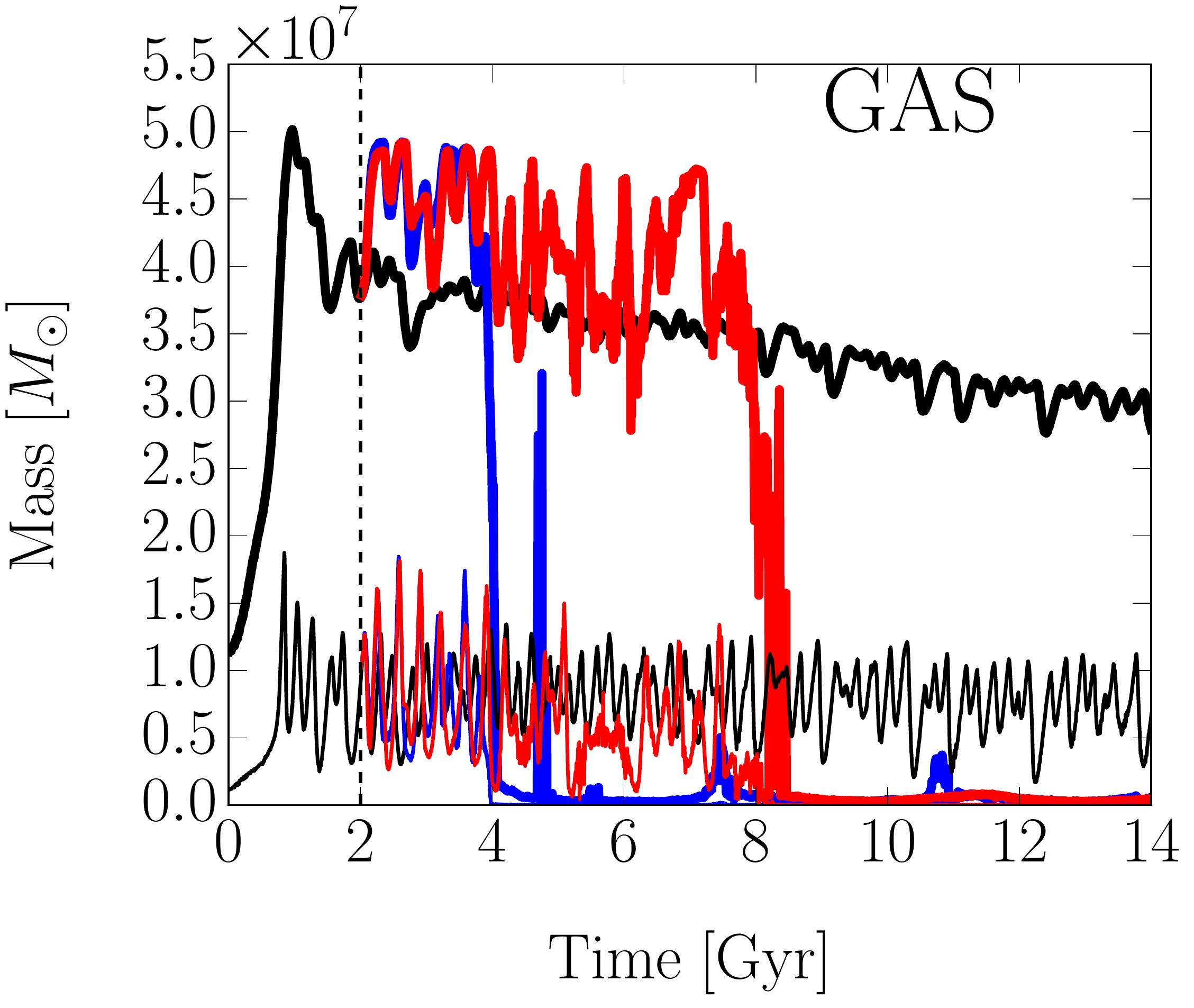}}
        \subfigure[NFW gas mass]{\includegraphics[width=0.495\textwidth]{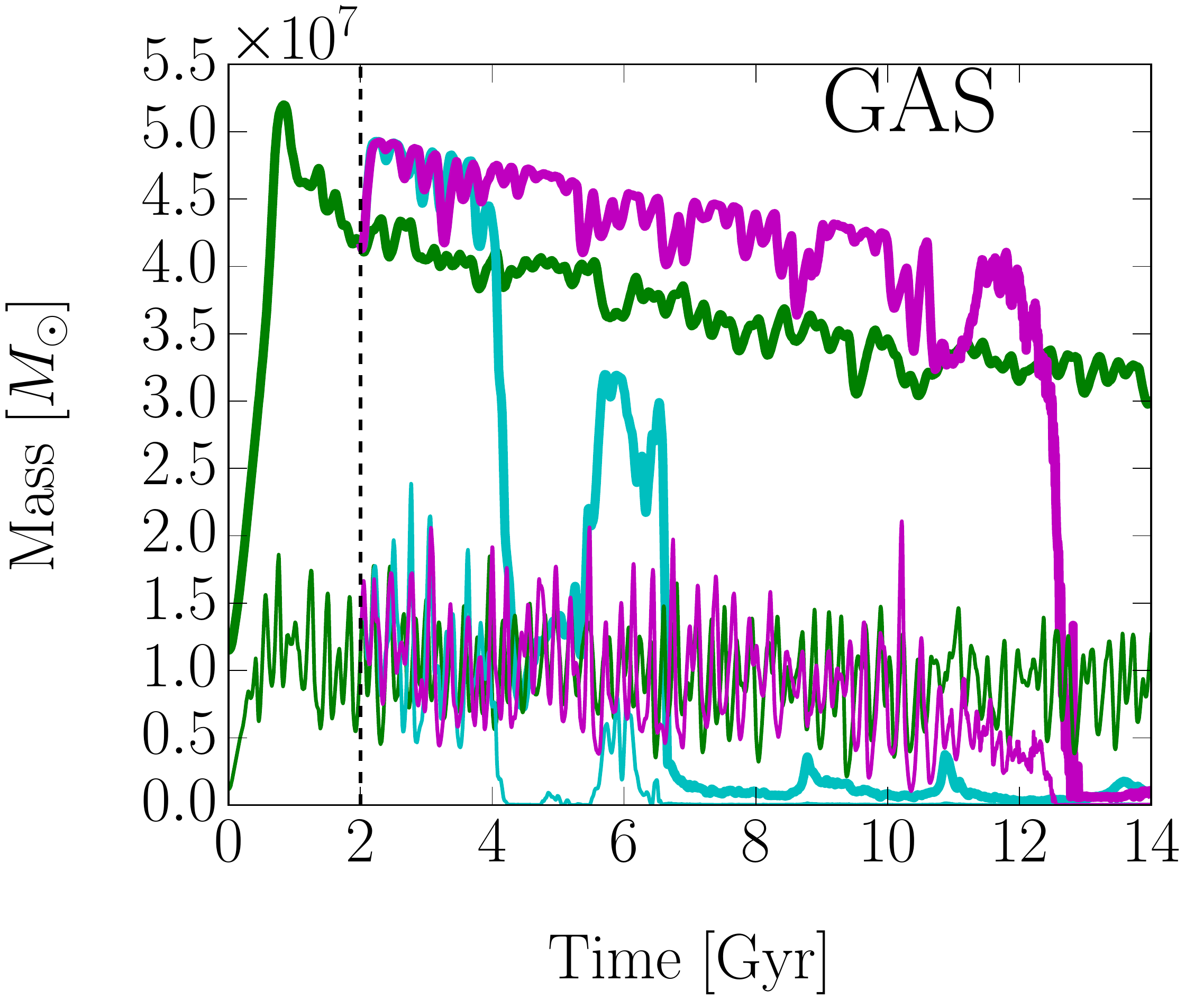}}
        \subfigure[Pseudo-isothermal stellar mass]{\includegraphics[width=0.495\textwidth]{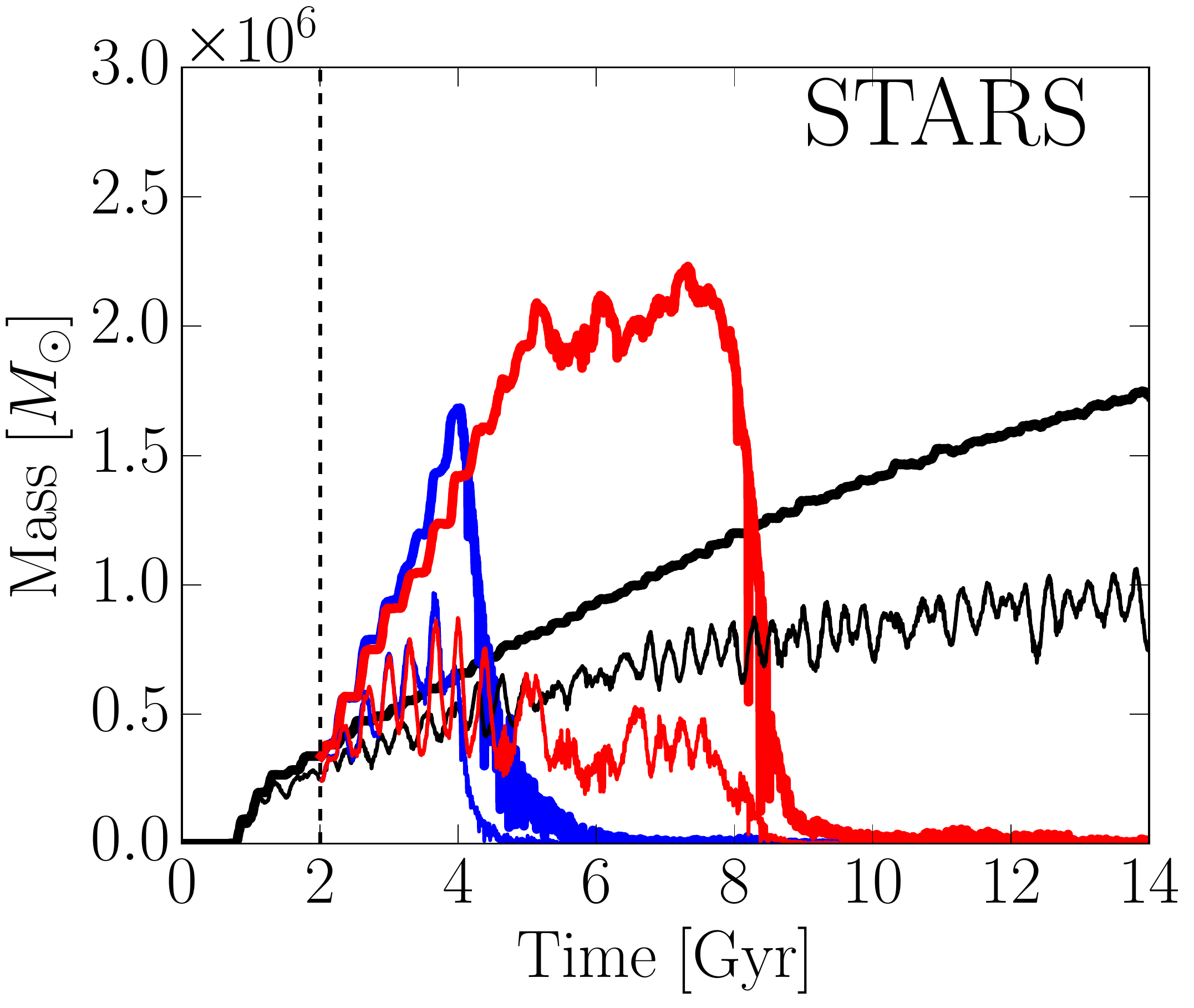}}
        \subfigure[NFW stellar mass]{\includegraphics[width=0.495\textwidth]{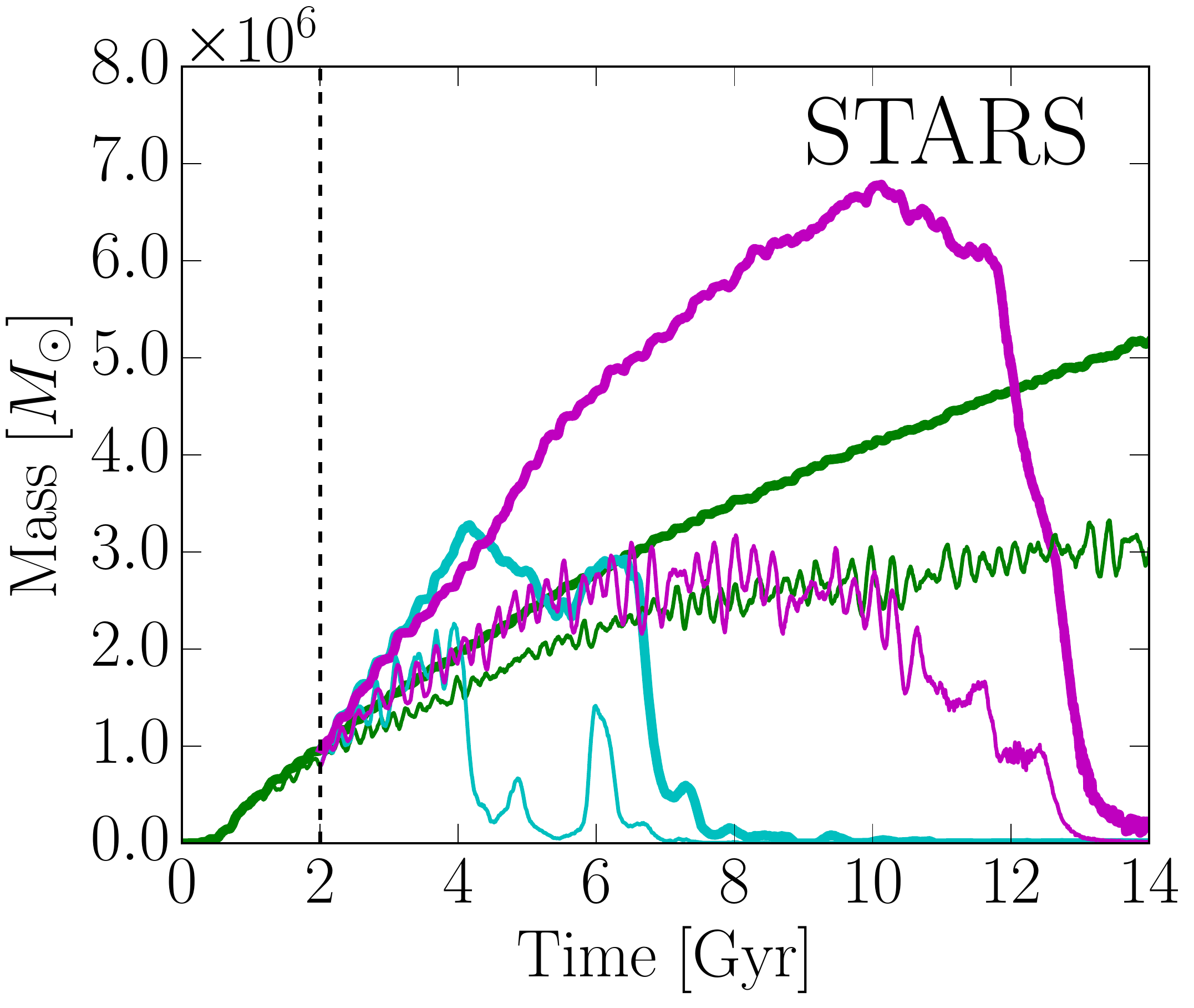}}

        \caption{Mass of dark matter (top row), stars (middle row), and gas (bottom row) within the central $1$ and $3$~kpc of the pseudo-isothermal and NFW dwarfs. The ISO model is shown with a black line, RPS-50 with a blue line, and RPS-100 with a red line. The NISO model is shown with a green line, NRPS-50 with a cyan line, and NRPS-100 with a magenta line. The $3$~kpc line is shown with thick lines and $1$~kpc with thin lines. The infall time is shown with a vertical black dashed line.}
        \label{fig:mass}
\end{figure*}

At the same time that the gas is moving within the dwarf, the mass contained within the central regions of the dwarf changes dramatically.
Figs \ref{fig:mass} shows the mass inside the central $1$ and $3$~kpc for the pseudo-isothermal and NFW models versus time.

\subsubsection{Dark matter mass}
The dark matter mass is stable (after initial settling) in the isolation model shown in panel (a), with slight oscillations in the mass as gas is blown out and reaccretes.
As tidal forces begin to affect the dwarf in RPS-50 and RPS-100 near the first perigalacticon, a {\ebf dramatic} drop in dark matter mass is seen.
{\ebf The drop in mass at the same time as baryons is lost is not surprising having been explored in \citet{Arraki2014}, who found that the removal of baryons from a dwarf tends to drop the density (and therefore mass) by a factor up to $(1+f_{\rm b})^4$. The effects here however are much stronger than the $\sim$$50\%$ mass loss predicted by \citeauthor{Arraki2014} for total baryon loss, and may be the consequence of all the mass being lost near perigalacticon.}

This reduction is enormous when considering the closest either model gets is $50$~kpc, a distance far from sufficient to tidally destroy dwarfs {\ebf when ram pressure stripping is not included} \citepalias{Nichols2014}.
Here the RPS-50 model is effectively destroyed after one passage, and the RPS-100 model is reduced by almost an order of magnitude and exists as only a stream, as indicated by the dark matter mass being near zero within the central $3$~kpc.

{\ebf The synergy between baryon loss and dark matter loss has also been noted in \citet{Zolotov2012}, where dwarfs with baryons lost a greater mass fraction than those in a simulation that only includes dark matter, particularly in those cases where dwarfs have a baryonic disk.
Here, with no disk present we see similar synergy, and this drop in density occurs due to the alteration in the potential, resulting from gas being lifted to higher orbits from supernova feedback \citep{Governato2010,Pontzen2012} and ram pressure stripping.\footnote{Although the mechanisms from supernova feedback are likely to lower the central density independently, in these simulations they do not occur rapidly enough to result in a net change. However even with an adiabatic change, a core can be created because the gas is prevented from reaccreting by ram pressure stripping that prevents  the dark matter from adiabatically contracting and consequently the density is still reduced.}}
That this tidal effect is so large, {\ebf completely removes} the pseudo-isothermal dwarfs resemblance to a classical dwarf spheroidal today, suggesting that Sextans at least looked remarkably different at infall to its present day structure.

Compared to ISO, NISO, shown in panel (b), shows a higher mass of dark matter in the central $3$~kpc, but the same fluctuations, corresponding to bursts of star formation and outflows of gas, are seen as in the ISO case.
As with the pseudo-isothermal models, dramatic drops in dark matter mass are seen when the dwarfs approach perigalacticon, in addition to the normal $1$~kpc fluctuations.
Because of the denser inner region of the dwarf, these dwarfs are more resistant to tidal stripping and, in both cases, take almost twice as long to be transformed into streams.

{\ebf To ensure the mass loss is a consequence of the ram pressure stripping and not a numerical artefact we have also simulated the evolution of a baryon free dark matter dwarf and a dark matter dwarf moving through a baryon free halo and therefore not experiencing ram pressure.
In both cases, these dwarfs survive with a noticeable dark matter core until the present day, suggesting that ram pressure exacerbates tidal stripping.
We describe these simulations in Appendix \ref{app:norms}.}

\subsubsection{Gas mass}\label{subsubsec:gasmass}
In contrast to the relatively stable nature of the dark matter in ISO, the gas mass in all pseudo-isothermal simulations, shown in panel (c), changes dramatically with bursts of star formation, particularly in the central $1$~kpc, which changes with star formation by a factor of two.
In the ram pressure simulations, the central $3$~kpc also experiences dramatic changes of $30$--$40$\%, with the central $1$~kpc experiencing changes by more than a factor of three.
Although only a few bursts take place, a general trend is seen, with slightly less material reaccreted each time, particularly to the inner $1$~kpc before finally being stripped near the first perigalacticon.

In the RPS-50 model, a spike is seen around $5$~Gyr ($3$~Gyr post infall), which is caused by stripped gas re-entering via the side of the box and within the central $3$~kpc of the dwarf.
As this gas has a large velocity, it never forms stars that remain bound to the dwarf, and subsequently does not impact any future analysis.
Around $8$~Gyr ($6$~Gyr post infall) and $11$~Gyr ($9$~Gyr post infall) a slight increase in the mass is seen because the dense galactic corona has enough mass within the central $3$~kpc to be noticeable and the rise in gas mass at this time is not an accretion event.

For the NFW models, the gas mass, shown in panel (d), shows similar trends as the pseudo isothermal models,
Once again, the compressive force increases the star formation and the oscillations of this gas.
In the NRPS-50 model, a clear drop in the gas mass is seen at perigalacticon, soon after a large amount of gas is then reaccreted at apogalacticon ($\sim$$6$~Gyr).
This gas is held slightly off centre throughout the dwarf's passage before the lower ram pressure forces allow it to recall into the central potential well of the dwarf, a mechanism for star formation bursts suggested previously in \citet{Nichols2012}.
This is the only model (including those not shown in this paper, with variations in star formation efficiency, supernova feedback, initial conditions etc.) in which this kind of mechanism appeared, and therefore any reaccretion event like this is likely to be rare.
The small increases seen at the perigalacticon passages  are also visible.
The NRPS-100 model has a comparatively simple history, gradually decreasing in mass as ram pressure strips the outer edges of the gas until the dwarf is tidally destroyed and all gas is loss.
The complicated nature of the gas removal suggests that any simple method to remove gas in semi-analytic simulations will likely end up as a poor approximation.

\subsubsection{Stellar mass}
The stellar mass in the pseudo-isothermal models, shown in panel (e), increases with bursts of star formation, discussed in \S\ref{subsec:SF}, with some change within the central $1$~kpc as gas expulsion drags some of the stellar mass to higher orbits, before the reaccretion brings material back.
The stellar mass in the central $3$~kpc for the ISO model grows monotonically with time, with the gas expulsion failing to drag any stars further out than this.
As the ram pressure models approach perigalacticon, a large stellar mass is lost from the inner central kiloparsec.
For RPS-50, these stars become unbound as the dwarf is rapidly transformed into a stream, but for the other ram pressure models most of these stars move to higher orbits outside the inner kiloparsec, with the stellar mass inside only suffering a small reduction until the dwarf is completely destroyed.

Because of the denser dark matter profile, the stellar mass within $1$ and $3$~kpc in the NFW dwarfs, shown in panel (f), tends to be higher than in pseudo-isothermal dwarfs.
As tidal forces begin to act on the dwarf, many stars move to higher orbits within the dwarf, reducing the mass within $1$~kpc even while the total stellar mass within $3$~kpc continues to increase.
This is partially counteracted in NRPS-50 by the reaccretion event dragging some stars back to lower orbits, in addition to the formation of new stars.
As in the pseudo-isothermal models, the dwarfs eventually become streams and the stellar mass within the central $3$~kpc drops to $\sim$$0$ as a consequence of this.

Ultimately all pseudo-isothermal and NFW models become tidal streams, however, the NFW profiles survive for a much longer time, surviving past the first perigalacticon with a large amount of gas.

\subsection{Spatial gaseous and stellar distribution}
\begin{figure*}
        \centering
        \includegraphics[height=0.9\textheight]{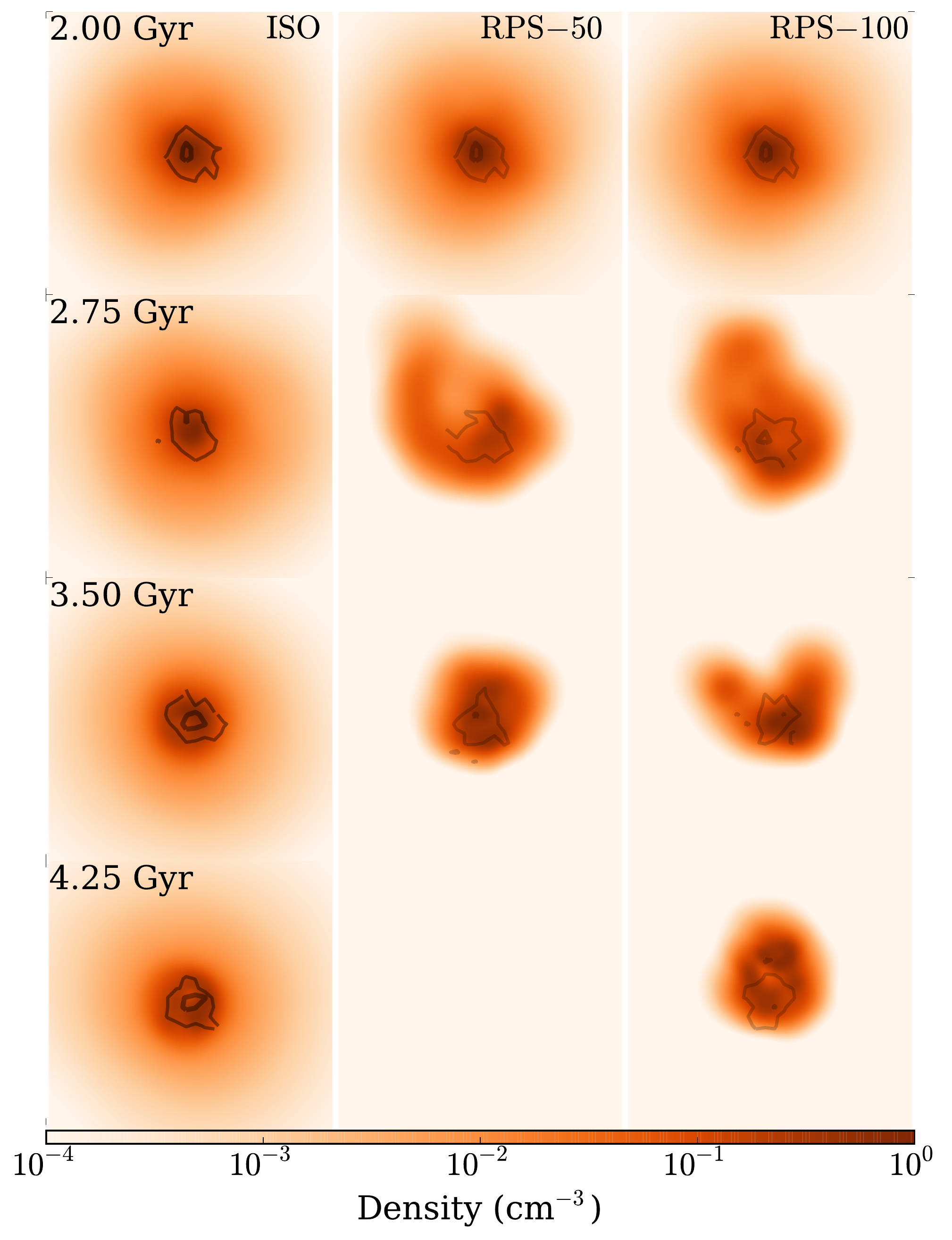}
        \caption{Density of gas (image) and dark matter (contours) at $750$~Myr intervals for pseudo-isothermal models from infall; beginning at infall in a $0.5$~kpc slice through the centre of the dwarf.
        Only gas below $10^6$~K is shown.
        Each box is $15$~kpc per side.
        The ISO model is shown in the left column, RPS-50 in the middle column, and RPS-100 in the right column.
        The gas density is listed on the colour bar and the dark matter contours (from thick to light) are $10^{6.5}$, $10^{6}$, and $10^{5.5}$~M$_\odot$~kpc$^{-3}$.}\label{fig:rps-dens}
\end{figure*}

\begin{figure*}
        \centering
        \includegraphics[height=0.9\textheight]{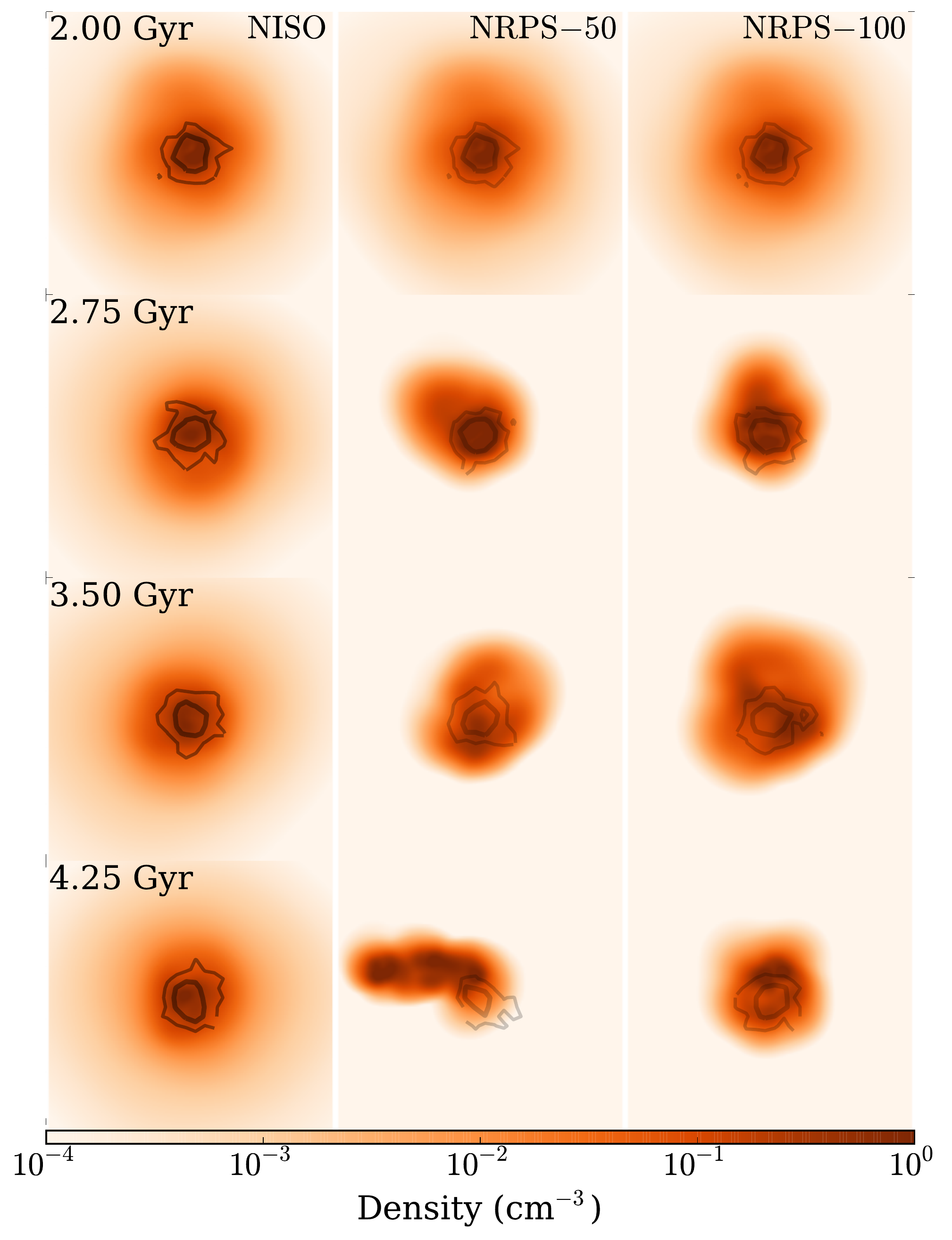}
        \caption{Density of gas (image) and dark matter (contours) at $750$~Myr intervals for NFW models from infall until around the first perigalacticon, beginning at infall in a $0.5$~kpc slice through the centre of the dwarf.
        Only gas below $10^6$~K is shown.
        Each box is $15$~kpc per side.
        The NISO  model is shown in the left column, NRPS-50 in the middle column and NRPS-100 in the right column.
        The gas density is listed on the colour bar and the dark matter contours (from thick to light) are $10^{6.5}$, $10^{6}$, and $10^{5.5}$~M$_\odot$~kpc$^{-3}$.}\label{fig:rps-dens-nfw}
\end{figure*}

As gas is stripped from the dwarf, it goes through numerous forms, which is shown for the pseudo-isothermal dwarf in Fig. \ref{fig:rps-dens} and for the NFW dwarfs in Fig. \ref{fig:rps-dens-nfw}. In Fig. 9, we show gas with a temperature below $10^6$~K through a $0.5$~kpc slice of the dwarf.
As would be expected, the loosely bound outer gas of the dwarf is quickly removed within a few hundred megayears.
The increase in star formation and related feedback, resulting from the confining pressure,  begins to blow out gas quite quickly. However, because of the relative velocity of the medium, this gas is not reaccreted in spherical shells, and instead forms a flattened bullet-like shape (see e.g. RPS-100 at $3.5$~Gyr or NRPS-100 at $2.75$~Gyr), with the leading edge of gas compressed by the ambient medium.
By $1.5$~Gyr after infall ($3.5$~Gyr of evolution), the dark matter is warped by tides and the removal of gas, becoming elongated in the direction of motion.
At the same time, the gas is no longer centred on the dwarf, but is offset to the side that is opposite the direction of motion.

This off-centre behaviour results in a noticeable change in the location of star formation (discussed below) and at the same time slows down the dwarf altering its orbit (this interesting feature is discussed later in \S\ref{subsec:orbchange}).

In the NFW models, a similar evolution is seen, except that the central gas is denser, and fails to experience the massive blowouts.
Similar to the pseudo-isothermal models, slight distortions in the outer dark matter profile are seen, and the gas is held slightly off-centre.
As the dwarf begins to approach perigalacticon, all the gas in RPS-50 is removed (last row), while in NRPS-50, RPS-100, and NRPS-100, some gas is able to survive and is held off-centre by the dwarf.

\begin{figure*}
        \centering
        \includegraphics[height=0.86\textheight]{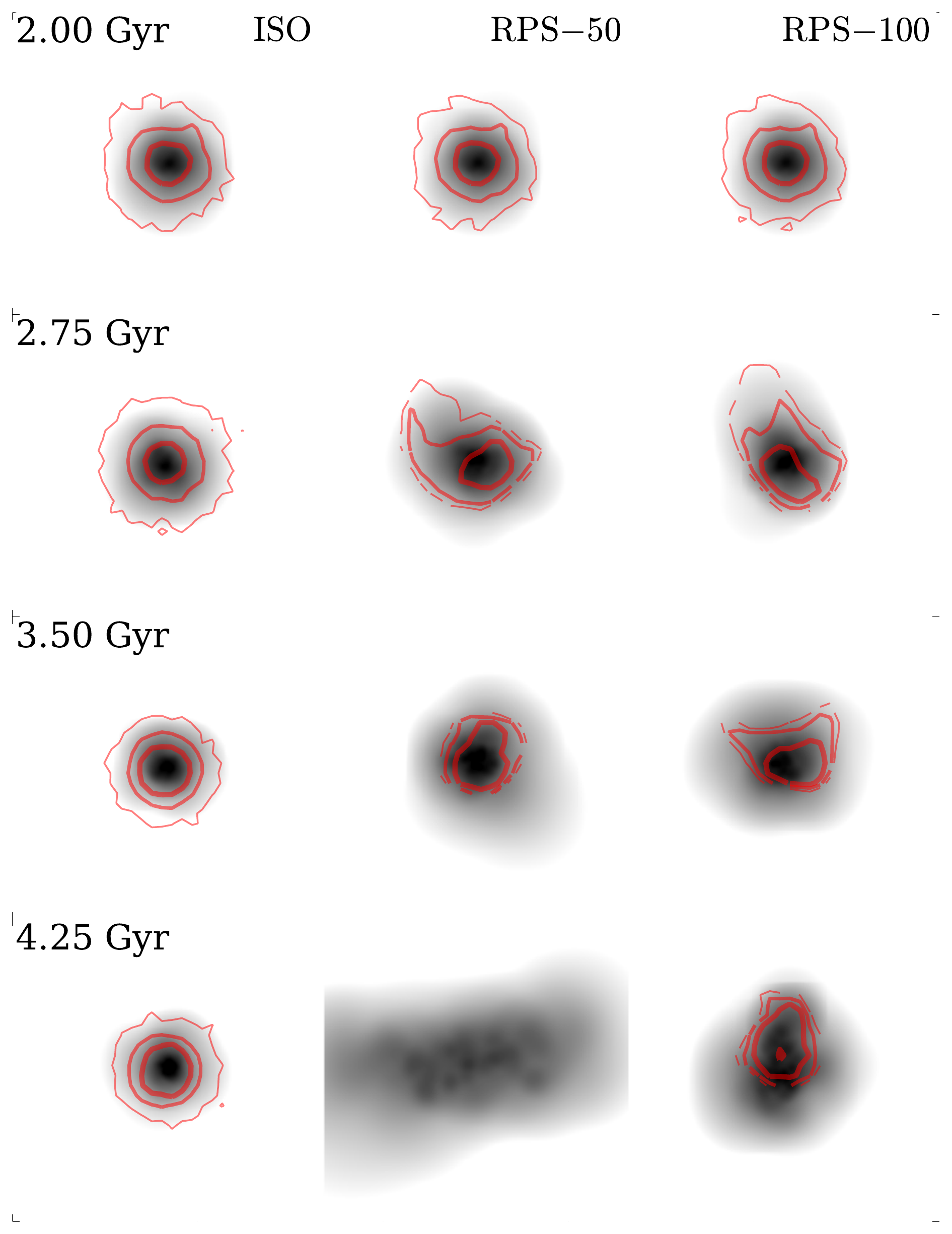}
        \caption{Luminosity of stars at $750$~Myr intervals from infall for the pseudo-isothermal models. The stellar particles are smoothed to encapsulate the $16$ nearest neighbours, as per \citet{Torrey2015}. Cool gas ($T<10^5$~K) column density contours are shown in red.}
        \label{fig:stardens}
\end{figure*}

The fact that the gas is no longer distributed in a roughly spherical fashion has a large impact on the positions of star formation.
In Fig. \ref{fig:stardens} we show the luminosity for the pseudo-isothermal models at the same times as the gas density in Fig. \ref{fig:rps-dens}.
As the luminosity is dominated by young stellar populations, this traces the star-forming regions within the dwarf.
As always, star formation takes place in regions where dense gas resides, in dwarfs undergoing ram pressure, which has the effect of creating obvious distortions from sphericity in the luminosity profile, particularly in the pseudo-isothermal models.
However, the dwarf's gravity and tidal effects tend to hide these distortions with time.

\begin{figure*}
        \centering
        \includegraphics[height=0.86\textheight]{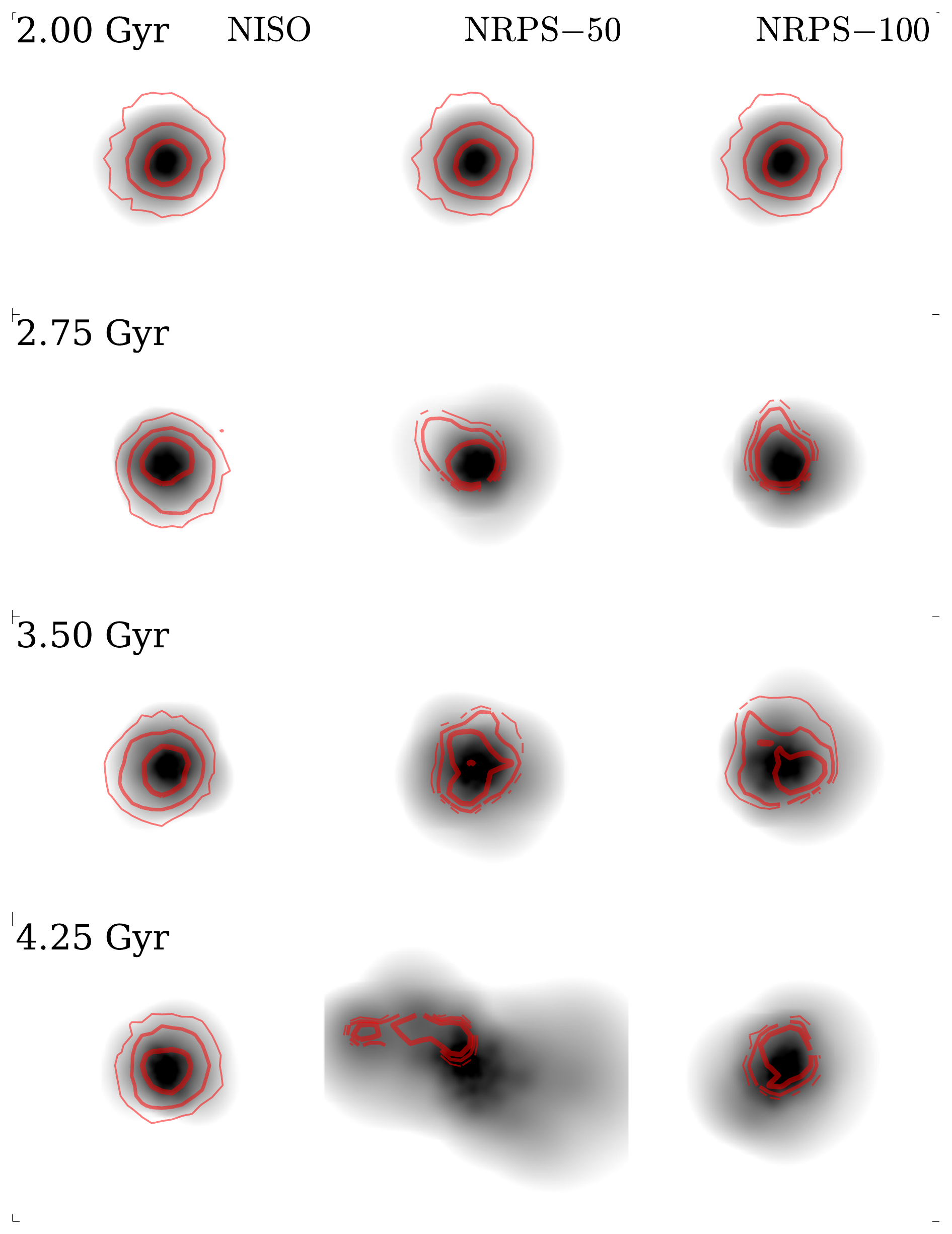}
        \caption{Luminosity of stars at $750$~Myr intervals from infall for the NFW. The stellar particles are smoothed to encapsulate the $16$ nearest neighbours, as per \citet{Torrey2015}. Cool gas ($T<10^5$~K) column density contours are shown in red. }\label{fig:stardens-nfw}
\end{figure*}

In the NFW models, shown in Fig. \ref{fig:stardens-nfw}, this effect is much more reduced with nearly all the stars held in the central region, with only a low brightness halo of star formation, as would be expected from the reduced blowouts.
These blowouts still result in a slight distortion in the NRPS-100 model as in RPS-100, and similarly, the tidal tails of RPS-50 also appear inside NRPS-50.
Of particular note is that as the last gas is expelled, stars continue to form in the clouds, resulting in a (temporary) bridge of stellar material outside the dwarf, particularly visible in the last row of Fig. \ref{fig:stardens-nfw} for the NRPS-50 model.

\subsection{Metallicity}

\begin{figure*}
        \centering
        \leavevmode
        \subfigure[Pseudo-isothermal]{\includegraphics[width=0.495\textwidth]{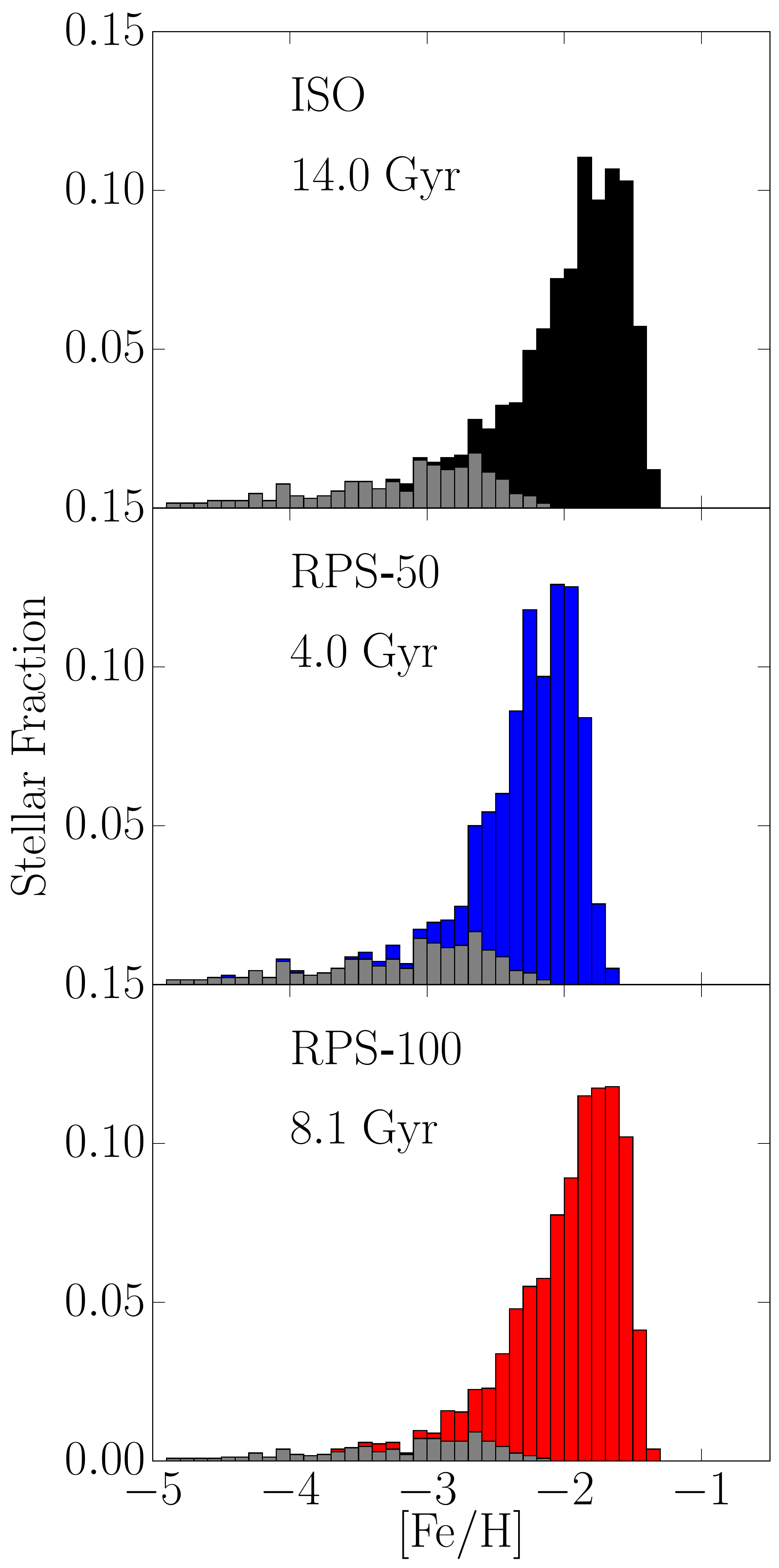}}
        \subfigure[NFW]{\includegraphics[width=0.495\textwidth]{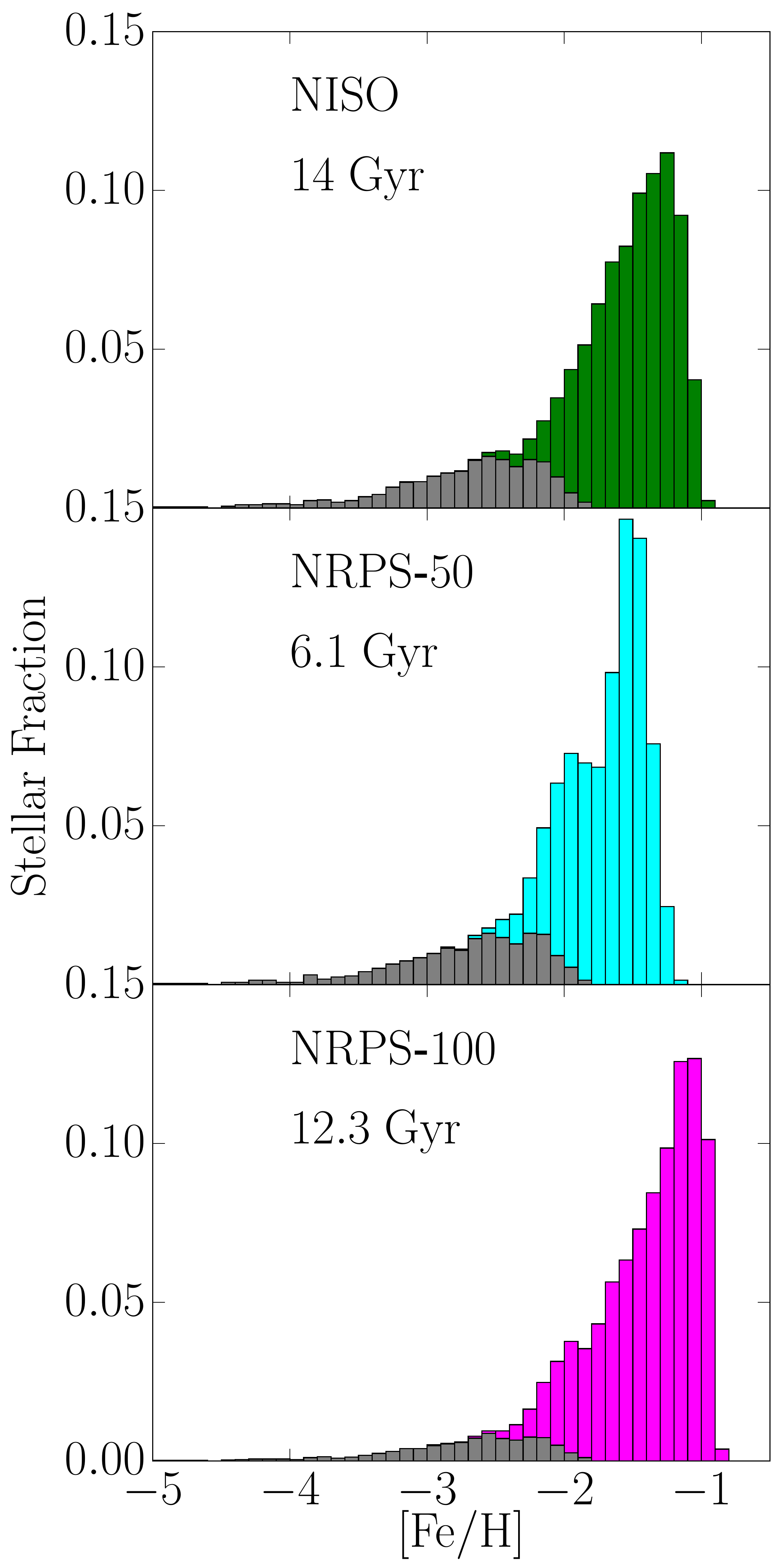}}
        \caption{$[$Fe$/$H$]$ distribution of stars in the pseudo-isothermal simulations at the point of maximum stellar mass. Time of maximum stellar mass is shown. Stars formed before infall are shown in each plot in grey.}\label{fig:FeH}
\end{figure*}

The increase in star formation that takes place upon infall would normally suggest a quick evolution of metallicity as generations of stars pollute the interstellar medium.
However, when a dwarf falls into a hot halo, its metallicity no longer necessarily undergoes this rapid evolution due to star formation taking place in low-metallicity gas  originally from the exteriors of the dwarf (discussed further in \S\ref{subsubsec:SFgas}).
We show the $[$Fe$/$H$]$ in Fig. \ref{fig:FeH}, with the pseudo-isothermal models in the left panel and NFW models on the right.
For each model, we consider the metallicity at the point of maximum stellar mass.

While the ISO and RPS-100 models both peak around $[$Fe$/$H$]$$\sim$$-1.7$, the RPS-50 model peaks at $[$Fe$/$H$]$$\sim$$-2$ despite both the RPS-50 and RPS-100 models possessing a greater stellar mass than the isolation model and star formation taking place over a number of gigayears.

For the NFW isolation model, NISO, the $[$Fe$/$H$]$ peaks at higher metallicity $[$Fe$/$H$]$$\sim$$-1.3$, a consequence of the greater star formation.
However once again the ram pressure model with a perigalacticon of $50$~kpc peaks at $[$Fe$/$H$]$$\sim$$-2$ and at the time of the reaccretion event (slightly later than the peak of stellar mass\ but representing a more evolved dwarf) has a peak of only $[$Fe$/$H$]$$\sim$$-1.6 $, which is exaggerated by the loss of stars due to tidal stripping.
With its much higher star formation, the NRPS-100 model is able to evolve further in metallicity, reaching a peak of  $[$Fe$/$H$]$$\sim$$-1.1,$ and forms  a significant fraction of stars above $[$Fe$/$H$]$$=-1$.

\begin{figure*}
        \centering
        \leavevmode
        \subfigure[Pseudo-isothermal]{\includegraphics[width=0.495\textwidth]{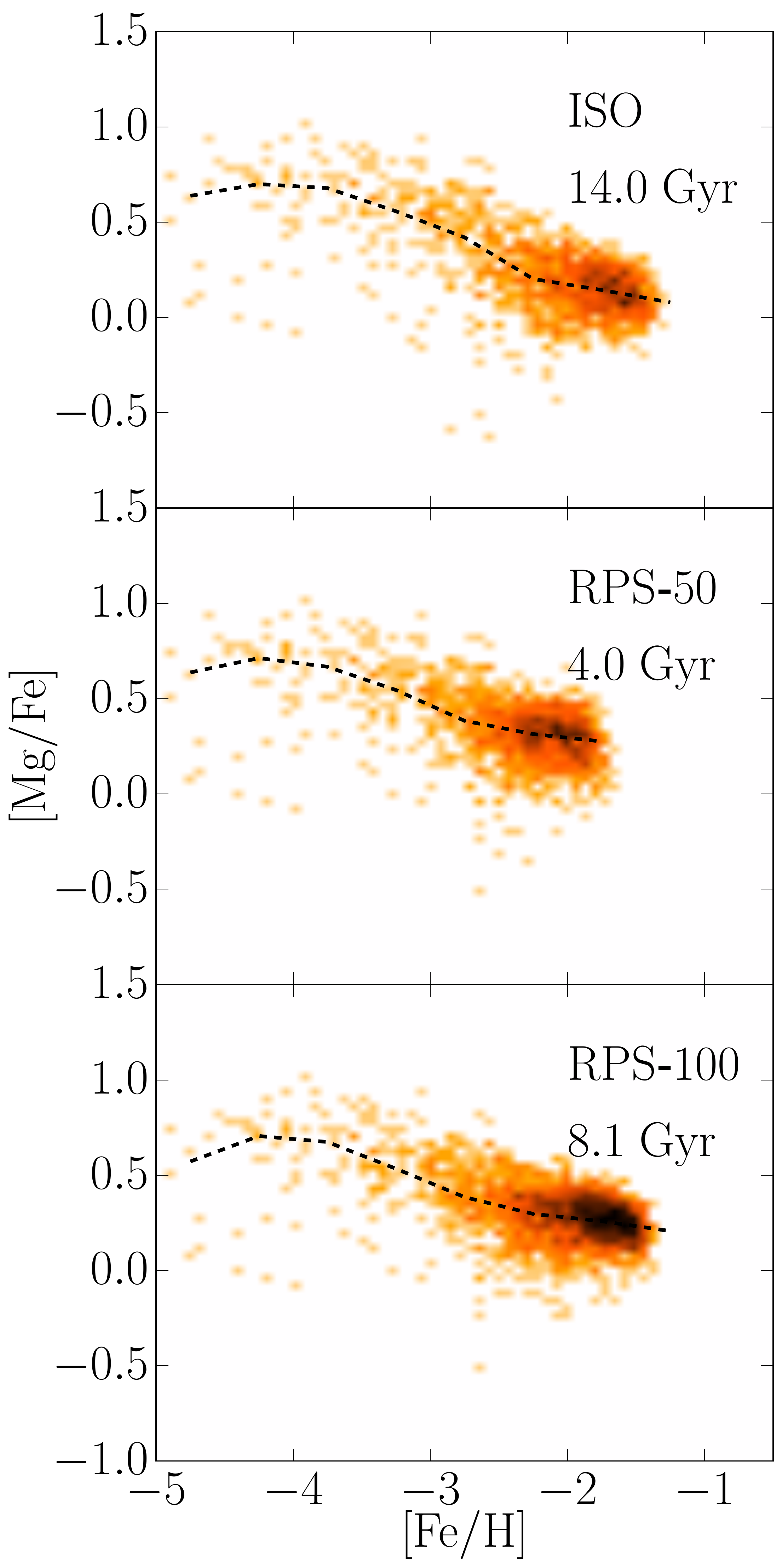}}
        \subfigure[NFW]{\includegraphics[width=0.495\textwidth]{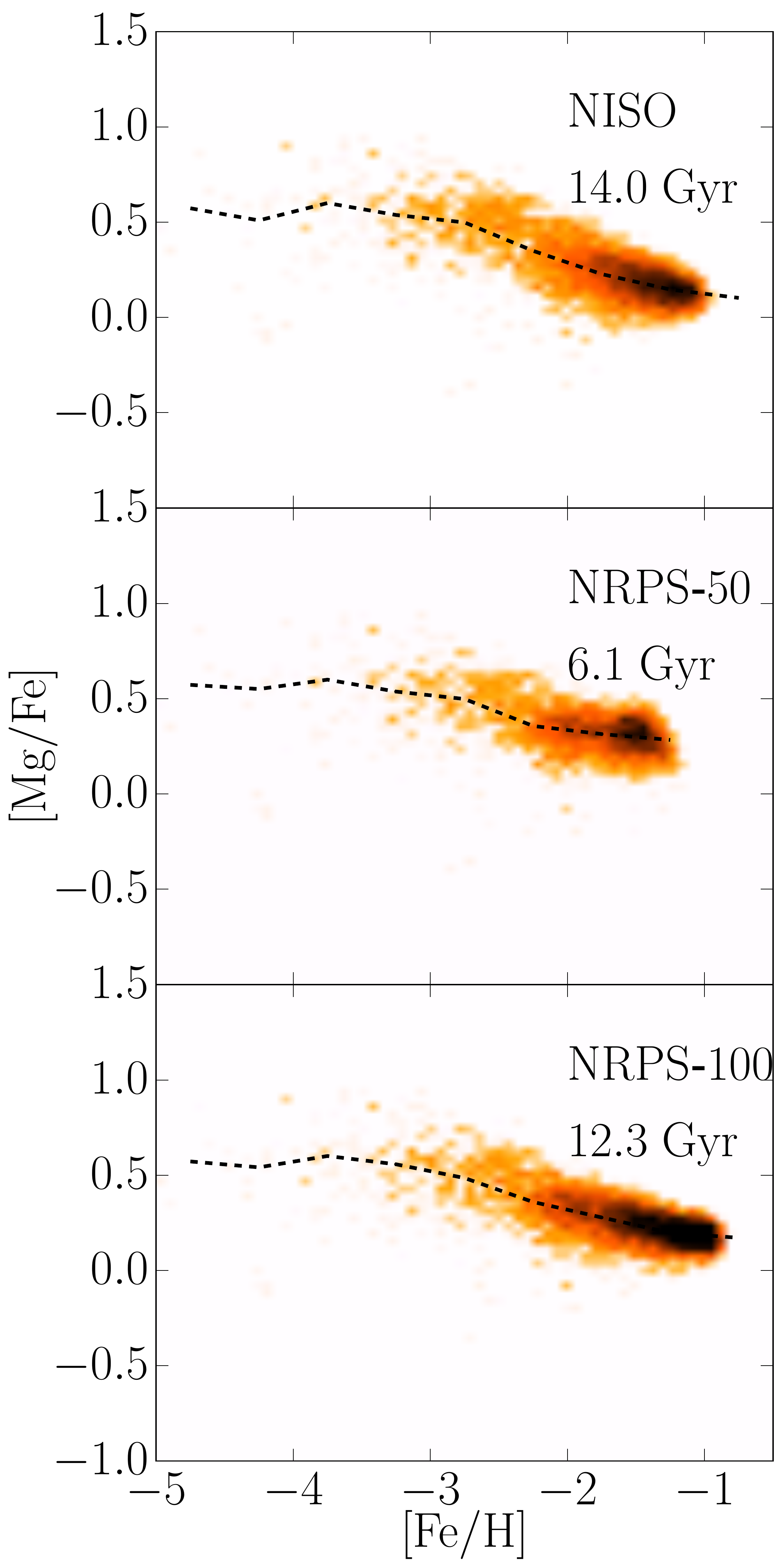}}
        \caption{$[$Mg$/$Fe$]$ of stars weighted by mass of each stellar particle in the pseudo-isothermal simulations at the point of maximum stellar mass. The black dashed line shows a running median over each $0.5$~dex bin in $[$Fe$/$H$]$.}\label{fig:MgFe}
\end{figure*}

This halting in $[$Fe$/$H$]$ evolution is also seen in the $[$Mg$/$Fe$]$ evolution shown in Fig. \ref{fig:MgFe}.
Because of  the increase in star formation, Type II supernova continue to enrich the medium in alpha elements, resulting in a higher $[$Mg$/$Fe$]$ past the knee.
This feature is common between the pseudo-isothermal and NFW ram pressure models and results in the isolation models having a larger fraction of low alpha stars ($[$Mg$/$Fe$]$ < 0) than any of the ram pressure models regardless of the $[$Fe$/$H$]$ distribution.

\subsubsection{Star-forming gas}\label{subsubsec:SFgas}

\begin{figure*}
        \centering
        \leavevmode
        \subfigure[Pseudo-isothermal]{\includegraphics[width=0.495\textwidth]{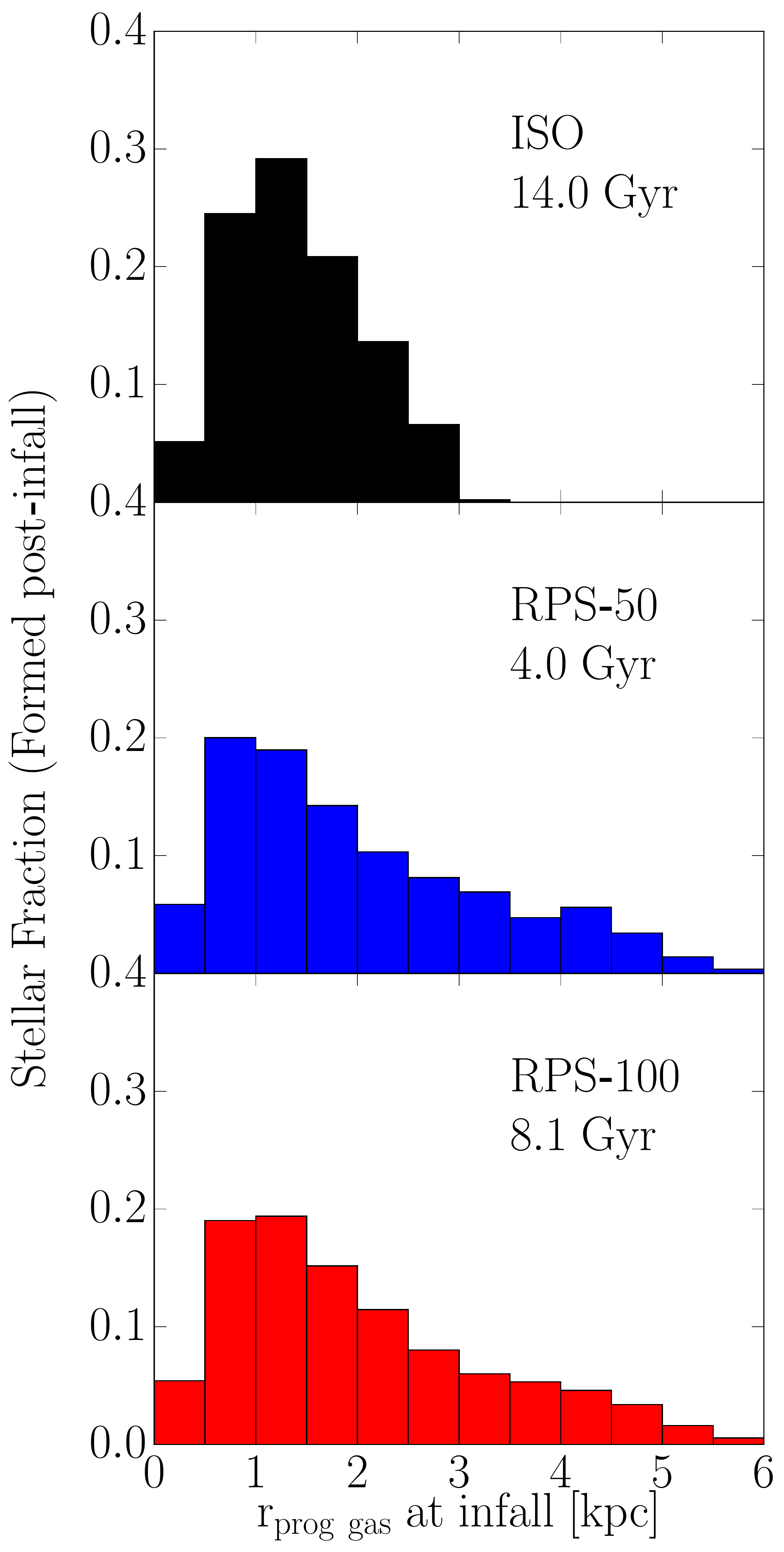}}
        \subfigure[NFW]{\includegraphics[width=0.495\textwidth]{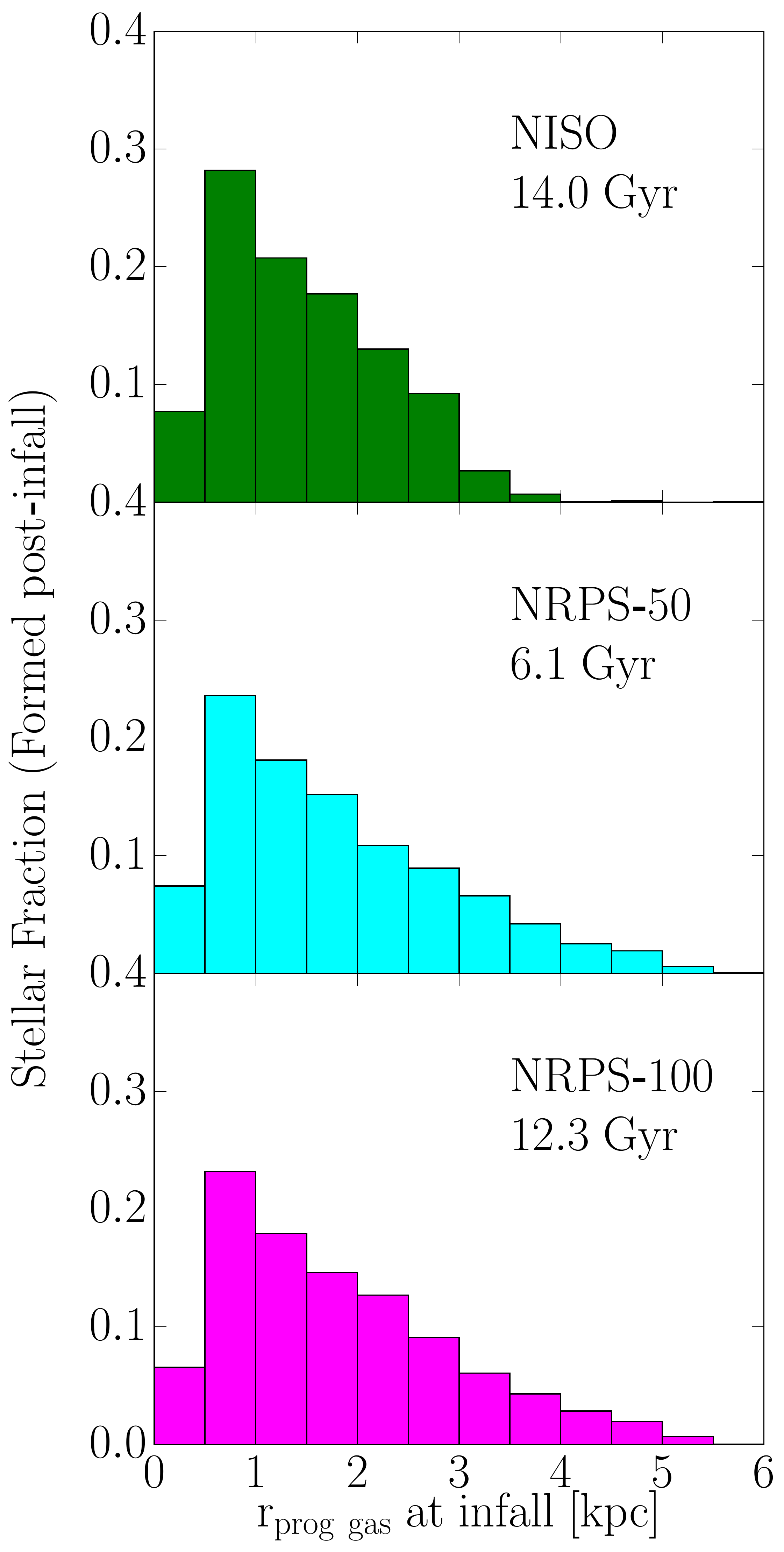}}
        \caption{Distribution of the radius of gas at infall inside the simulations, which goes on to form stars by the point of maximum stellar mass.}\label{fig:Rstars}
\end{figure*}

Although only a minor component (particularly for the NFW models), it can be seen in Fig. \ref{fig:FeH} that stars of low-metallicity $[$Fe$/$H$]<-3$ continue to form after infall.
As all the central gas on infall possess a metallicity above this, at least some stars must be formed from gas originating in the outskirts of the pseudo-isothermal dwarf, and this is a likely cause for the halting in metallicity evolution.
In Fig. \ref{fig:Rstars} we show the radius of gas at $2$~Gyr that goes on to form stars in the future.
It is immediately clear that nearly all gas that goes on to form stars in the isolation case is located within the central $2.5$~kpc of the dwarf.
However, for the RPS-50 and RPS-100 model $\sim$$25\%$ of the stellar population is formed from gas outside this region.
For the NFW models, the formation of stars of low metallicity is less common, with only a few low-metallicity stars formed post infall.
Once again, the radius of gas that goes on to form stars is higher for the ram pressure models than the isolation model.
Here, the NRPS-50 model is similar to the NRPS-100 model despite the reaccretion event that occurs, which is indicative of most of the gas at large radii already being lost before perigalacticon.

This gas from the outskirts appears to form gas through two different mechanisms. The primary mechanism is gas in the direction of travel is compressed, increasing in density and allowing it to eventually form stars.
A secondary mechanism, however, is that because of the dwarf's
gravity field, gas that  is expelled towards the front is perturbed enough to collide at the rear of the dwarf, allowing the gas to be reaccreted at the rear, effectively forming Bondi-Hoyle accretion.

\subsection{Orbital changes}\label{subsec:orbchange}

As evident from \S\ref{subsubsec:gasmass}, the stripping of the gas from the dwarf is not instantaneous, and for a large period of time, the gas is held off-centre as the dwarf's gravity balances the forces resulting from the ram pressure (see in particular the third row of Fig. \ref{fig:rps-dens} for an example of this).
{\ebf
This gravitational restoring force results in the gas of the dwarf acting like a parachute slowing down the dark matter and stellar content of the dwarf over many gigayears.}

The net impact of this off-centre gas is in the rotating frame of reference, a transfer of momentum and energy from the hot medium to the gas (gas to hot medium in the host galaxy reference frame), and then to the dark matter and stellar component of the dwarf through the gravitational restoring force, allowing the gas of the dwarf to act as a parachute slowing the entire dwarf down.
This transfer of energy results in a compression and heating of the hot medium \citep[see also][]{Roediger2015}, but of larger impact here, a change in the orbit of the dwarf {\ebf an effect also seen in \citet{Smith2012}.}
{\ebf Because of  the large mass contrast between the dwarf and the galaxy and the static nature of the potential used, this effect is purely due to the ram pressure force.
We verify that this effect is not due to the rotation scheme or  the creation/deletion scheme in Appendix \ref{app:norms} by performing two tests: firstly a dwarf in the hot coronae of the host halo, but without any baryons of its own; and secondly a dwarf in the rotation scheme without this hot coronae, which experiences only tidal forces.
In both cases, the dwarf's orbit remains constant with time, suggesting the effect here is due to the interaction between the hot coronae and the gas of the dwarf.}

\begin{figure*}
        \centering
        \leavevmode
        \subfigure[Pseudo-isothermal]{\includegraphics[width=0.5\textwidth]{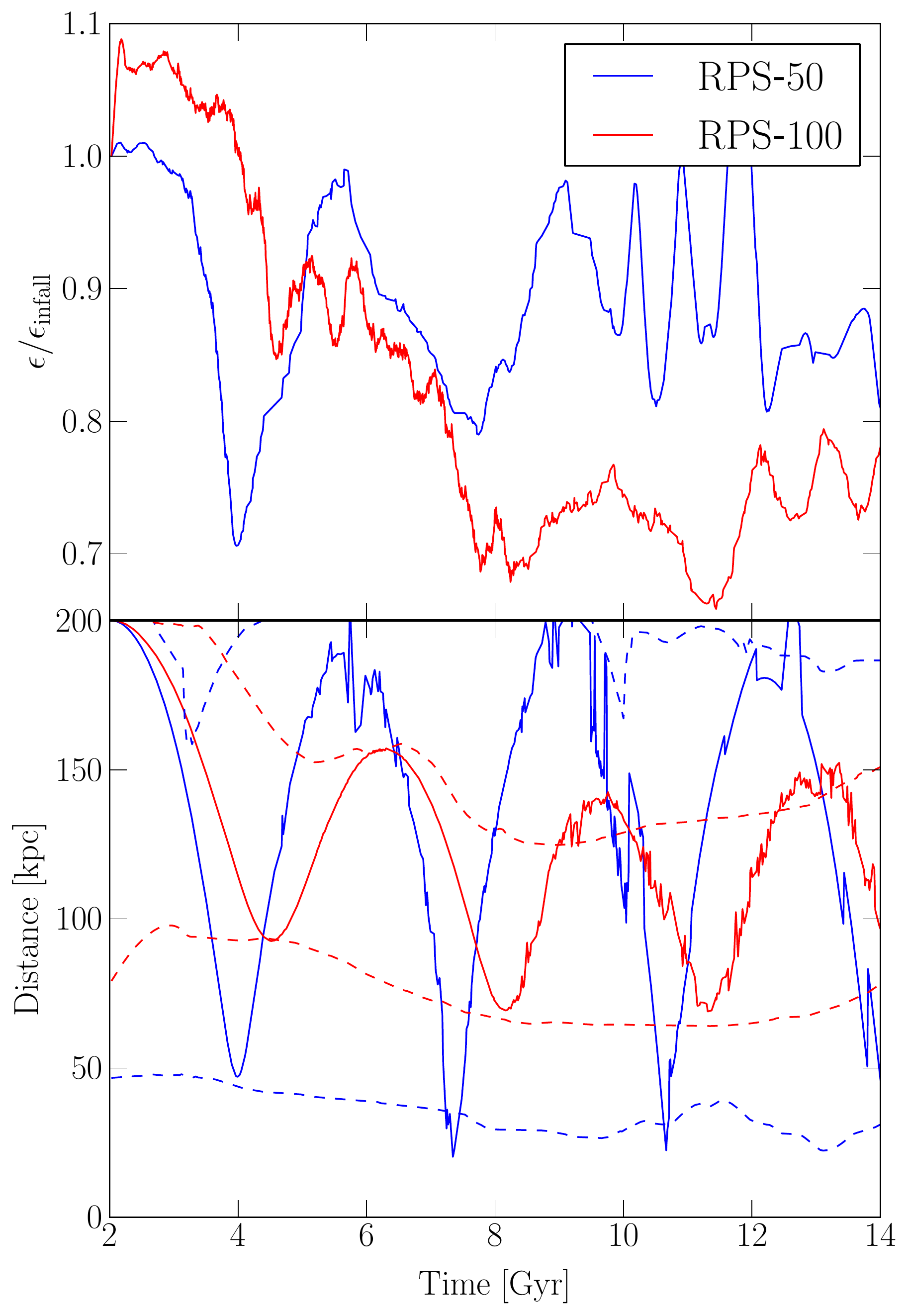}}
        \subfigure[NFW]{\includegraphics[width=0.495\textwidth]{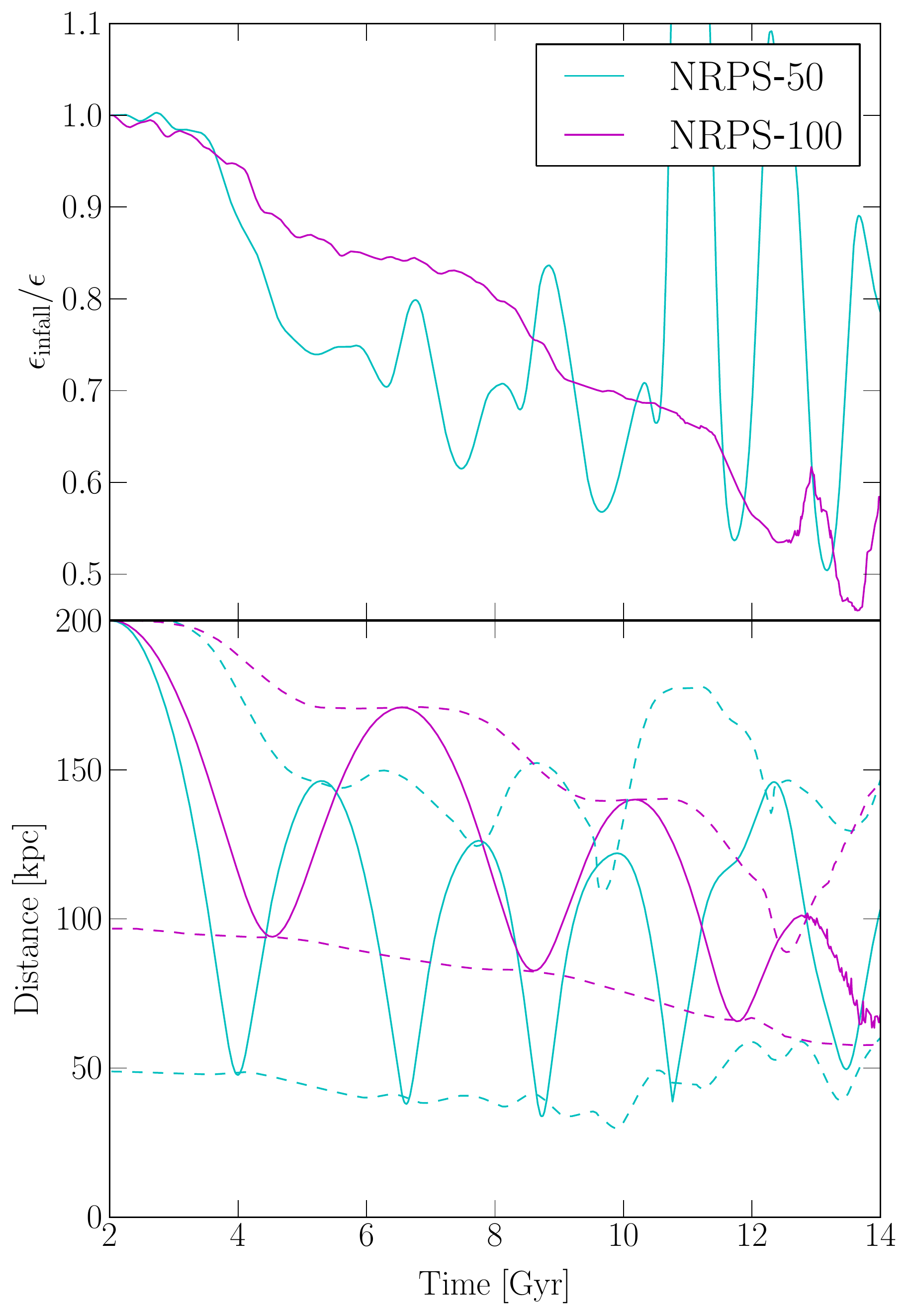}}
        \caption{Specific energy (kinetic plus potential energy) relative to infall and orbital evolution of the ram pressure models. RPS-50 is shown with\ a blue line, RPS-100 with red, NRPS-50 with cyan, NRPS-100 with magenta. For the orbital evolution, at each snapshot we calculate the peri- and apogalacticon of the orbit of the dwarf, based on its current position and energy.
        The galactocentric distance of the dwarf at any given time is shown with a solid line, while the peri- and apogalacticons are shown with dashed lines.
        }\label{fig:orbenergy}
\end{figure*}

We show the change in specific energy (kinetic plus potential energy) of the centre of the dwarf and how it affects the orbit of a dwarf as a function of time in Fig. \ref{fig:orbenergy}.
{\ebf The energy is calculated by examining the velocity and potential energy of the centre of the dwarf, excluding rapid changes due to the jumps in the centre determination, and is smoothed with a window function over $20$ snapshots ($94$ Myr).} 
Selecting the centre of the dwarf means we do not have to worry about particle deletion over time, but this choice increases the noise at which the part of the dwarf that is considered the centre changes.
Here, a general downwards trend is visible, as the dwarf loses energy and becomes more bound to the host halo.

For RPS-50 and NRPS-50, which lose all their gas at the first perigalacticon passage, this loss of energy mostly occurs just prior to this passage, suggesting the dominant effect is  reducing the apogalacticon.
The longer period over which gas is removed from RPS-100 and NRPS-100 is seen as the continual drop in energy over $\sim$$8$~Gyr.
As the dwarfs become streams, the calculation of the centre becomes quite uncertain. Resulting in spurious jumps in energy, these final jumps greatly exaggerate the (slight) gain in energy that can occur in the centre due to reaccretion of gas after a burst of star formation. A slight gain is seen as reaccreted gas was preferentially pushed in the direction of motion.

At each snapshot, we also calculate the peri- and apogalacticon of the orbit of the dwarf, based on its current position and energy.
Here the perigalacticon is slightly decreased, however, the apogalacticon experiences a $\sim$$10\%$--$\sim$$25\%$ drop.
This is particularly evident in the NRPS-100 model, where the extended loss of gas results in a gradual decline in both perigalacticon and apogalacticon and downwards jumps in apogalacticon near perigalacticon passages.
Once the gas is stripped, the lack of interaction with the hot medium results in minimal change to the orbit, although once again the tidally disrupted dwarf lacks a clear centre, resulting in spurious jumps in perigalacticon and apogalacticon as the calculated centre rapidly shifts.

These changes are above the accuracy expected to be determined by Gaia surveys \citep[approximately $14\%$ accurate in determinations of peri- and apogalacticon]{Lux2010}, and hence the orbital changes due to ram pressure stripping is important in tracing the orbits of satellite galaxies and streams backwards in time.

{\ebf The change in energy allows for the calculation of the coefficient of drag at each time step. As something potentially useful for semi-analytic models but not a focus of this paper, we discuss this in Appendix \ref{app:drag}. In general, the coefficient of drag, using a reference area of the region where $N_{\rm HI} > 10^{21}$~cm$^{-2}$, is a factor of a few $\sim$$3$--$4$ that of a sphere.}

\section{Conclusions}\label{sec:conclusion}

For the first time, a full chemodynamical model of a dwarf galaxy is examined as it infalls into a the hot corona of another galaxy.
Using an improved version of \texttt{GEAR,} we  simulated the infall of a Sextans-like dwarf with a pseudo-isothermal and an NFW profile through utilisation of a rotating frame of reference with variable external gas density.
The methods used here are novel and allow quick calculation of the ram pressure forces upon a dwarf.

If the dwarf was already extended upon infall, {\ebf e.g. having low-density cores due to early internal processes}, even at large galactocentric distances the pressure of the hot halo more than doubles the star formation inside the dwarf.
If the dwarf is denser, however, and retains an NFW profile, this effect is lessened although still noticeable.
This star formation blows out large amounts of gas, greatly assisting ram pressure in stripping the dwarf, an effect also seen in \citet{Gatto2013}.

As gas is stripped from the dwarfs, it is pushed off centre against the direction of motion, resulting in a gradual energy loss of the dwarf altering the orbit.
This gas is then finally lost around perigalacticon for most of the dwarfs considered here, after the dense medium causes a final burst of star formation.
This final burst drags some dark matter with it due to the mutual gravitational attraction between the gas and dark matter.
At perigalacticon, this has the affect of greatly increasing the impact of tides, resulting in the pseudo-isothermal dwarf losing all its mass even at large perigalacticons ($r_{\rm peri}$$\sim$$100$~kpc), and transforming the dwarf into a stream.
When gas is not lost at perigalacticon, it can be held off-centre until star formation eventually forces the gas outside of the dwarf, even as the ram pressure forces acting upon it drop.

That the dense NFW models also succumb to these tidal effects suggests that the Sextans dwarf spheroidal was markedly different from its present day form upon infall, and has been altered to its present day extended low-density form through the synergy of tidal and ram pressure forces.
As the star formation is required to experience shut down within $6$~Gyr, higher orbits are unlikely to solve these problems.

The present day form of Sextans dSph could be achieved through it beginning with a denser dark matter profile at infall, or by having a much lower baryon fraction (well below the universal ratio, e.g. through UV heating) to minimise these synergistic effects.

The metallicity of the dwarf is also affected, with the rapid star formation being inefficient at  recycling of metals, resulting in a lower peak in metallicity than would be expected by stellar mass alone.
Although dwarfs are expected to be inefficient regardless \citepalias{Nichols2014}, the evolution here is effectively halted for the pseudo-isothermal dwarfs.
In addition to this poor recycling, the gas that goes on to form stars upon infall is not the same gas that\ would form stars in isolation.
The combination of compression due to ram pressure and Bondi-Hoyle-like accretion results in a large fraction of gas from the outskirts going on to form stars.
This star formation from metal poor gas makes dwarf galaxies whose kinematics suggest an early infall one of the best targets for searches of extremely metal poor stars.

Even if the $[$Fe$/$H$]$ evolution can continue, and in dwarfs that orbit at high-perigalacticon it may even be enhanced, the increase in star formation alters the ratio of Type II supernova to Type Ia supernova, resulting in slightly higher $[$Mg$/$Fe$]$ at high $[$Fe$/$H$]$.
This higher $[$Mg$/$Fe$]$ is likely not directly observable, but notably, the dispersion is reduced with the ram pressure models tending to be concentrated above $[$Mg$/$Fe$]=0$, while the isolation models have a larger spread at high metallicity.

When considering semi-analytic models of galaxy formation that does not include ram pressure, the most obvious changes to approximate ram pressure stripping are:
\begin{itemize}
\item Dwarf galaxies that show signs of tidal disruption without considering ram pressure will likely not survive the combination of tidal and ram pressure forces.
\item $[$Fe$/$H$]$ evolution is slowed, moving a dwarf upwards on the $[$Fe$/$H$]$ v $L$ relation.
\item At the high-metallicity end (past the ``knee"), the $[$Mg$/$Fe$]$ relation is marginally higher than would otherwise be expected.
\item The orbital properties of the dwarf change, depending on the timescale of gas removal. Energy losses of $10$--$30\%$ are readily possible, disproportionately resulting in a drop in apogalacticon.
\end{itemize}

Slight errors in the ram pressure force are introduced throughout the calculation because of the method ($10\%$--$30\%$ at perigalacticon, discussed further in appendix \ref{app:halo}).
These errors are still well within the order of magnitude uncertainty in the observational values of the density and orbital velocity of the dwarfs at perigalacticon, and in particular seem to result in an underestimate of the ram pressure force at perigalacticon, as such any quantitative values (such as time of stripping) may be off.
Nevertheless, the large impact of ram pressure stripping seen here suggests that these quantitative values are much better approximations than those taken from dwarf galaxies modelled in isolation or in constant density wind tunnels and hence through the modelling of ram pressure stripped dwarfs, we can gain valuable insight into the nearby satellite galaxies.

\begin{acknowledgements}
The data reduction has been performed using the parallelised Python \texttt{pNbody} package
(\texttt{http://lastro.epfl.ch/projects/pNbody/}). This work was supported
by the Swiss National Science Foundation (FNS $200021\_153234$).
This work also benefitted from the International Space Science Institute (ISSI) in Bern, thanks to the funding of the team, ``The first stars in dwarf galaxies".
\end{acknowledgements}

\appendix
\section{Rotation scheme integration and tests}\label{app:tide-test}
Because of the non-symplectic nature of the leapfrog integrator, care needs to be taken when dealing
with the rotating reference frame to advance the velocity at each step due to the accelerations dependence on the velocity.

We test a few schemes through two tests: first , a simple system of rotating rings, and also  a dwarf that only experiences  tidal forces as in \citetalias{Nichols2014}.

\begin{figure}
\centering
\includegraphics[width=0.5\textwidth]{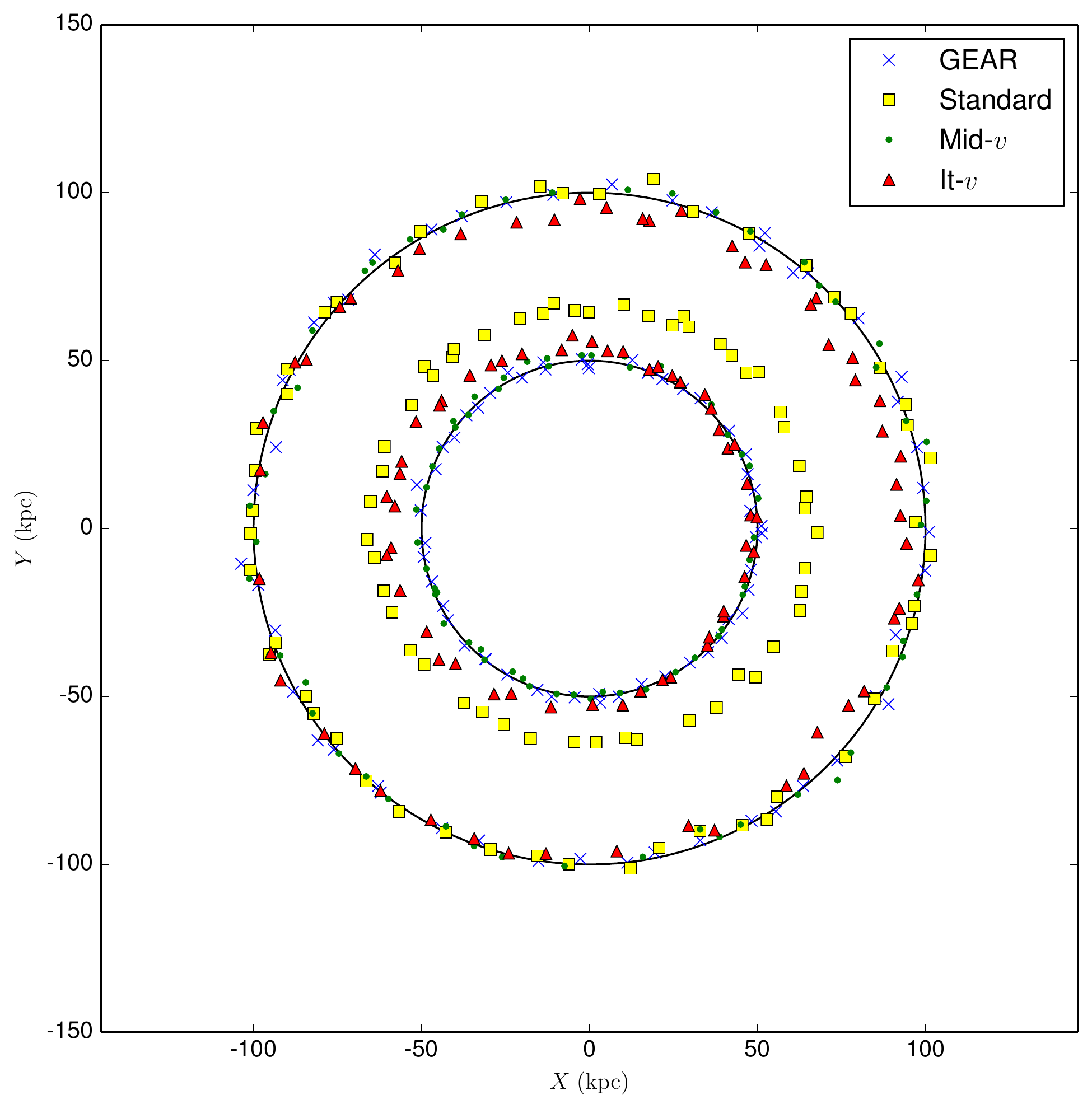}
\caption{Results at $23.3$~Gyr for the two-ring test. Two rings rotate in opposite directions around a $1\times10^{11}$~M$_\odot$ mass located at $(0,0)$. When the frame of reference rotates (all but the standard \texttt{GEAR} code), it rotates around the point $(100,0)$.
The points should continue to orbit at $50$~kpc and $100$~kpc represented by two black rings.
The standard \texttt{GEAR} code in a non-rotating frame is shown with blue crosses.
In a rotating frame of reference, the results of calculating the forces at every time step is shown with yellow squares, using the half time step velocity (Mid-$v$) as green points, and the iteration scheme (It-$v$) with red triangles.
Only every $20$th point is shown for clarity.}\label{fig:ringtest}
\end{figure}

First, we consider a system of two simple rings, each comprising a number of particles of negligible mass (we use $1000$ particles of mass $0.01$~M$_\odot$) in circular orbits at $50$~kpc and $100$~kpc around a $1\times10^{11}$~M$_\odot$ point mass (modelled as a Plummer sphere with a softening of $0.5$~kpc).
If the central mass is considered the origin, then the rotating box is positioned at $(100,0)$ and is set to orbit in the same direction as the outer ring (clockwise), giving the outer ring zero velocity in the rotating frame of reference.
The inner ring is set to rotate in the opposite direction (anti-clockwise in this case).
This system is then simulated for $23.3$~Gyr ($5000$ time units in our set up), just shy of 2.5 rotations of the outer ring. 
We show the results in Fig. \ref{fig:ringtest}.
We note that all errors here are time step dependent, with all methods performing well at small time steps, however, large differences are seen at the time steps selected by the code.

As the periodic box is in a circular orbit, this test only evaluates the Coriolis and centrifugal forces.
In the ideal case, the two rings (shown with black circles)  maintains their orbits for all time, and this appears in the standard \texttt{GEAR} code (blue crosses; we note that this is identical to the standard \texttt{Gadget-2} code, with an external potential force added) without the rotating reference frame.
Inside the rotating reference frame, calculating the forces at the standard time step (yellow squares) and using the half-time step velocity inside the fictitious forces gradually introduces errors that build in time, with the inner ring expanding to a radius of $\sim65$~kpc and the outer ring staying constant (a sign that the centrifugal force is being evaluated accurately).
We find better results (green points) if we calculate the fictitious forces using an approximated future velocity (Mid-$v$) generated by the following:
\begin{eqnarray}
        \vec{v}_{t,{\rm pred}} &=& \vec{v}_{t-1/2} + \vec{a}(\vec{q}_{t-1/2},\vec{x}_{t},\vec{v}_{t-1/2},\vec{\omega}_{t-1/2})\Delta{}t/2,\\
        \vec{v}_{t+1/2} &=& \vec{v}_{t-1/2} + \vec{a}(\vec{q}_{t-1/2},\vec{x}_{t},\vec{v}_{t,{\rm pred}},\vec{\omega}_{t-1/2})\Delta{}t.
\end{eqnarray}
Updating $\vec{\omega}$ or $\vec{q}$ to an approximation does not greatly change the result here. 
In this case, the rings seem to suffer minimal contraction or expansion.

We display one further case (It-$v$, red triangles).This form performs better in other tests, although it is seemingly more sensitive to step size (at low-step size all methods perform similarly) and in pure dark matter simulations (which tend towards larger step sizes) should be avoided.
Here, the mid point is approximated as per above, but then it is approximated again using this calculated value.
Here, we also assume that the angular velocity changes slightly across a time step, but that the acceleration in the angular velocity is negligible. We find
\begin{eqnarray}
        \vec{v}_{t,{\rm pred}} &=&  \vec{v}_{t-1/2} + \vec{a}(\vec{q}_{t-1/2},\vec{x}_{t},\vec{v}_{t-1/2},\vec{\omega}_{t-1/2})\Delta{}t/2,\\
        \vec{\omega}_{t,{\rm pred}} &=& \vec{\omega}_{t-1/2} + \dot{\vec{\omega}}_{\rm old}\Delta{t}/2,\\
        \vec{v}_{t,{\rm pred}} &=& \vec{v}_{t-1/2} + \vec{a}(\vec{q}_{t-1/2},\vec{x}_{t},\vec{v}_{t,{\rm pred}},\vec{\omega}_{t,{\rm pred}})\Delta{}t/2,\\
        \vec{v}_{t+1/2} &=& \vec{v}_{t-1/2} + \vec{a}(\vec{q}_{t-1/2},\vec{x}_{t},\vec{v}_{t,{\rm pred}},\vec{\omega}_{t,{\rm pred}})\Delta{}t.
\end{eqnarray}

A second test is to compare the results with just tidal forces, particularly focusing on the star formation present within the dwarf.
We follow the same method as in \citetalias{Nichols2014}, simulating a dwarf in isolation, before inserting it within the potential of a host galaxy.
We choose an orbit with a perigalacticon of $80$~kpc and an apogalacticon of $200$~kpc, injecting the dwarf after $2$~Gyr of evolution in isolation. This is the same simulated dwarf we use below, and its properties may be found in more detail in \S\ref{sec:sims}.
The host halo parameters are the same as used in the main simulations and are shown in Table \ref{table:params}.
The effects of just tides are discussed in detail in \citetalias{Nichols2014},  and here we just investigate the density profile, which provides resistance against ram pressure stripping, and star formation, which affects the chemical evolution of the dwarf,  to compare the codes.
Here we implement the particle deletion scheme discussed below in \S\ref{subsec:PD}, but this has minimal effects on the dynamics of the dwarf (we do not follow it all the way to destruction).
We compare the standard \texttt{GEAR} code \citepalias[with slight updates from][]{Nichols2014}, with the mid-point velocity scheme and the iterated velocity scheme for each property with the normal  \texttt{GEAR} selected step sizes and the iterated velocity scheme limited to the typical step size \texttt{GEAR} chooses for the main simulations ($\sim$$0.094$~Myr).

\begin{figure}
\centering
\includegraphics[width=0.5\textwidth]{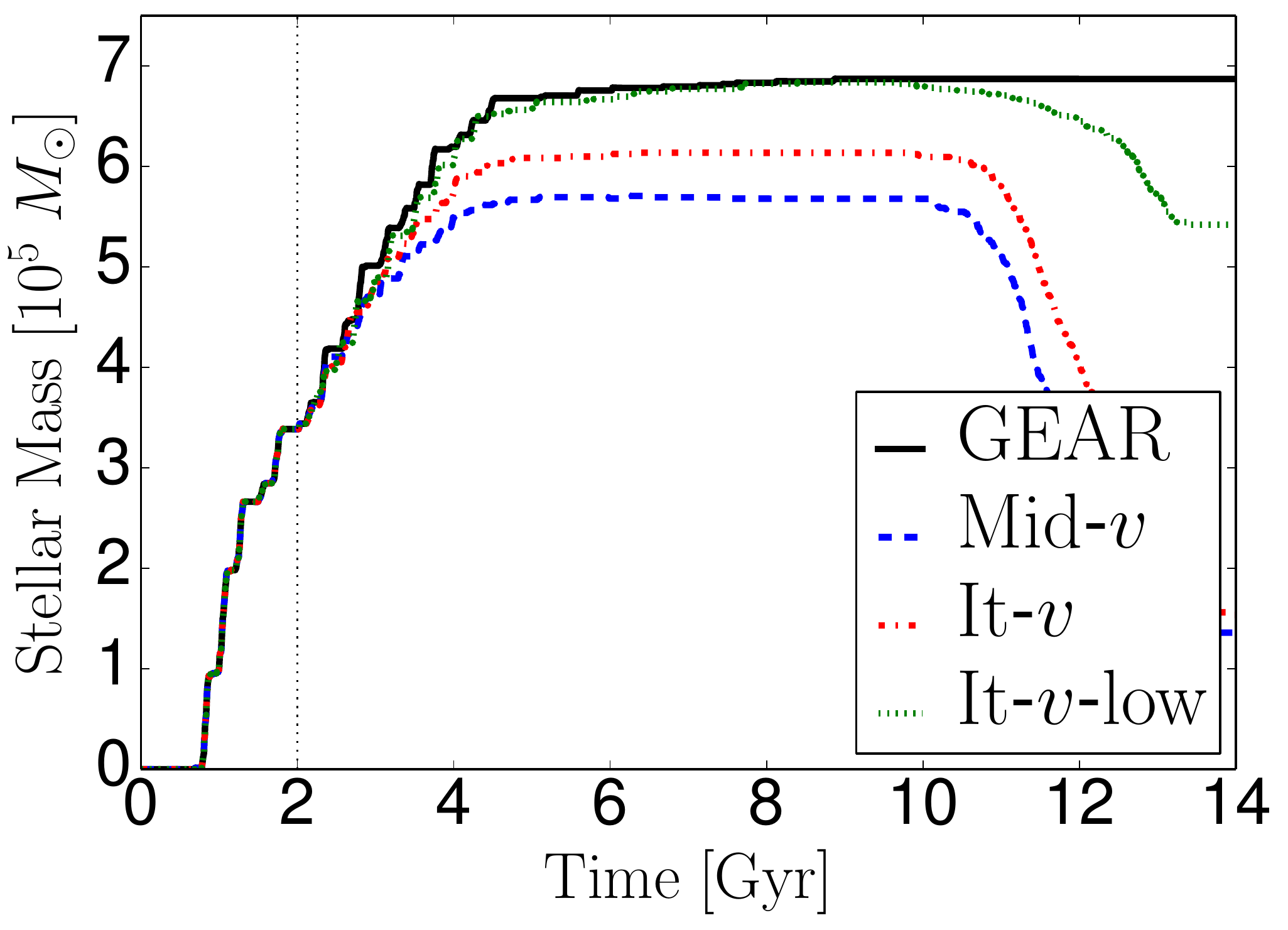}
\caption{Stellar mass for the dwarf only experiencing tidal forces in the standard \texttt{GEAR} implementation (blue) and in the rotating reference with the Mid-$v$ scheme (green dash) and with the It-$v$ scheme (red dash-dot) and the It-$v$ scheme with a low step size (green dotted).
The beginning of the orbit is shown with a vertical dotted line.
For the first $\sim1.5$~Gyr, the stellar mass is basically the same with similar bursts (see Fig. \ref{fig:tidal-SFR}), beyond this, however, the Mid-$v$ scheme flattens out, still producing stars for another $\sim1.5$~Gyr, but at a lower rate.
The drop in stellar mass at the end in the rotation schemes is due to stellar particles being deleted as they cross the boundaries.
This drop is lower in the low step size due to better simulation of the dwarf core (see Fig. \ref{fig:tidal-dens}).}\label{fig:tidal-SFH}
\end{figure}

\begin{figure}
\centering
\includegraphics[width=0.5\textwidth]{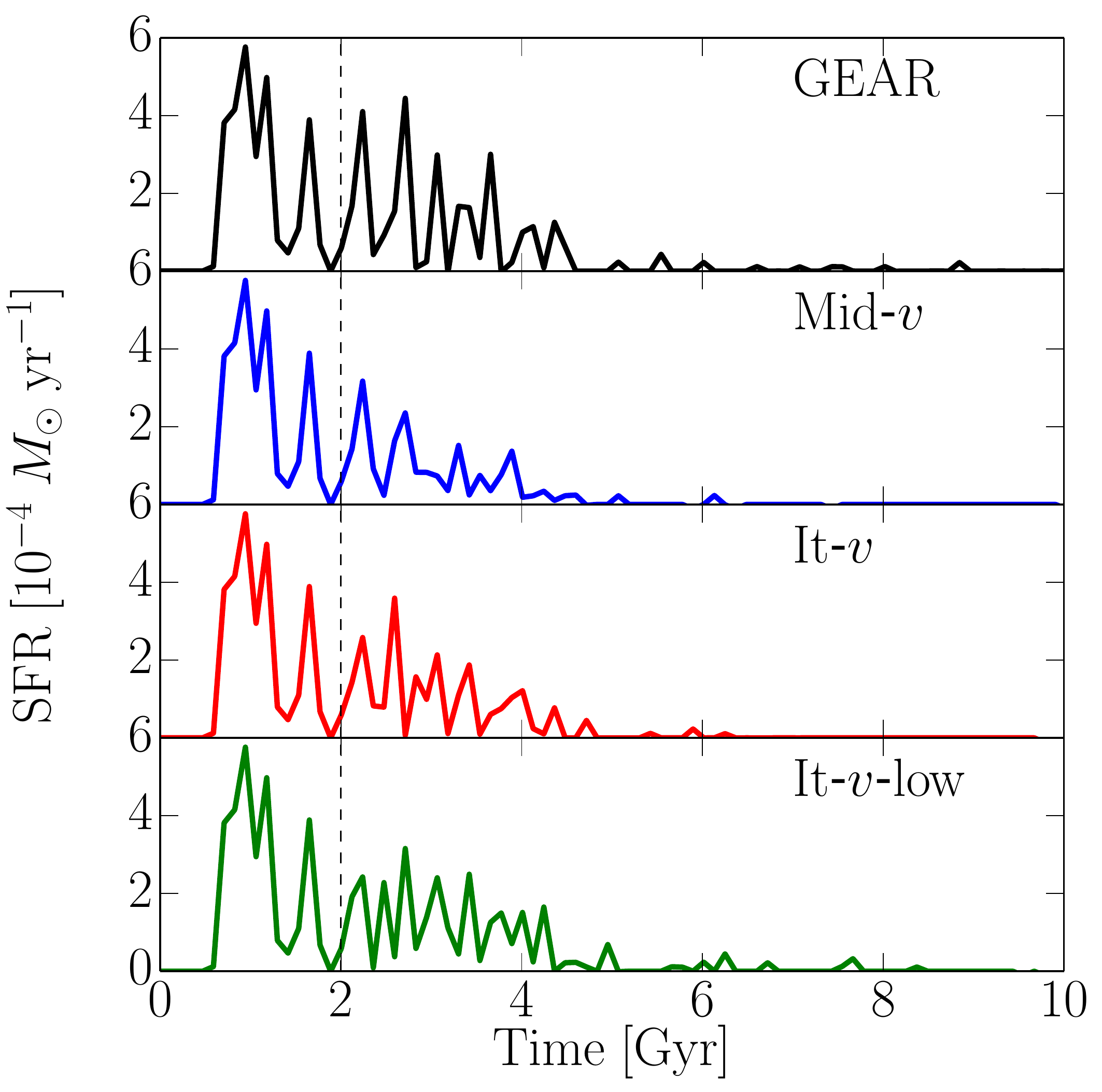}
\caption{Star formation rate for the dwarf only experiencing tidal forces in the standard \texttt{GEAR} implementation (blue, top) and in the rotating reference with the Mid-$v$ scheme (green, middle) and with the It-$v$ scheme (red, bottom).
The beginning of the orbit is shown with a vertical dotted line.
For the first $\sim1.5$~Gyr, the bursts are similar between all cases with short bursts of $\sim4\times10^{-4}~$M$_\odot$ (and a slightly extended one in the normal \texttt{GEAR} code), which leads to a similar chemical evolution.
Beyond this point however, Mid-$v$ run drops to $\sim1\times10^{-4}~$M$_\odot$ before becoming extremely low levels around $6$~Gyr.
Suggesting that the change in density profile impacts the star formation rate, and subsequently the feedback and chemical evolution within the dwarf.
}\label{fig:tidal-SFR}
\end{figure}

We show the stellar mass in Fig. \ref{fig:tidal-SFH} and the star formation rate in Fig. \ref{fig:tidal-SFR}.
It is immediately obvious that the various schemes produce different star formation histories; while the standard \texttt{GEAR} code (top panels in both) experiences a period of extended star formation. In both of the other cases, this continues the bursty nature of the isolated dwarf (before the dashed line).
That the \texttt{GEAR} code changes from a star formation history characterised by bursts to one with a gradual period of star formation (at least initially) is likely related to the lack of pause in cooling that occurs when beginning a simulation from a snapshot \citepalias[see][for a discussion of this]{Nichols2014}, why this does not occur with the other two is not known, but the star formation histories are similar apart from
this initial region and that suggests it will not greatly affect the final results.

\begin{figure*}
\centering
\includegraphics[width=\textwidth]{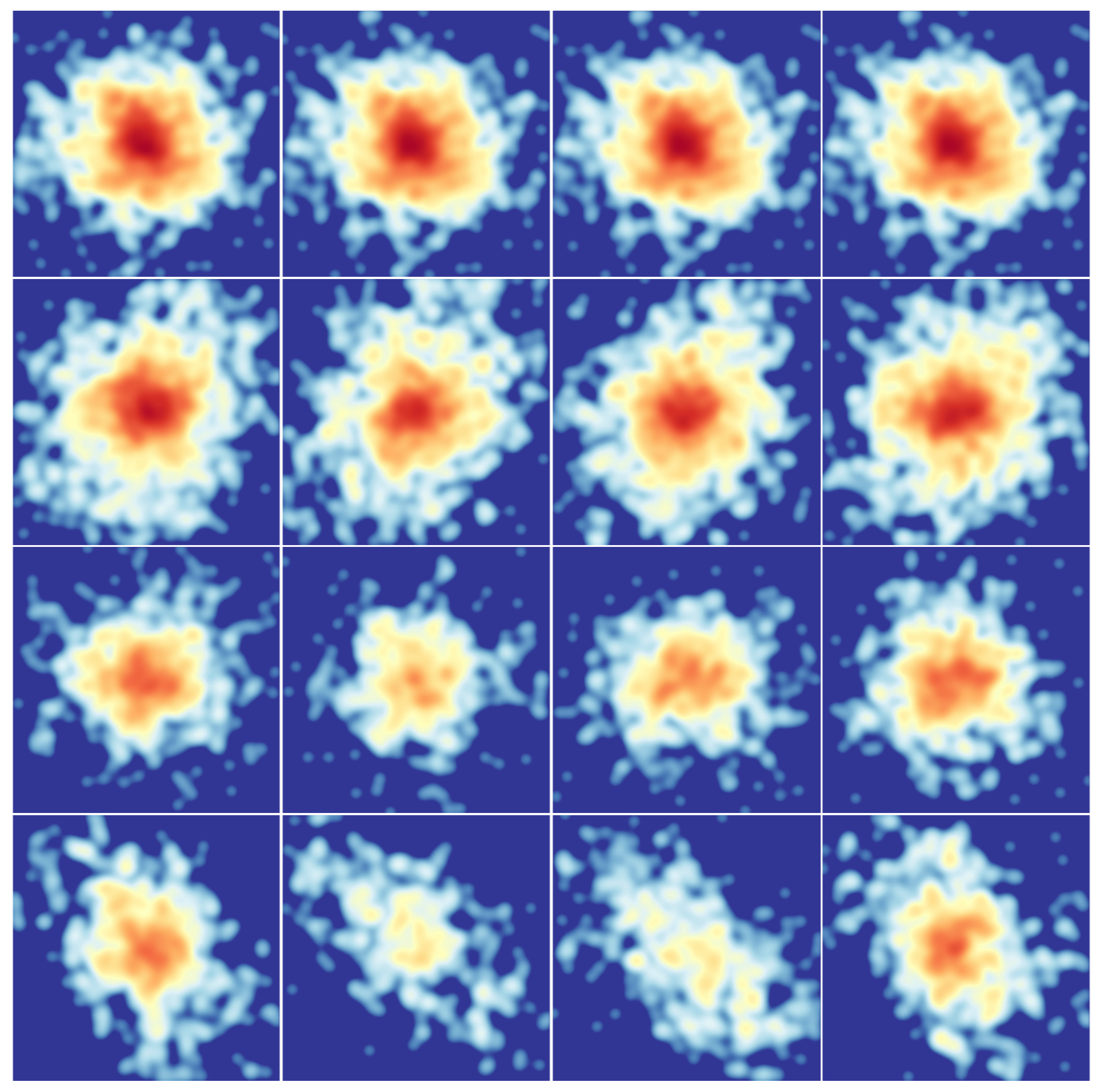}
\caption{Smoothed density in a central $1$ kpc slice (logarithmic from maximum density at simulation start to $1\%$ of this) through the centre of the dwarf in the three schemes.
Each box is $10$~kpc in width and height.
The standard \texttt{GEAR} scheme is on the left, Mid-$v$ scheme in second column, It-$v$ scheme with natural step sizes in the third, and the It-$v$ scheme with limited step sizes in the right column.
From top to bottom, the slices take place at $2$~Gyr (simulation start), $4.8$~Gyr (perigalacticon),  $7.6$~Gyr (apogalacticon), and $10.4$~Gyr (perigalacticon).
In the rotating frame schemes, the dwarf is rotated back to the host galaxy frame of reference to be directly comparable with the standard \texttt{GEAR} scheme.
As should be the case, the density profiles are identical at the simulation start, by the first perigalacticon, however (second row), the Mid-$v$ scheme is less dense in the centre and slightly over dense in the tidal tails; this central under density is carried over to the first apogalacticon (third row) before the Mid-$v$ dwarf is effectively destroyed at the next perigalacticon (bottom panel). The It-$v$ dwarf survives these passages in slightly better shape, particularly at the second perigalacticon where it still retains a slightly denser core.
When limiting step sizes to those typical of the RPS simulations, the It-$v$ scheme performs quite well, possessing a similar core to the normal \texttt{GEAR} scheme at each point.}\label{fig:tidal-dens}
\end{figure*}

A more noticeable change occurs in the density profile of the dwarf as shown in Fig. \ref{fig:tidal-dens}.
Here the mid-$v$ scheme (middle column) has a lower central density from the first perigalacticon passage (second row), which gets worse as time goes on.
That much of the star formation occurs soon after input (and the dwarf is quenched in all cases) suggests that the schemes  differ more in their estimation of the dynamical effects than in the chemical evolution of the systems.
The changes in the star formation (of order $20\%$) between the simulations is not a great concern when simulating dwarfs and is roughly the amount produced by changing random seeds.

\section{Particle deletion/creation scheme}\label{app:halo}
The creation and deletion of particles is not without error.
The creation of particles ignores things like pressure when deciding where particles are placed and their velocity, and the deletion relies on experimental values which are context dependent.

The simplest test case to examine what errors we can expect is a simple wind tunnel.
Even here however there are a  number of free parameters to consider: the velocity, pressure, and internal energy of the particles all influence the final error.
We consider a wind tunnel of length $50$ ($50$~kpc in physical units) and inject particles with a velocity of 0.5 ($103$~km~s$^{-1}$).
All particles have a density of $9\times10^{-8}$ ($3.7\times10^{-5}$~cm$^{-3}$) and an internal energy generated from a temperature of $2\times10^6$~K.
These conditions and velocity are similar to a dwarf orbiting at the outer edge of the halo.
Although this box is not rotating, we delete particles if they come within $0.1$ of the outer edge (this is smaller than the value used for the rotating box, but as discussed below, the rotation minimises the backwards wave).

\begin{figure}
\centering
\includegraphics[width=0.5\textwidth]{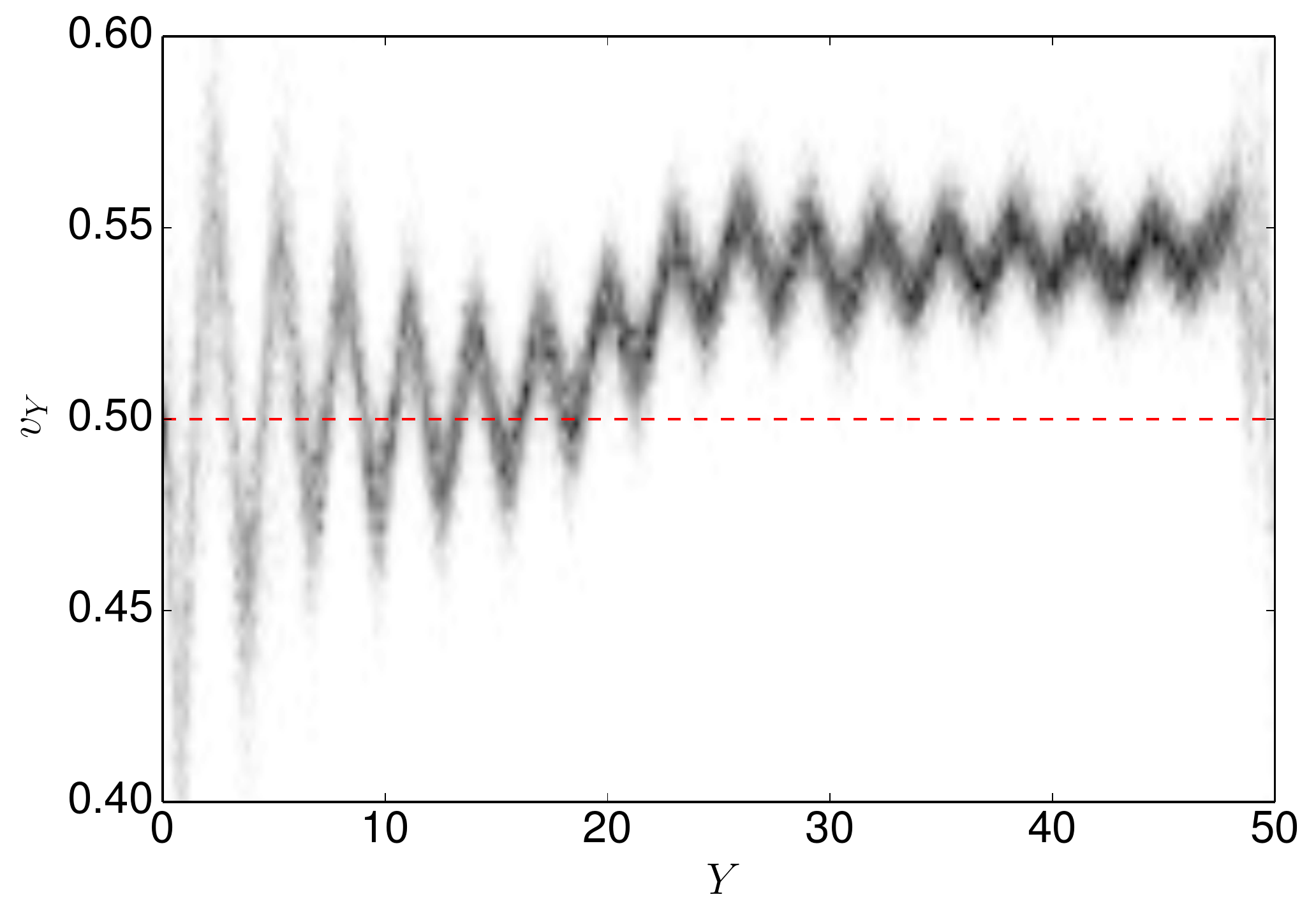}
\caption{Two-dimensional histogram of the velocity and position of the particles in a wind tunnel (darker shades represent more particles there). The expected velocity is shown with a red dashed line at $0.5$.
The creation/deletion of particles introduces waves into the box, which are damped with position, and the deletion of the particles leads to a rising velocity towards the outer edge of the box.}
\label{fig:Cdens-vel}
\end{figure}

\begin{figure}
\centering
\includegraphics[width=0.5\textwidth]{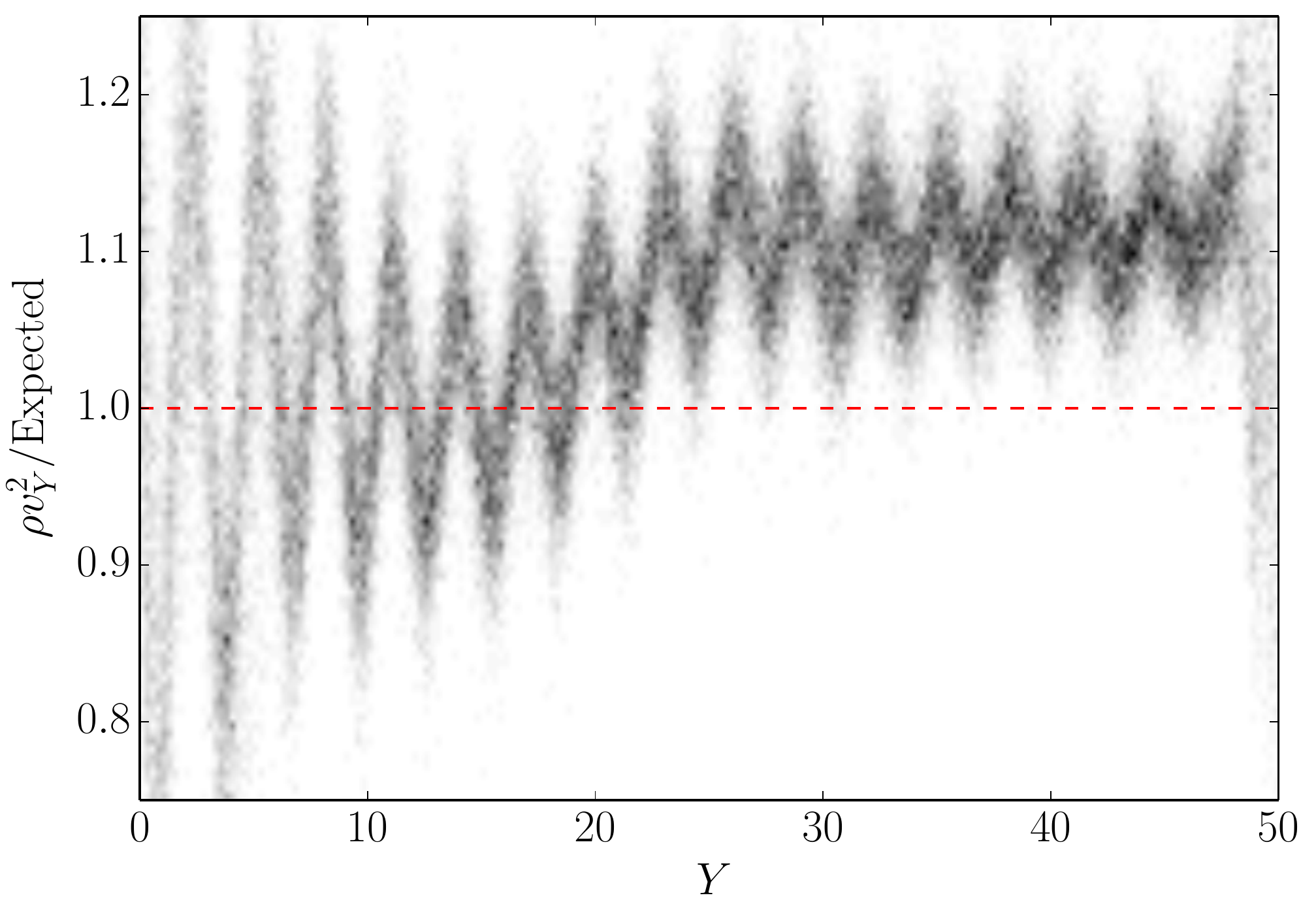}
\caption{Two-dimensional histogram of the ram pressure force (divided by the expected) and position of the particles in a wind tunnel (darker shades represent more particles there). The expected ram pressure force is shown with a red dashed line at $1.0$.
Here, we see that the creation/deletion of particles introduces waves into the box, with a rising velocity towards the outer edge of the box.}
\label{fig:Cdens-rps}
\end{figure}

We display the velocity of the particles as a function of position in Fig. \ref{fig:Cdens-vel}, and the ram pressure force divided by the true velocity in Fig. \ref{fig:Cdens-rps} at a time of $10$.
The creation and deletion of the particles introduce standing waves into the box, which are damped in amplitude towards the middle.
Beyond the middle, the deletion of particles (and the subsequent vacuum it creates) increases the velocity beyond the centre.  This creates a slight increase in velocity, which travels backwards before settling at a value of $\sim$$0.55\pm0.02$.
If a deletion region of size $0.5$ is used in the wind tunnel,  the velocity instead settles at a value of $\sim$$0.8$.
The change in velocity and changes in density due to the waves introduces a spread in ram pressure at the centre from about $90\%$  of the true value to about $130\%$.

\begin{figure}
\centering
\includegraphics[width=0.5\textwidth]{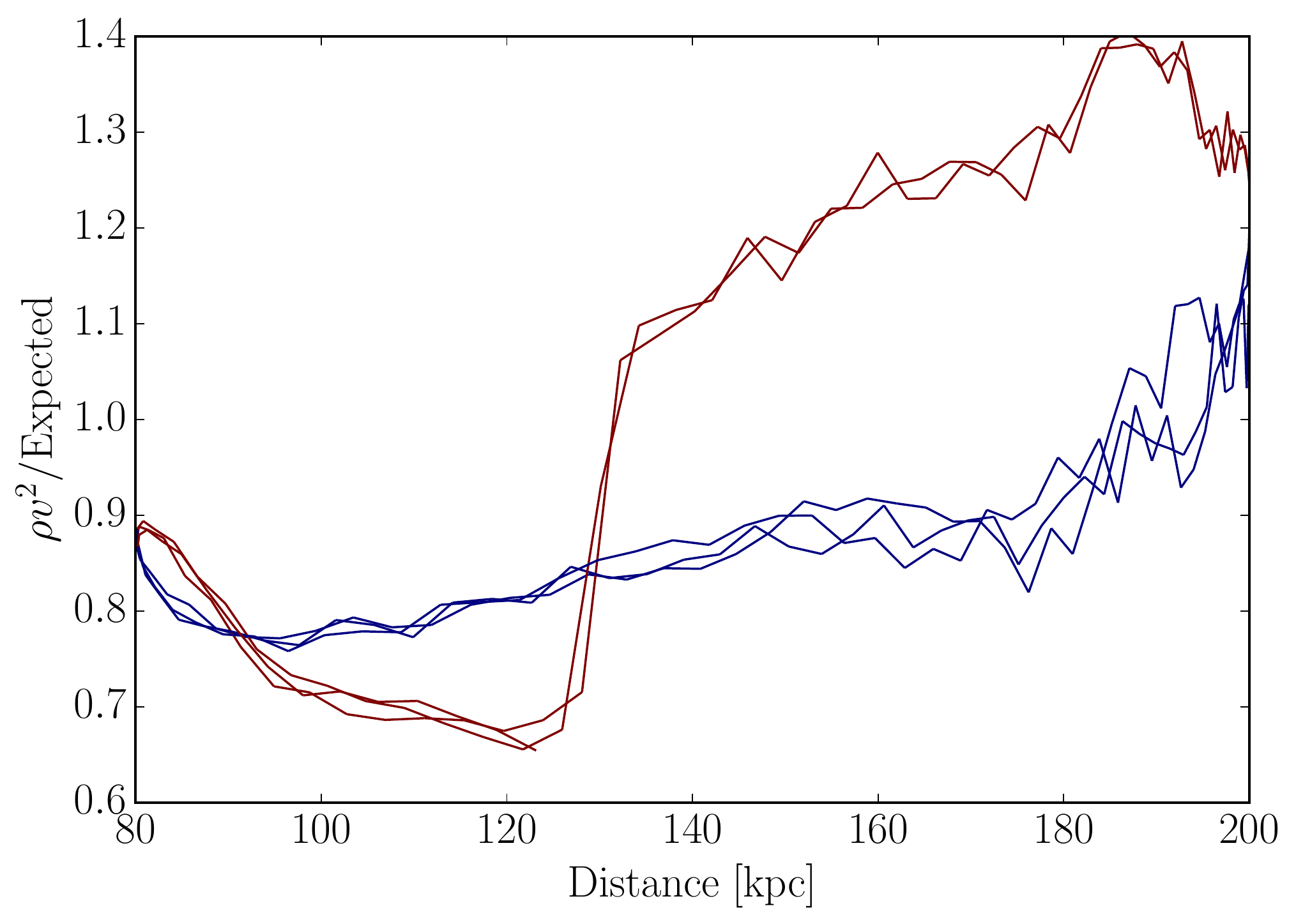}
\caption{Ratio of the ram pressure force near the centre of the box to the expected versus radius for a sample orbit. The region of the orbit decreasing in galactocentric distance is shown in blue, while the region of increasing distance is shown in red.}
\label{fig:rps-R}
\end{figure}

We also consider the effects of the orbit changing, which introduces a pressure gradient across the box.
The periodic box is placed in an orbit with a perigalacticon of $80$~kpc and apogalacticon of $200$~kpc, identical to that used in Appendix \ref{app:tide-test}, in a box of side $50$~kpc, a gas particle mass of $6.0\times10^3$~M$_\odot,$ and a halo density profile as per \S\ref{subsec:host}.
We show the results in Fig. \ref{fig:rps-R} as a function of the radius of the orbit.
Here, the pressure gradient results in a slight underestimation of the ram pressure at perigalacticon (by about $\sim$$10\%$), while being slightly above at apogalacticon.
It is clear that over the densest fastest region of the orbit, any dwarf  likely experiences a slight underestimation of the ram pressure, with the overestimation that occurs towards apogalacticon.

Notably, the larger error appears to be as the dwarf increases in galactocentric distance (i.e. approaching apogalacticon), and suggests that the scheme has more trouble accounting for a decrease in pressure than the increase in pressure associated with infall.

The choice to base it on the current positions of the dwarf results in creating an effective delay in the density of the halo reaching the dwarf (since the particles have to travel half the box).
As the orbit can change on these scales, the solution to this problem would require integrating the pivot point forward and inserting the particles at its future density.
Given the correction factor, this is not a trivial solution, particularly at perigalacticon where the impact is biggest, however, future versions of the code will attempt to incorporate this solution.

\section{Particle creation}\label{app:partcre}

\begin{figure*}
\centering
\includegraphics[width=\textwidth]{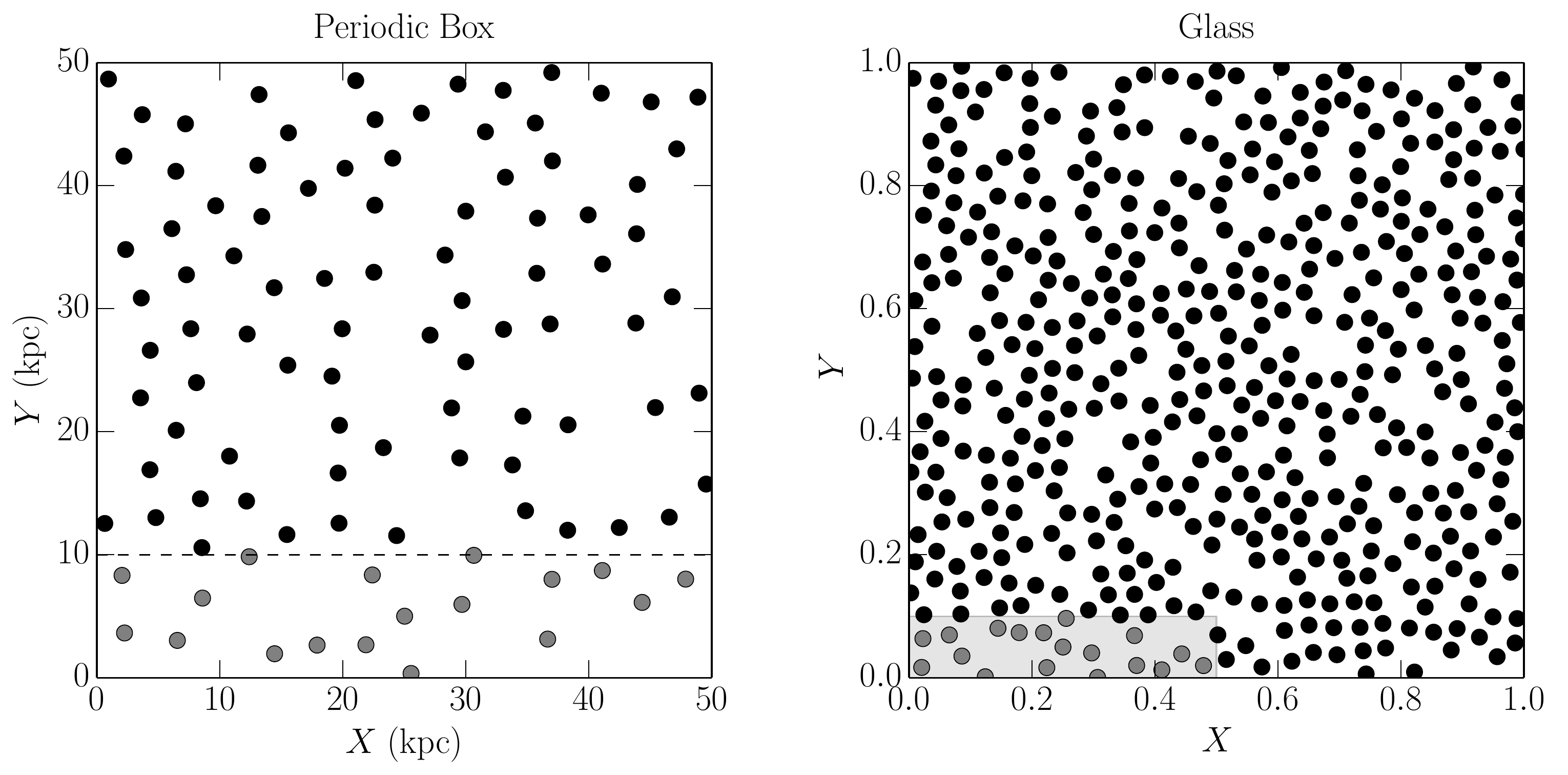}
\caption{Particle creation schematic. The periodic box is shown in the left panel. The SPH glass is shown on the right. When particles need to be created (below the dashed line), a region of the SPH glass is selected (right panel, grey shading) according to equation (\ref{eq:partcreation}).
The positions of the particles within this region is then scaled to match the boundaries of the particle creation region, resulting in the grey points.
The velocity of the particles is then set according to equation (\ref{eq:partvel}), and the other properties based on the assumed halo profile.
At the next time step, the region of glass selected would start at the top of the shaded region (0.1 in the right-hand box) repeating the glass in any dimension when necessary.}\label{fig:partcreation}
\end{figure*}

At simulation start, the positions of particles of an SPH glass (of $N_{\rm glass}$ particles) in a unit cube are passed through to memory.
At each time step, of size $\Delta{}t$, particles  of mass $m_{\rm gas}$ are injected with a density corresponding to the halo halo at density $\rho_{\rm halo}$.
Particles are inserted by mapping the volume required to be filled by the particles to the glass cube.
We show a schematic of the particle creation in Fig. \ref{fig:partcreation} to assist in the understanding.

If the number of particles required to fill the entire box (of side $L$) is $N_{\rm halo} = \rho_{\rm halo}L^3m_{\rm gas}$, then the volume at each time step that needs to be filled corresponds to a rectangular prism $(x\times{}y\times{}z)$ on the glass of size,
\begin{equation} \left(\frac{N_{\rm halo}}{N_{\rm glass}}\right)^{1/3}\times \frac{||\vec{V}_p||\Delta{}t}{L}\left(\frac{N_{\rm halo}}{N_{\rm glass}}\right)^{1/3} \times \left(\frac{N_{\rm halo}}{N_{\rm glass}}\right)^{1/3}. \label{eq:partcreation}
\end{equation}
All particles within this prism, and moving the prism along the velocity axis at each time step, are ``accreted", that is, a new gas particle is created at the position $(N_{\rm halo}/N_{\rm glass})^{-1/3}(x_g,y_g,z_g)$, where $(x_g,y_g,z_g)$ is the position of the particle on the glass relative to the origin of the prism.
As $(N_{\rm halo}/N_{\rm glass})^{-1/3}$ is not, in general, equal to one the periodicity is lost, 
but we find using a glass is still considerably better than randomly distributing halo particles in the new volume.

{\ebf 
\section{Comparison of baryon free satellites and halos}\label{app:norms}

We compare here the mass and orbital evolution of a satellite with dark matter only and one with baryons and dark matter that orbits in the rotating reference frame without the external host coronae being present.
The baryon free model, NBF-$50,$ is created by evolving the same original dark matter potential without dwarfs and inserting it on the same orbit as NRPS-$50,$ which has a perigalacticon of $50$~kpc and an apogalacticon of $200$~kpc.
The tidal model, NTO-$50,$ is simply the NISO model inserted at $2$~Gyr on the same orbit of $50$~kpc perigalacticon and $200$~kpc apogalacticon, but with no Milky Way-like hot halo.
In both cases, the models were limited to a maximum step size of $\sim$$0.094$~Myr, and the step size GEAR quickly selects for the ram pressure simulations; all other parameters are the same as for the RPS cases.

These two models provide complementary tests.
NBF-$50$ ensures any orbital evolution is not due to dynamical friction or  errors in the particle creation/deletion, creating spurious gravitational attractions behind the dwarf.
NTO-$50$ checks that the mass evolution is not a by-product of the rotation scheme or only due to tidal influences.
Both models, without any form of hydrodynamic drag, also provide a test that the orbital evolution is not a by-product of the rotation scheme.

\subsection{Mass evolution}

We show the dark matter mass of the NTO-$50$ and NBF-$50$ models and NRPS-$50$ as a function of time in Fig. \ref{fig:testmass}.
Similar to \citet{Arraki2014}, the removal of baryons in NBF-$50$ produces a dramatic loss in dark matter density (and therefore mass within the $1$~kpc and $3$~kpc boundaries), although the mass loss is slightly under that found by \citeauthor{Arraki2014} possibly attributable to the instantaneous removal of the baryons here.
The tidal model NTO-$50$ retains the central mass density of NRPS-$50$, and tracks it closely until the first perigalacticon.

At the first perigalacticon, mass is lost from all models, however, it is dramatically more from NRPS-$50$ than NBF-$50$ or NTO-$50$, which lose $\sim$$30\%$--$50\%$ within $3$~kpc.
This is compared to NRPS-$50$ which because of the ram pressure  loses more than $70\%$.
By the next perigalacticon, the NRPS-$50$ model is destroyed into a stream, while both the NBF-$50$ and NTO-$50$ models survive until the present day.

That both models undergo tidal stripping but retain their general structure confirms that the ram pressure stripping of gas acts synergistically with tidal stripping to remove mass from the dwarf, including its dark matter mass.

\begin{figure}
\includegraphics[width=0.5\textwidth]{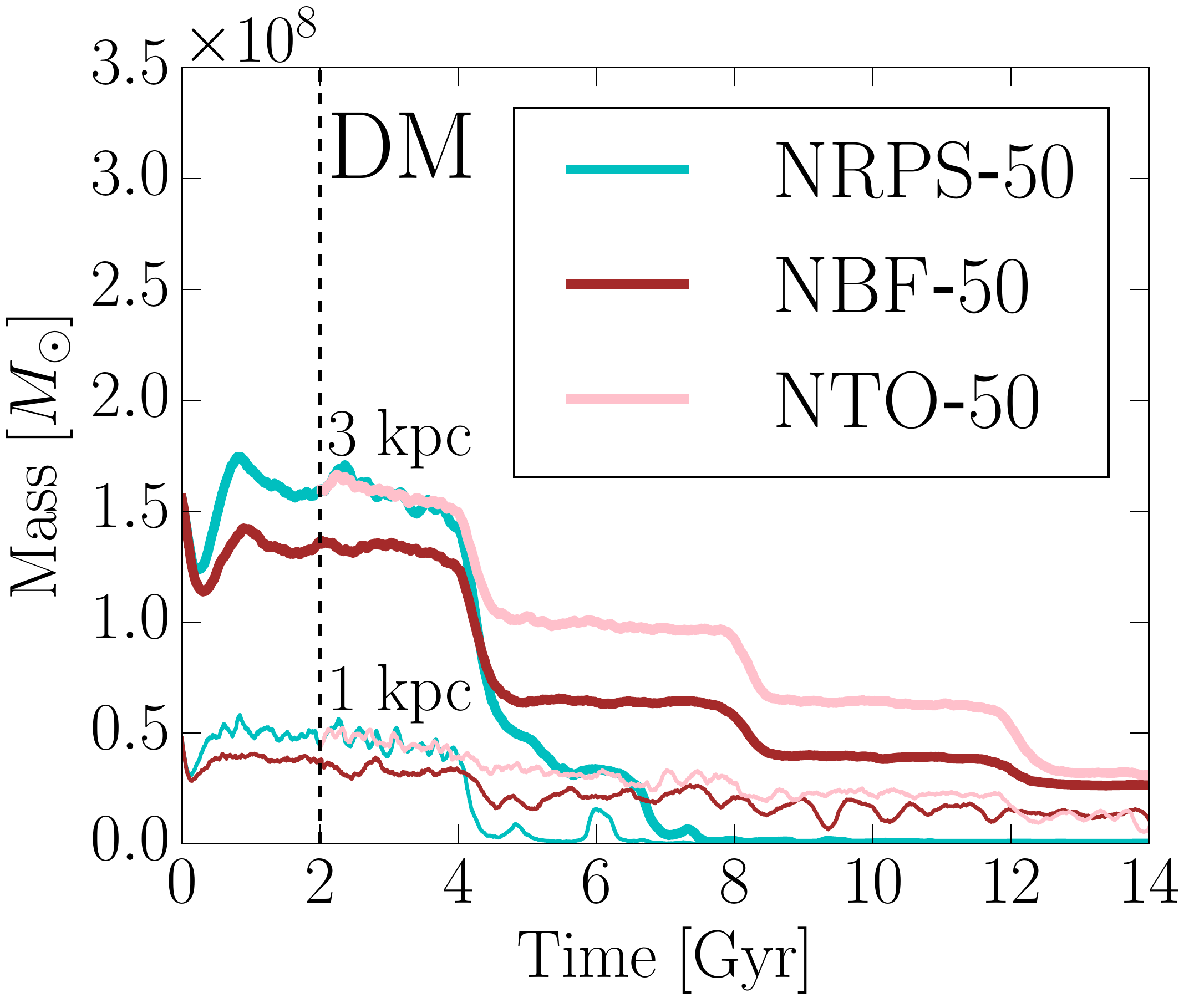}
\caption{Mass of dark matter within $1$~kpc (thin lines) and $3$~kpc (thick lines) for the NRPS-$50$ (cyan), NBF-$50$ (brown), and NTO-$50$ (pink) models.}\label{fig:testmass}
\end{figure}

\subsection{Orbital evolution}

We compare the orbital evolution (see \S\ref{subsec:orbchange}) between the three models in Fig. \ref{fig:testorbchange}.
Notably, neither NBF-$50$ or NTO-$50$ show a change in orbital energy, as these orbital energies are  always  approximately the equal to the  infall energy.
Similarly, their peri- and apogalacticons remain approximately constant.
NBF-$50$ does show a minor drop in apogalacticon of order $5$~kpc, which can be regarded as the error arising because of the halo generation method.
Notably, this is still much smaller than the $50$~kpc drop in apogalacticon NRPS-$50$ suffers at the first perigalacticon.
The constant nature of these peri- and apogalacticons increases the confidence that the drag experienced on the dwarf is due to a hydrodynamic interaction between the gas of a dwarf and the hot halo, which is then gravitationally coupled to the dark matter of the dwarf.

\begin{figure}
\includegraphics[width=0.5\textwidth]{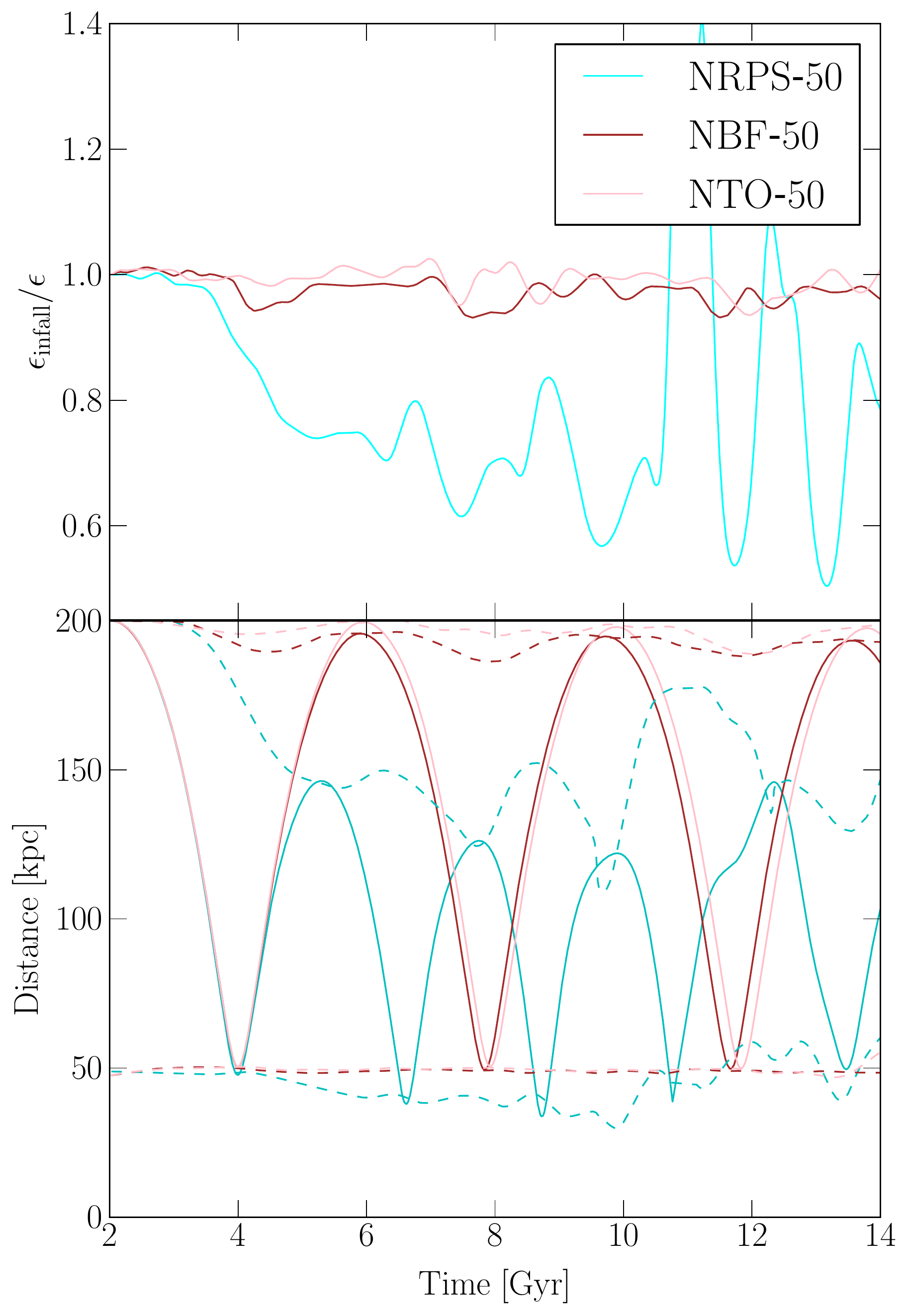}
\caption{Orbital evolution of NRPS-$50$ (cyan), NBF-$50$ (brown), and NTO-$50$ (pink). Line styles are as per Fig. \ref{fig:orbenergy}.}\label{fig:testorbchange}
\end{figure}

}

\section{Drag of a dwarf galaxy}\label{app:drag}
{\ebf

The drop in energy throughout a dwarf's orbit (Fig. \ref{fig:orbenergy}) and subsequent orbital change suggests that hydrodynamic drag is non-negligible when considering the evolution of a dwarf.
By comparing the energy change to the mass nearby the coefficient of drag of the galaxy can be calculated.
If the kinetic energy at time step $i$ is $E_{{\rm k},i}$ and potential energy $E_{{\rm p},i}$ then the coefficient of drag at the time step $i$ is given by
\begin{equation}
C_{{\rm d},i} = \frac{2\left(\sqrt{2E_{{\rm k},i}-2E_{{\rm p},i}}-\sqrt{2E_{{\rm k},i+1} }\right)}{\rho{}_iA_iv_i^2M_i(t_{i+1}-t_i)}
.\end{equation}

We consider the mass of the dwarf within $10$~kpc of the centre, and assume the energy change is simply the energy per unit mass of the centre multiplied by the mass.
The choice of area is much more ambiguous, consisting of a clumpy medium, and it is not clear what reference should be chosen as hot gas may stream through parts of the structure.
At each snapshot\ we select the total area of the dwarf, which exceeds a column density in the direction of motion of $n_{\rm H}=1\times10^{21}$~cm$^{-2}$ averaged over $0.25$~kpc$^{2}$ cells.
This boundary is not the only potential choice, but seems to correspond to the area of the dwarf through which hot gas cannot easily flow.

As the coefficient of drag generally depends on the Reynolds number of the flow, we calculate the Reynolds number\footnote{We note that the artificial viscosity affects the Reynolds number of the flow and is different between each particle pair. We ignore these differences and use the viscosity of the hot gas halo rather than the numerical viscosity.} at each time step as well, assuming the reference length is the same as the radius of a circle with equivalent area to the reference area.
We plot the mean coefficient of drag versus Reynolds number in Fig. \ref{fig:drag} for the NRPS-50 model for the first $1.5$~Gyr (450 snapshots) and the NRPS-100 model for $\sim$$8$~Gyr (2000 snapshots). This time selection avoids the recreation event in NRPS-50, which leads to significant errors and stream formation in NRPS-100.
We also show a fit to the coefficient of drag to a smooth sphere versus Reynolds number \citep[equation 7 of][]{Mikhailov2013} and see that although the coefficient of drag on the dwarf is much higher (as could be expected from the shape of the dwarf), it follows a similar trait of rapidly dropping until Re$\sim$$200$ where $C_{\rm d}$$\sim$$3$--$10$ before dropping down to $C_{\rm d}$$\sim$$1$.

\begin{figure}
\includegraphics[width=0.5\textwidth]{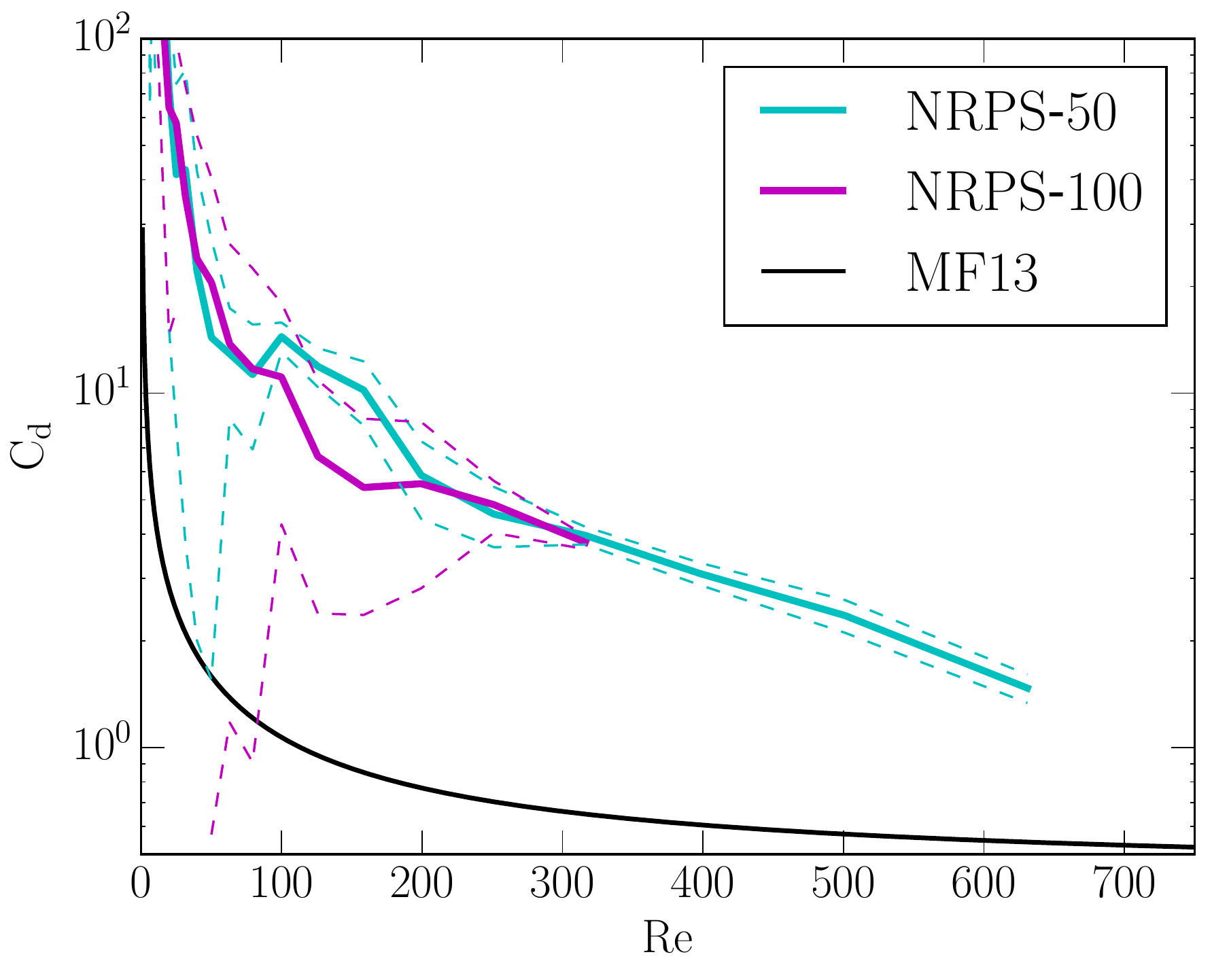}
\caption{Coefficient of drag of a dwarf galaxy throughout its orbit against the Reynolds number of the flow (calculated using the hot gas viscosity, not the numerical). The solid lines show the mean of the coefficient of drag for NRPS-$50$ (cyan) and NRPS-$100$ (magenta) within each (logarithmic) Reynolds number bin. Dashed lines show the standard deviation on each bin. A fit to the changing coefficient of drag of a sphere versus Reynolds number from \citet{Mikhailov2013} is shown with a black line.}
\label{fig:drag}
\end{figure}

The large amount of scatter at low Reynolds number is not in itself surprising, with the dwarf changing shape numerous times due to internal feedback, which is minimised at high Reynolds number because of the higher containing pressure.
This large degree of scatter (and the dependence of $C_{\rm d}$ on area) makes it difficult to approximate the orbital change in semi-analytic models.
}


\begin{thebibliography}{49}
\expandafter\ifx\csname natexlab\endcsname\relax\def\natexlab#1{#1}\fi

\bibitem[{{Agertz} {et~al.}(2007){Agertz}, {Moore}, {Stadel}, {Potter},
  {Miniati}, {Read}, {Mayer}, {Gawryszczak}, {Kravtsov}, {Nordlund}, {Pearce},
  {Quilis}, {Rudd}, {Springel}, {Stone}, {Tasker}, {Teyssier}, {Wadsley}, \&
  {Walder}}]{Agertz2007}
{Agertz}, O., {et~al.} 2007, \mnras, 380, 963

\bibitem[{{Aoki} {et~al.}(2009){Aoki}, {Arimoto}, {Sadakane}, {Tolstoy},
  {Battaglia}, {Jablonka}, {Shetrone}, {Letarte}, {Irwin}, {Hill}, {Francois},
  {Venn}, {Primas}, {Helmi}, {Kaufer}, {Tafelmeyer}, {Szeifert}, \&
  {Babusiaux}}]{Aoki2009}
{Aoki}, W., {et~al.} 2009, \aap, 502, 569

\bibitem[{{Arraki} {et~al.}(2014){Arraki}, {Klypin}, {More}, \&
  {Trujillo-Gomez}}]{Arraki2014}
{Arraki}, K.~S., {Klypin}, A., {More}, S., \& {Trujillo-Gomez}, S. 2014,
  \mnras, 438, 1466

\bibitem[{{Battaglia} {et~al.}(2011){Battaglia}, {Tolstoy}, {Helmi}, {Irwin},
  {Parisi}, {Hill}, \& {Jablonka}}]{Battaglia2011}
{Battaglia}, G., {Tolstoy}, E., {Helmi}, A., {Irwin}, M., {Parisi}, P., {Hill},
  V., \& {Jablonka}, P. 2011, \mnras, 411, 1013

\bibitem[{{Bland-Hawthorn} {et~al.}(2007){Bland-Hawthorn}, {Sutherland},
  {Agertz}, \& {Moore}}]{JBH2007}
{Bland-Hawthorn}, J., {Sutherland}, R., {Agertz}, O., \& {Moore}, B. 2007,
  \apjl, 670, L109

\bibitem[{{Boylan-Kolchin} {et~al.}(2011){Boylan-Kolchin}, {Bullock}, \&
  {Kaplinghat}}]{Boylan-Kolchin2011a}
{Boylan-Kolchin}, M., {Bullock}, J.~S., \& {Kaplinghat}, M. 2011, \mnras, 415,
  L40

\bibitem[{{Breddels} \& {Helmi}(2013)}]{Breddels2013}
{Breddels}, M.~A., \& {Helmi}, A. 2013, \aap, 558, A35

\bibitem[{{Chiboucas} {et~al.}(2013){Chiboucas}, {Jacobs}, {Tully}, \&
  {Karachentsev}}]{Chiboucas2013}
{Chiboucas}, K., {Jacobs}, B.~A., {Tully}, R.~B., \& {Karachentsev}, I.~D.
  2013, \aj, 146, 126

\bibitem[{{de Blok} {et~al.}(2008){de Blok}, {Walter}, {Brinks},
  {Trachternach}, {Oh}, \& {Kennicutt}}]{deBlok2008}
{de Blok}, W.~J.~G., {Walter}, F., {Brinks}, E., {Trachternach}, C., {Oh},
  S.-H., \& {Kennicutt}, Jr., R.~C. 2008, \aj, 136, 2648

\bibitem[{{Gatto} {et~al.}(2013){Gatto}, {Fraternali}, {Read}, {Marinacci},
  {Lux}, \& {Walch}}]{Gatto2013}
{Gatto}, A., {Fraternali}, F., {Read}, J.~I., {Marinacci}, F., {Lux}, H., \&
  {Walch}, S. 2013, \mnras

\bibitem[{{Governato} {et~al.}(2010){Governato}, {Brook}, {Mayer}, {Brooks},
  {Rhee}, {Wadsley}, {Jonsson}, {Willman}, {Stinson}, {Quinn}, \&
  {Madau}}]{Governato2010}
{Governato}, F., {et~al.} 2010, \nat, 463, 203

\bibitem[{{Grcevich} \& {Putman}(2009)}]{Grcevich2009}
{Grcevich}, J., \& {Putman}, M.~E. 2009, \apj, 696, 385

\bibitem[{{Hopkins}(2013)}]{Hopkins2013}
{Hopkins}, P.~F. 2013, \mnras, 428, 2840

\bibitem[{{Huber} \& {Pfenniger}(2001)}]{Huber2001}
{Huber}, D., \& {Pfenniger}, D. 2001, \aap, 374, 465

\bibitem[{{Kafle} {et~al.}(2014){Kafle}, {Sharma}, {Lewis}, \&
  {Bland-Hawthorn}}]{Kafle2014}
{Kafle}, P.~R., {Sharma}, S., {Lewis}, G.~F., \& {Bland-Hawthorn}, J. 2014,
  \apj, 794, 59

\bibitem[{{Kazantzidis} {et~al.}(2013){Kazantzidis}, {{\L}okas}, \&
  {Mayer}}]{Kazantzidis2013}
{Kazantzidis}, S., {{\L}okas}, E.~L., \& {Mayer}, L. 2013, \apjl, 764, L29

\bibitem[{{Kirby} {et~al.}(2011){Kirby}, {Cohen}, {Smith}, {Majewski}, {Sohn},
  \& {Guhathakurta}}]{Kirby2011}
{Kirby}, E.~N., {Cohen}, J.~G., {Smith}, G.~H., {Majewski}, S.~R., {Sohn},
  S.~T., \& {Guhathakurta}, P. 2011, \apj, 727, 79

\bibitem[{{Klypin} {et~al.}(1999){Klypin}, {Kravtsov}, {Valenzuela}, \&
  {Prada}}]{Klypin1999}
{Klypin}, A., {Kravtsov}, A.~V., {Valenzuela}, O., \& {Prada}, F. 1999, \apj,
  522, 82

\bibitem[{{Kroupa}(2001)}]{Kroupa2001}
{Kroupa}, P. 2001, \mnras, 322, 231

\bibitem[{{Lux} {et~al.}(2010){Lux}, {Read}, \& {Lake}}]{Lux2010}
{Lux}, H., {Read}, J.~I., \& {Lake}, G. 2010, \mnras, 406, 2312

\bibitem[{{Maio} {et~al.}(2007){Maio}, {Dolag}, {Ciardi}, \&
  {Tornatore}}]{Maio2007}
{Maio}, U., {Dolag}, K., {Ciardi}, B., \& {Tornatore}, L. 2007, \mnras, 379,
  963

\bibitem[{{Mayer} {et~al.}(2006){Mayer}, {Mastropietro}, {Wadsley}, {Stadel},
  \& {Moore}}]{Mayer2006}
{Mayer}, L., {Mastropietro}, C., {Wadsley}, J., {Stadel}, J., \& {Moore}, B.
  2006, \mnras, 369, 1021

\bibitem[{Mikhailov \& Freire(2013)}]{Mikhailov2013}
Mikhailov, M., \& Freire, A.~S. 2013, Powder Technology, 237, 432

\bibitem[{{Miller} \& {Bregman}(2013)}]{Miller2013}
{Miller}, M.~J., \& {Bregman}, J.~N. 2013, \apj, 770, 118

\bibitem[{{Moore} {et~al.}(1999){Moore}, {Ghigna}, {Governato}, {Lake},
  {Quinn}, {Stadel}, \& {Tozzi}}]{Moore1999}
{Moore}, B., {Ghigna}, S., {Governato}, F., {Lake}, G., {Quinn}, T., {Stadel},
  J., \& {Tozzi}, P. 1999, \apjl, 524, L19

\bibitem[{{Navarro} {et~al.}(1997){Navarro}, {Frenk}, \& {White}}]{Navarro1997}
{Navarro}, J.~F., {Frenk}, C.~S., \& {White}, S.~D.~M. 1997, \apj, 490, 493

\bibitem[{{Nichols} \& {Bland-Hawthorn}(2011)}]{Nichols2011}
{Nichols}, M., \& {Bland-Hawthorn}, J. 2011, \apj, 732, 17

\bibitem[{{Nichols} \& {Bland-Hawthorn}(2013)}]{Nichols2013}
---. 2013, \apj, 775, 97

\bibitem[{{Nichols} {et~al.}(2012){Nichols}, {Lin}, \&
  {Bland-Hawthorn}}]{Nichols2012}
{Nichols}, M., {Lin}, D., \& {Bland-Hawthorn}, J. 2012, \apj, 748, 149

\bibitem[{{Nichols} {et~al.}(2014){Nichols}, {Revaz}, \&
  {Jablonka}}]{Nichols2014}
{Nichols}, M., {Revaz}, Y., \& {Jablonka}, P. 2014, \aap, 564, A112

\bibitem[{{Oh} {et~al.}(2015){Oh}, {Hunter}, {Brinks}, {Elmegreen}, {Schruba},
  {Walter}, {Rupen}, {Young}, {Simpson}, {Johnson}, {Herrmann}, {Ficut-Vicas},
  {Cigan}, {Heesen}, {Ashley}, \& {Zhang}}]{Oh2015}
{Oh}, S.-H., {et~al.} 2015, \aj, 149, 180

\bibitem[{{Ploeckinger} {et~al.}(2014){Ploeckinger}, {Hensler}, {Recchi},
  {Mitchell}, \& {Kroupa}}]{Ploeckinger2014}
{Ploeckinger}, S., {Hensler}, G., {Recchi}, S., {Mitchell}, N., \& {Kroupa}, P.
  2014, \mnras, 437, 3980

\bibitem[{{Pontzen} \& {Governato}(2012)}]{Pontzen2012}
{Pontzen}, A., \& {Governato}, F. 2012, \mnras, 421, 3464

\bibitem[{{Revaz} {et~al.}(2015){Revaz}, {Arnaudon}, {Bonvin}, {Nichols}, \&
  {Jablonka}}]{Revaz2015}
{Revaz}, Y., {Arnaudon}, A., {Bonvin}, V., {Nichols}, M., \& {Jablonka}, P.
  2015, in prep

\bibitem[{{Revaz} \& {Jablonka}(2012)}]{Revaz2012}
{Revaz}, Y., \& {Jablonka}, P. 2012, \aap, 538, A82

\bibitem[{{Rocha} {et~al.}(2012){Rocha}, {Peter}, \& {Bullock}}]{Rocha2012}
{Rocha}, M., {Peter}, A.~H.~G., \& {Bullock}, J. 2012, \mnras, 425, 231

\bibitem[{{Roediger} {et~al.}(2015){Roediger}, {Kraft}, {Nulsen}, {Forman},
  {Machacek}, {Randall}, {Jones}, {Churazov}, \& {Kokotanekova}}]{Roediger2015}
{Roediger}, E., {et~al.} 2015, \apj, 806, 104

\bibitem[{{Sawala} {et~al.}(2012){Sawala}, {Scannapieco}, \&
  {White}}]{Sawala2012}
{Sawala}, T., {Scannapieco}, C., \& {White}, S. 2012, \mnras, 420, 1714

\bibitem[{{Shetrone} {et~al.}(2001){Shetrone}, {C{\^o}t{\'e}}, \&
  {Sargent}}]{Shetrone2001}
{Shetrone}, M.~D., {C{\^o}t{\'e}}, P., \& {Sargent}, W.~L.~W. 2001, \apj, 548,
  592

\bibitem[{{Smith} {et~al.}(2012){Smith}, {Fellhauer}, \& {Assmann}}]{Smith2012}
{Smith}, R., {Fellhauer}, M., \& {Assmann}, P. 2012, \mnras, 420, 1990

\bibitem[{{Springel}(2005)}]{Springel2005a}
{Springel}, V. 2005, \mnras, 364, 1105

\bibitem[{{Sutherland} \& {Dopita}(1993)}]{Sutherland1993}
{Sutherland}, R.~S., \& {Dopita}, M.~A. 1993, \apjs, 88, 253

\bibitem[{{Tafelmeyer} {et~al.}(2010){Tafelmeyer}, {Jablonka}, {Hill},
  {Shetrone}, {Tolstoy}, {Irwin}, {Battaglia}, {Helmi}, {Starkenburg}, {Venn},
  {Abel}, {Francois}, {Kaufer}, {North}, {Primas}, \&
  {Szeifert}}]{Tafelmeyer2010}
{Tafelmeyer}, M., {et~al.} 2010, \aap, 524, A58

\bibitem[{{Theler} {et~al.}(2015){Theler}, {Theler}, {Theler}, {Theler},
  {Theler}, {Theler}, {Theler}, {Theler}, {Theler}, {Theler}, {Theler},
  {Theler}, {Theler}, {Theler}, \& {Theler}}]{Theler2015}
{Theler}, R., {et~al.} 2015, in prep

\bibitem[{{Torrey} {et~al.}(2015){Torrey}, {Snyder}, {Vogelsberger}, {Hayward},
  {Genel}, {Sijacki}, {Springel}, {Hernquist}, {Nelson}, {Kriek}, {Pillepich},
  {Sales}, \& {McBride}}]{Torrey2015}
{Torrey}, P., {et~al.} 2015, \mnras, 447, 2753

\bibitem[{{Walker} \& {Pe{\~n}arrubia}(2011)}]{Walker2011}
{Walker}, M.~G., \& {Pe{\~n}arrubia}, J. 2011, \apj, 742, 20

\bibitem[{{Wiersma} {et~al.}(2009){Wiersma}, {Schaye}, {Theuns}, {Dalla
  Vecchia}, \& {Tornatore}}]{Wiersma2009}
{Wiersma}, R.~P.~C., {Schaye}, J., {Theuns}, T., {Dalla Vecchia}, C., \&
  {Tornatore}, L. 2009, \mnras, 399, 574

\bibitem[{{Xue} {et~al.}(2008){Xue}, {Rix}, {Zhao}, {Re Fiorentin}, {Naab},
  {Steinmetz}, {van den Bosch}, {Beers}, {Lee}, {Bell}, {Rockosi}, {Yanny},
  {Newberg}, {Wilhelm}, {Kang}, {Smith}, \& {Schneider}}]{Xue2008}
{Xue}, X.~X., {et~al.} 2008, \apj, 684, 1143

\bibitem[{{Zolotov} {et~al.}(2012){Zolotov}, {Brooks}, {Willman}, {Governato},
  {Pontzen}, {Christensen}, {Dekel}, {Quinn}, {Shen}, \&
  {Wadsley}}]{Zolotov2012}
{Zolotov}, A., {et~al.} 2012, \apj, 761, 71

\end{thebibliography}
\end{document}